\documentclass{jpp}
\usepackage{wrapfig}
\usepackage{amsmath}
\usepackage{amsbsy}
\usepackage{amssymb}
\usepackage{graphicx}
\usepackage{natbib}
\defcitealias{ITER99_c2}{ITER Physics Basis 1999}
\usepackage[colorlinks=true, citecolor=blue,linkcolor=blue]{hyperref}

\usepackage[utf8]{inputenc}
\usepackage[T1]{fontenc}
\usepackage{cleveref}
\usepackage{cancel}
\usepackage{xcolor} 
\usepackage{array}

\newcommand{\mcP}{\mathcal{P}}

\newcommand{\bx}{\boldsymbol{x}}

\newcommand{\bu}{\boldsymbol{u}}
\newcommand{\bv}{\boldsymbol{v}}

\newcommand{\bQ}{\boldsymbol{Q}}

\newcommand{\mcE}{\mathcal{E}}
\newcommand{\const}{\mathrm{const}}

\shorttitle{Metastability of stratified MHD equilibria and their relaxation}
\shortauthor{D. N. Hosking, D. Wasserman and S.~C. Cowley}
\title{Metastability of stratified magnetohydrostatic equilibria and their relaxation}

\author{D.~N. Hosking\aff{1,}\aff{2}\corresp{\email{dhosking@princeton.edu}}, D.~Wasserman\aff{3} \and S.~C. Cowley\aff{4}}

\affiliation{\aff{1}Princeton Center for Theoretical Science, Princeton, NJ 08540, USA
\aff{2}Gonville \& Caius College, Trinity Street, Cambridge, CB2 1TA, UK
\aff{3}Northeastern University, Boston, MA 02115, USA
\aff{4}Princeton Plasma Physics Laboratory, Princeton, NJ 08540, USA}

\begin{document}

\maketitle

\begin{abstract}
Motivated by explosive releases of energy in fusion, space and astrophysical plasmas, we consider the nonlinear stability of stratified magnetohydrodynamic (MHD) equilibria against two-dimensional interchanges of straight magnetic-flux tubes. We demonstrate that, even within this restricted class of dynamics, the linear stability of an equilibrium does not guarantee its nonlinear stability: equilibria can be \textit{metastable}. We show that the minimum-energy state accessible to a metastable equilibrium under non-diffusive 2D dynamics can be found by solving a combinatorial optimisation problem. These minimum-energy states are, to good approximation, the final states reached by our simulations of destabilised metastable equilibria for which turbulent mixing is suppressed by viscosity. To predict the result of fully turbulent relaxation, we construct a statistical mechanical theory based on the maximisation of Boltzmann's mixing entropy. This theory is analogous to the Lynden-Bell statistical mechanics of collisionless stellar systems and plasma, and to the Robert-Sommeria-Miller (RSM) theory of 2D vortex turbulence. Our theory reproduces well the results of our numerical simulations for sufficiently large perturbations to the metastable equilibrium.
\end{abstract}

\section{Introduction\label{sec:introduction}}

Both in the laboratory and in nature, plasmas that host strong magnetic fields sometimes exist in slowly evolving quasi-equilibrium states. Such plasmas may have a dynamical timescale ${\tau_A \sim L/v_A}$ ($L$ is the system's length scale and $v_A$ the Alfv\'{e}n speed) that is much smaller than the timescales associated with energy injection and transport. In the solar corona, for example, the footpoints of magnetic-flux loops evolve on the photospheric driving timescale of~${\sim10~\mathrm{min}}$, while $\tau_A\sim 10~\mathrm{s}$~\citep{CranmerWinebarger19}. Likewise, the transport timescale for magnetic-confinement-fusion devices is typically~$\sim 0.1 \,\mathrm{s}$, much larger than $\tau_A\sim 10^{-6} \,\mathrm{s}$~\citepalias{ITER99_c2}. Occasionally, these plasmas depart suddenly and violently from their quasi-equilibria. Explosive releases of energy---\textit{eruptions}---involving substantial reconfiguration of the magnetic field happen both in the corona [coronal mass ejections; see, e.g.,~\citet{PFChen11} for a review] and fusion experiments [disruption events; see, e.g.,~\citet{ITER07_c3}]. Evidently, the quasi-equilibria can become unstable during their evolution. When eruptions occur, the system relaxes towards a new state that is a lower minimum of the potential energy in configuration space. For a significant amount of potential energy to be liberated, the new minimum must be distant from the original one. The instability must therefore be nonlinear: eruptions happen when \textit{metastable} states are destabilised.

Ideal magnetohydrodynamic (MHD) instabilities may be categorised into two types: kink and interchange (or ballooning) instabilities. Kink instabilities are global, occurring at the system scale, and are characterised by significant variation along magnetic field lines ($k_{\parallel}\sim k_{\perp}$, where $k_{\parallel}$ and $k_{\perp}$ are characteristic wavenumbers along and across the magnetic field, respectively). In contrast, interchange instabilities~\citep{ConnorHastieTaylor79} are local and scale-independent, with elongation along magnetic field lines ($k_{\parallel} \ll k_{\perp}$). In most known cases, a supercritical bifurcation occurs when an MHD equilibrium crosses the linear threshold for kink instability; two new stable equilibria that are nearby in configuration space become realisable~\citep{Friedrichs60, Rutherford71, White77, Lorenzini09}. Because these equilibria are nearby, no significant release of potential energy is possible if the system is pushed out of one and into the other. On the other hand, \textit{subcritical} bifurcation is possible at the linear threshold for interchange instability. In this case, the system is nonlinearly unstable at marginal linear stability. Any stable equilibrium to which the system can relax is distant in configuration space, meaning that a finite amount of potential energy can be liberated. 

Previous studies have elucidated certain properties of the relaxation of MHD equilibria from states that are metastable to interchange-type dynamics. \citet{CowleyArtun97} studied the case of a stratified equilibrium with initially horizontal magnetic field embedded in conducting walls (fixed field-line endpoints). They showed that gradients of thermal or magnetic pressure (balanced in equilibrium by gravity) can provide sources of free energy for a buoyancy instability that is stabilised by magnetic tension only for linear perturbations---not nonlinear ones. By solving the weakly nonlinear equations of motion numerically, \citet{CowleyArtun97} observed a phenomenon that they termed \textit{detonation}: progressive destabilisation of the metastable equilibrium by erupting finger-like magnetic structures. These results were extended to more general geometry by \citet{Hurricane97}, \citet{WilsonCowley04} and~\citet{Ham18}. \citet{CowleyCowley15} showed that, with fast thermal conduction along field lines, erupting flux tubes have two possible fates: either they find a new equilibrium position or they reach a singular state with zero magnetic-field strength (\textit{flux expulsion}).

Despite the successes of these studies, certain fundamental questions remain difficult to answer accurately because of the geometrical complexity that arises from the bending of magnetic field lines. Such questions include: To what state does a metastable equilibrium relax when it is destabilised? What fraction of its energy is available for liberation? Is that energy always liberated in practice, i.e., is relaxation complete?

\begin{figure*}
    \centering
    \includegraphics[width=.9\textwidth]{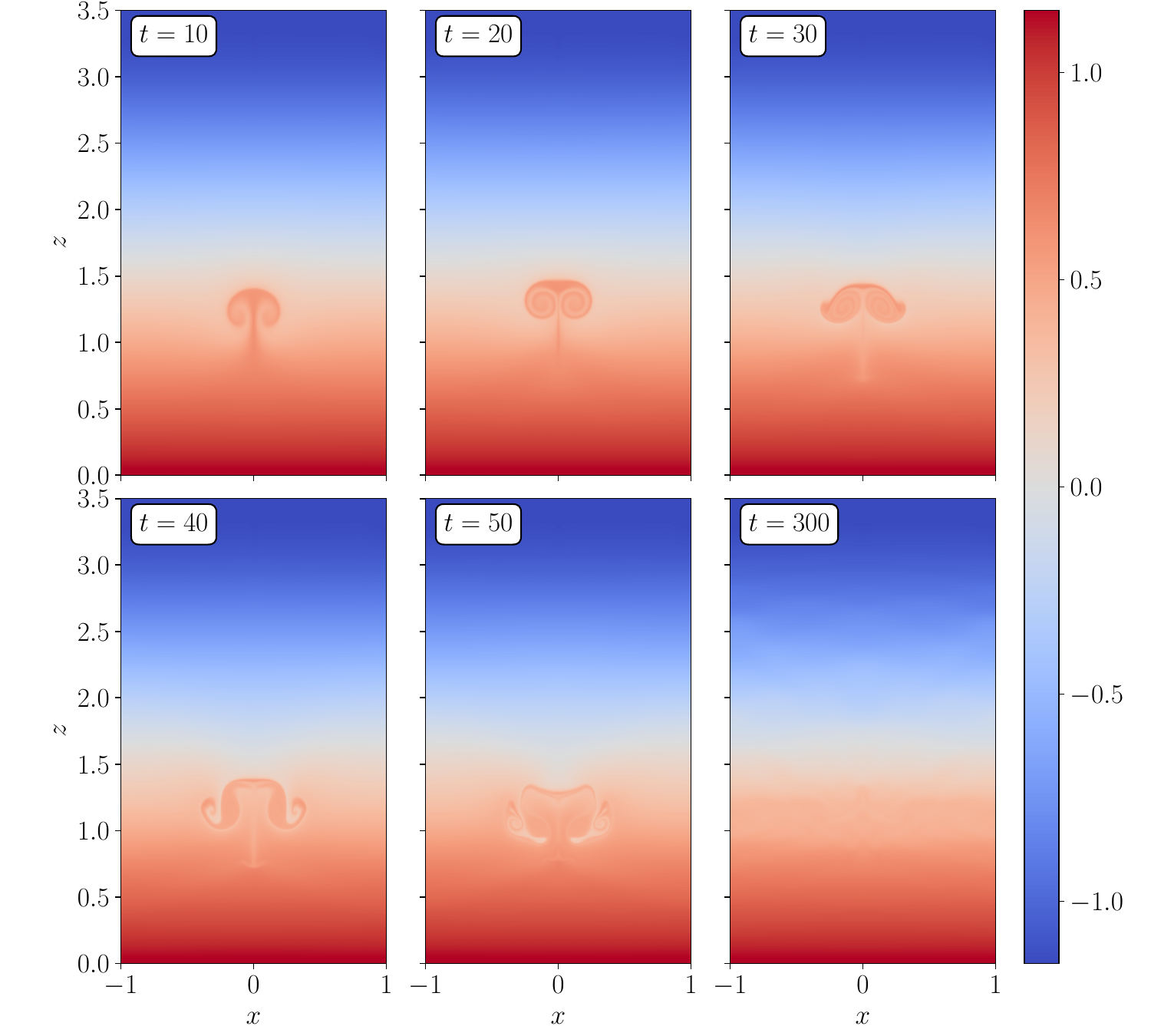}
    \caption{2D simulation of an MHD atmosphere subjected to an impulse that does not trigger instability. The atmosphere relaxes to a final state that approximates the initial condition. The quantity visualised is the natural logarithm of the ratio of the entropy function \eqref{entropyfn} to the specific magnetic flux \eqref{specificmagflux}. This quantity is conserved in a Lagrangian sense in the absence of diffusion and controls the compressibility of the fluid, with larger values being more compressible (see Section~\ref{sec:metastabilityMHD}). The initial velocity field is ${\bu = u_0 \boldsymbol{\hat{z}}\exp(-[x^2+(z-1.0)^2]/0.1^2)}$ with $u_0=0.2$. The equilibrium is defined by~\eqref{m_profile} with $\epsilon_0=10^{-2}$ in~\eqref{epsilon_buffer}. The co-ordinates $x$ and $z$ are measured in units of the total-pressure scale height at $z=0$ and the time $t$ is measured in units of the sound-crossing time of the total-pressure scale height at $z=0$ (see Section~\ref{sec:examples} for details). A movie version of this figure is available in the published version of this article.}
    \label{fig:fig1}
\end{figure*}

\begin{figure*}
    \centering
    \includegraphics[width=.9\textwidth]{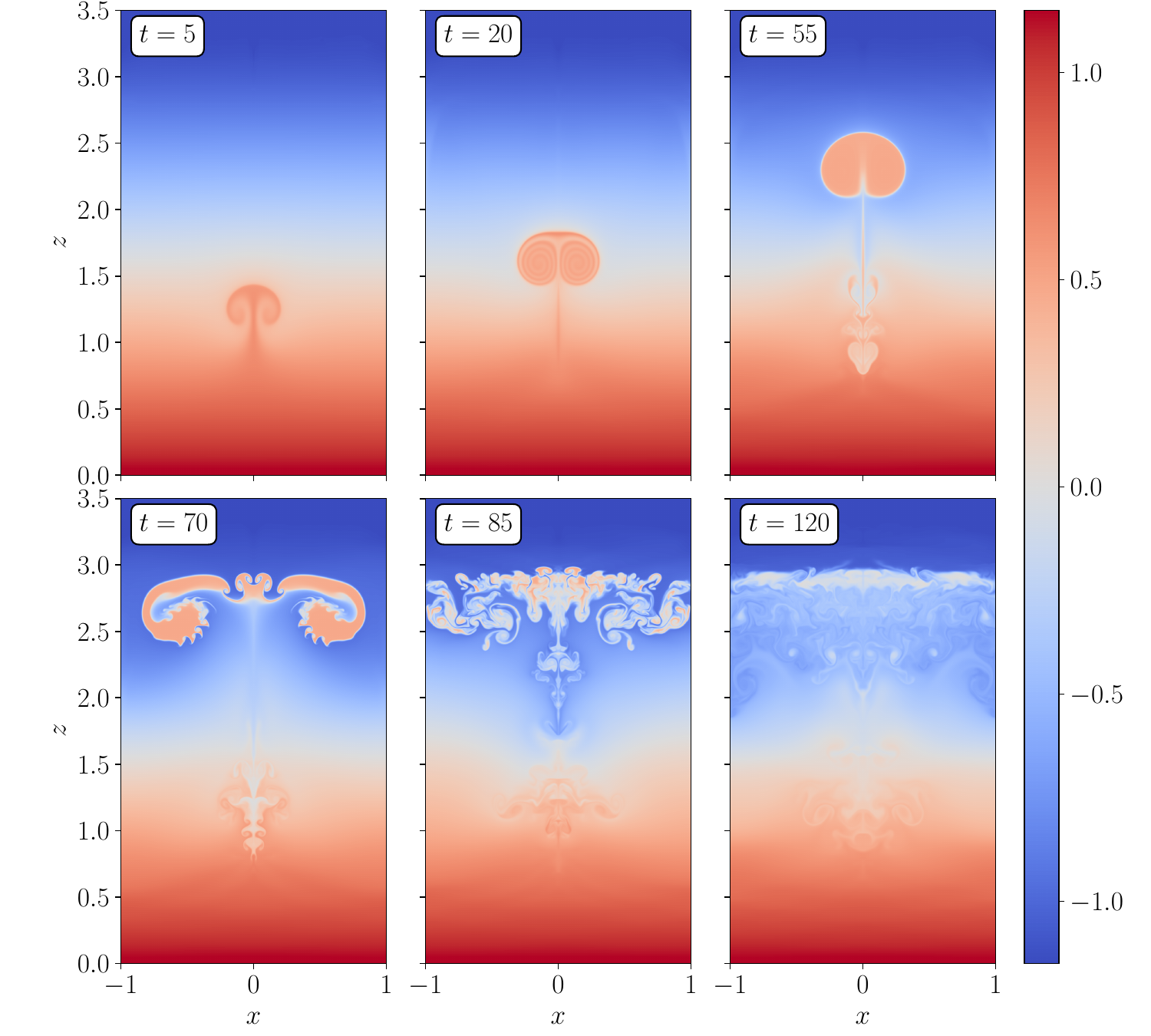}
    \caption{As figure~\ref{fig:fig1}, but for an initial velocity field that is twice as large ($u_0=0.4$). The equilibrium is destabilised. A movie version of this figure is available in the published version of this article.}
    \label{fig:fig2}
\end{figure*}

The aim of this paper is to consider metastability in a simpler setting where these questions can be answered. A key result (which, to our knowledge, has not been recognised previously) is that bending of magnetic field lines, while certainly a feature of eruptions in the corona and fusion devices, is not a requirement for the existence of nonlinear instability. We demonstrate this fact in figures~\ref{fig:fig1} and~\ref{fig:fig2}, which visualise numerical simulations that are 2D with out-of-plane magnetic field only. In figure~\ref{fig:fig1}, a small perturbation to a stratified equilibrium or ``atmosphere'' (gravity is in the negative-$z$ direction) does not lead to destabilisation: the original state is restored. In figure~\ref{fig:fig2}, a larger perturbation destabilises the same  equilibrium: a rising plume of material does not return to its original position, but establishes a new equilibrium state (see figure captions for further information about the physical setup and Appendix~\ref{app:numerical} for details of numerical methods). 

In the 2D context (upon which we focus exclusively in this paper) metastability is enabled by the fact that less magnetised fluid (i.e., fluid with a larger ratio~$\beta$ of the thermal to magnetic pressures) is more compressible than fluid that is more magnetised (smaller~$\beta$).  A flux tube may therefore experience a greater change in density in response to a large (nonlinear) displacement than the background density profile does over the same distance, provided its $\beta$ is larger than that of the surrounding flux tubes. This produces a destabilising buoyancy force, even though the equilibrium may be stable linearly. In fact, a precisely analogous phenomenon occurs in the Earth's atmosphere. This is because air that is saturated (i.e., sufficiently cool for water vapour to condense) is more compressible than unsaturated air. As a result, the atmosphere can be linearly stable to convection but unstable nonlinearly to displacement of saturated air through the so-called \textit{level of free convection}. The resulting updraughts lead to the formation of cumulonimbus clouds and, as a result, thunderstorms \citep[see, e.g.,][]{RogersYau96}. We present a general theoretical treatment of metastability due to composition-dependent compressibility in Section~\ref{sec:metastability_theory} and review its application to the terrestrial atmosphere in Appendix~\ref{app:moist}.

The rest of our paper is concerned with the question of how metastable MHD equilibria relax when they are destabilised. In Section~\ref{sec:ground_state_theory}, we show that the available energy under nonlinear interchanges of flux tubes may be determined accurately by solving a combinatorial optimisation problem (linear sum assignment)---although we believe that we are the first to use this approach for a magnetised system, we note that a similar one has been employed recently by \citet{Hieronymus15}, \citet{Su16} and \citet{Stansifer17} to calculate available potential energy in atmospheric and oceanic contexts. We show in Section~\ref{sec:HungarianResults} that the states that are global minima of the potential energy typically turn out to have horizontal structure over a finite range of the vertical coordinate $z$, even if the initial configurations are one-dimensional in $z$. This appears to be a new observation. We further find that the amount of energy that can be liberated from a metastable MHD equilibrium in 2D is always a small fraction (at most a few percent) of its total potential energy (Section~\ref{sec:smallE}). The reason is that flux tubes exclude each other: when a flux tube moves to the top or bottom of an atmosphere, it prevents others from doing the same.

The smallness of the available potential energy means that the kinetic energy that develops during relaxation is small compared with the internal energy of the fluid. Hence, relaxation is subsonic (for the vast majority of fluid). This has two important consequences. First, the decay of kinetic energy due to viscosity results in negligible heating, so that the thermal entropy of the fluid is unchanged during relaxation, provided that thermal and magnetic diffusion can be neglected. A corollary is that if viscosity is sufficiently large to suppress turbulent mixing, i.e., the Reynolds number is sufficiently small (we estimate precisely how small in Section~\ref{sec:conditions}), then relaxation terminates in a state close to the one with minimum energy with respect to ideal rearrangements. We confirm this expectation with direct numerical simulations in Section~\ref{sec:LowReNumerics}.

A second consequence of relaxation being subsonic is that fluid parcels maintain local pressure balance during relaxation: the fractional variation in pressure at fixed height is small compared with that of density. We utilise this fact in Section~\ref{sec:LyndenBell} to construct a statistical mechanical theory of turbulent relaxation at large Reynolds number. We conjecture that the state that develops as a result of turbulent mixing is the ``most mixed'' (in the sense of maximising the Boltzmann mixing entropy with respect to non-diffusive rearrangements of flux tubes) locally pressure-balanced state with the given total energy (Section~\ref{sec:LyndenBellEquations}). This idea is analogous to the \citet{LyndenBell67} statistical mechanics of collisionless stellar systems and plasma, and to the Robert-Sommeria-Miller (RSM) theory of 2D vortex turbulence~\citep{Miller90, Miller92, Robert91, RobertSommeria91}, which has found widespread application in geophysical fluid dynamics (see \citet{SinghONeill22} for a recent review). In our theory, the specific entropy $s$ and magnetic flux $\chi$, each of which is conserved in a fluid-element-wise sense under non-dissipative dynamics, play the role of the phase-space density or vorticity in the Lynden-Bell or RSM theories, respectively. 

Of course, neglect of diffusion is valid only until such time as the turbulent mixing develops sufficiently fine-scale structure in $s$ and $\chi$. In the Lynden-Bell (RSM) theory, diffusion due to collisions (viscosity) is straightforwardly modelled: when diffusion acts, the statistical mechanical probability distribution function for phase-space density (vorticity) collapses onto its expectation value. Our case is different: magnetic diffusion and thermal conduction increase the total thermal entropy (by a much larger amount than viscous heating does). Thus, we extract predictions for diffused states by assuming that our probability distribution function collapses not onto its expectation value but onto the value consistent with flux and energy conservation (Section~\ref{sec:diffusion}). As it turns out, the resulting 1D equilibrium state may itself be unstable, linearly or nonlinearly, because diffusion alters buoyancy.\footnote{A well-known example is the phenomenon of ``buoyancy reversal'' in the terrestrial atmosphere: when buoyantly rising dry air mixes with moist air, it may, after diffusion, become more dense than the unmixed moist air and sink again as a result [see, e.g., \citet{Stevens05}].} We consider taking the predicted unstable diffused state to be the starting point for a new relaxation (if necessary, this procedure can be repeated until a stable diffused state is reached, although we find that one iteration is often enough to do so). The profiles of $s$ and $\chi$ produced by this procedure are typically only slightly modified from the ones derived from the statistical mechanics in the first instance---qualitatively, the subsequent rearrangements and accompanied diffusion tend to produce local plateaus in the profiles of $s$ and $\chi$.

In Section~\ref{sec:numerics2}, we compare our theoretical predictions with the results of direct numerical simulations of relaxing metastable equilibria at large Reynolds number. The agreement turns out to be reasonably good, although less so for small initial perturbations. The reason for this appears to be that the detonation is incomplete in such cases: the system becomes trapped in a new metastable state and does not mix thoroughly. We conclude with a discussion of the possible implications and applications of our study in Section~\ref{sec:conclusion}.

\section{Theory of convective metastability\label{sec:metastability_theory}}

\subsection{Definitions}

In this paper, we shall be concerned with a fluid dynamics defined by the momentum equation
\begin{equation}
    \rho \frac{\dd \bu }{\dd t} = -\bnabla P - \rho g \hat{\boldsymbol{z}},\label{momentum}
\end{equation}where $\rho$ is density, $\bu$ fluid velocity, $t$ time, $P$ total pressure, $g$ the constant gravitational acceleration and $\dd/\dd t\equiv \p /\p t + \bu\bcdot \bnabla$ the material derivative. Density evolves according to the continuity equation,
\begin{equation}
    \frac{\dd\rho}{\dd t}=-\rho \bnabla \bcdot \bu,
\end{equation}and is related to $P$ by the equation of state
\begin{equation}
    \rho = \rho(P,\bQ)\implies P=P(\rho,\bQ)\label{eos}.
\end{equation}The vector $\bQ$ encodes the conserved material properties of the fluid (i.e., its Lagrangian invariants), which we allow to vary spatially. We shall primarily be concerned with the case of 2D MHD with out-of-plane magnetic field, for which the relevant components of~$\bQ$ are the specific entropy $s$ and specific magnetic flux $\chi\equiv B/\rho$, where $B$ is the magnetic-field strength; we shall specialise to this case in Section~\ref{sec:metastabilityMHD}. In different contexts, the components of $\bQ$ might instead (or additionally) include mixing ratios (e.g., salinity or specific humidity) or the entropies of a coupled radiation field or cosmic rays. Over sufficiently small time scales and at sufficiently large spatial scales that diffusion can be neglected, $\bQ$ is conserved following fluid particles, i.e.,
\begin{equation}
    \frac{\dd \bQ}{\dd t}=0.\label{adiabatic}
\end{equation}By neglecting diffusion, we exclude double-diffusive buoyancy instabilities [see, e.g., \citet{Garaud18, HughesBrummell21}, and references therein] from our analysis. We likewise exclude instabilities relating to anisotropic thermal conductivity \citep{Balbus00, Quataert08}. The application of our methods to the saturation of such instabilities is a topic to which we plan to return in future work (see the discussion in Section~\ref{sec:discussion}).

\subsection{Stability analysis\label{sec:generalEqState}}

We now consider the stability under equations~\eqref{momentum}-\eqref{adiabatic} of a one-dimensional static equilibrium state, i.e., one for which $\bu = 0$ and $P$, $\rho$ and $\bQ$ depend on $z$ only, with $\dd P/\dd z = -\rho g$.  
The net force $\boldsymbol{F}$ on a small parcel of fluid moved in pressure balance with its surroundings and without diffusion [i.e., satisfying~\eqref{adiabatic}] from initial height $z_1$ to new height $z_2$ is
\begin{equation}
    \boldsymbol{F} = -gV_2\hat{\boldsymbol{z}}\left[\rho(P_2,\bQ_1) - \rho(P_2,\bQ_2)\right],\label{F2pressure}
\end{equation}where $V_2$ is the volume of the parcel at $z_2$ and ${P_2 = P(z_2)}$, etc. Writing the difference in densities as an integral in $z$, we obtain
\begin{equation}
    \boldsymbol{F} = gV_2\hat{\boldsymbol{z}}\int^{z_2}_{z_1}\dd z \frac{\dd \bQ}{\dd z} \bcdot \frac{\p \rho(P_2,\bQ)}{\p\bQ}\label{F2pressure2}.
\end{equation}

\subsubsection{Linear stability}

The criterion for linear stability follows from taking ${\delta z \equiv z_2 -z_1\to 0}$ in~\eqref{F2pressure2}:
\begin{equation}
    \mathcal{L}\equiv -\frac{\dd \bQ}{\dd z} \bcdot \frac{\p \ln \rho(P,\bQ)}{\p\bQ}>0,\quad \forall z.\label{linearstability}
\end{equation}The function $\mathcal{L}$ may alternatively be written as
\begin{equation}
    \mathcal{L} = -\frac{\dd \ln \rho}{\dd z} + \frac{\p \ln \rho (P,\bQ)}{\p \ln P}\frac{\dd \ln P}{\dd z},\label{linearstability_alternative}
\end{equation}from which $\mathcal{L}>0$ is readily interpreted as the condition for the density of a fluid parcel displaced upwards (downwards) infinitesimally while conserving $\bQ$ to be greater (less) than the density of the background fluid in the new position of the parcel.

\subsubsection{Nonlinear stability}

We can use~\eqref{F2pressure2} to write the criterion for nonlinear stability as
\begin{equation}
    -\delta z\int^{z_2}_{z_1}\dd z \frac{\dd \bQ}{\dd z} \bcdot \frac{\p \rho(P_2,\bQ)}{\p\bQ} > 0, \quad \forall z_1,\,z_2.\label{nonlinearcriterion0}
\end{equation}If \eqref{nonlinearcriterion0} is satisfied, then the buoyancy force is always in the opposite direction to $\delta z$. In many cases of interest (including the examples listed above), the components $Q_i$ of $\bQ$ can each be chosen such that an increase in $Q_i$ at fixed pressure causes expansion:
\begin{equation}
    \frac{\p \ln\rho(P,\bQ)}{\p Q_i}<0, \quad \forall i, P.\label{expands}
\end{equation}In the case of $Q_i = s$, for example,~\eqref{expands} holds for any fluid that expands under heating at fixed pressure. We shall assume in what follows that~\eqref{expands} holds. In that case, a sufficient condition for \eqref{nonlinearcriterion0} to hold is
\begin{equation}
    \frac{\dd Q_i}{\dd z}>0, \quad \forall i,z.\label{nonlinearcriterion}
\end{equation}

Importantly,~\eqref{linearstability} and \eqref{nonlinearcriterion} are equivalent if the vector $\bQ$ has only one component [and~\eqref{expands} holds]. A fluid of this kind cannot exist in a metastable equilibrium, because it is always nonlinearly stable if linearly stable. The archetypical example is ordinary hydrodynamics, for which the only component of~$\bQ$ is~$s$, and therefore an equilibrium is nonlinearly stable to convection if it satisfies the Schwarzschild criterion~\citep{Schwarzschild06}
\begin{equation}
    \frac{\dd s}{\dd z}>0, \quad \forall z.\label{Schwarzschild}
\end{equation}

\subsubsection{Metastability}

For fluids for which $\bQ$ has more than one component,~\eqref{nonlinearcriterion} guarantees nonlinear stability, but is not necessary for the weaker condition of linear stability: if $\dd Q_i / \dd z<0$ for some~$i$, other components of $\bQ$ can compensate so that the linear-stability criterion~\eqref{linearstability} remains satisfied. In such cases, the equilibrium may be nonlinearly unstable despite being linearly stable, i.e., it may be metastable. In such cases, the condition for metastability may be deduced by recasting~\eqref{F2pressure} as the upwards force per unit mass of moved fluid, ${\boldsymbol{F}\bcdot \hat{\boldsymbol{z}}/\rho(P_2,\bQ_1)V_2 = (e^\mathcal{R}-1)g}$, where
\begin{equation}
    \mathcal{R} \equiv \ln \frac{\rho (P_2, \bQ_2)}{\rho (P_2, \bQ_1)} =-\int^{z_2}_{z_1} \dd z \mathcal{L} +\int^{z_2}_{z_1} \dd z \frac{\dd \ln P}{\dd z}[\kappa(P,\bQ)-\kappa(P,\bQ_1)],\label{metastability_compressibility}
\end{equation}and
\begin{equation}
    \kappa(P,\bQ)\equiv \frac{\p \ln \rho(P,\bQ)}{\p \ln P}\label{compressibility}
\end{equation}is the dimensionless compressibility. To obtain the second equality in~\eqref{metastability_compressibility}, we have used the fundamental theorem of calculus, i.e., we have differentiated $\mathcal{R}$ with respect to~$z_2$ and integrated the result from $z_1$ to $z_2$. Equation~\eqref{metastability_compressibility} separates the integrated linear buoyancy response (the first integral on the second line) with the nonlinear response (the second integral), revealing the latter to be determined by the path-integrated difference in~$\kappa$ between the moving parcel and its surroundings.\footnote{Note that the term in~\eqref{metastability_compressibility} that involves the compressibility of the surroundings cancels between the nonlinear response and the integral of \eqref{linearstability_alternative}; the nonlinear response may therefore be viewed as a correcting for the fact that the term involving $\mathcal{L}$ uses the ``wrong'' compressibility (that of the background, rather than that of the moving parcel) to determine the density change.} If the displaced fluid is more compressible than the fluid through which it moves, i.e., $\kappa(P,\bQ_1)>\kappa(P,\bQ)$, then the second integral on the right-hand side of~\eqref{metastability_compressibility} has the same sign as $\delta z$, so its contribution to the buoyancy force is destabilising. If the stabilising effect of the integral involving $\mathcal{L}$ is sufficiently small, then the second integral dominates in~\eqref{metastability_compressibility} for $\delta z$ larger than some critical value, $\delta z_c$. Such an equilibrium is metastable. 

\subsection{Direction and size of displacement required for nonlinear instability}

Because metastability requires more compressible fluid to be moved through less compressible fluid [see~\eqref{metastability_compressibility}], equilibria can be metastable only to perturbations in the direction in which the compressibility of the background fluid decreases.\footnote{The exception is when the compressibility has a local extremum. In that case, the equilibrium may be unstable to both upwards and downwards perturbations (corresponding to a local maximum of the compressibility) or to neither (a local minimum). We provide explicit examples of both cases in Section~\ref{sec:examples}.} This direction is given by the sign of the compressibility scale height
\begin{equation}
    H_{\kappa} \equiv -\left(\frac{\p \ln\kappa }{\p \bQ} \bcdot \frac{\dd \bQ }{\dd z}\right)^{-1}.\label{Hkappa}
\end{equation}

We can estimate the local value of $\delta z_c$ in the limit of $\mathcal{L}\to 0$ by balancing the sizes of the two integrals in~\eqref{metastability_compressibility} to find that
\begin{equation}
    \delta z_c \simeq \frac{ H_P H_{\kappa}{\mathcal{L}}}{\kappa},\label{deltazc}
\end{equation}where we have defined the pressure scale height $H_P \equiv |\dd \ln P/ \dd z|^{-1}$, neglected variation in $\mathcal{L}$ on the scale $\delta z_c$, and used the fact that $-\kappa H_{\kappa}^{-1}$ is the coefficient of $\delta z$ in the small-$\delta z$ expansion of the term in square brackets in~\eqref{metastability_compressibility}. According to~\eqref{deltazc}, $\delta z_c$ becomes arbitrarily small compared with any stratification scale height as $\mathcal{L}\to 0$. Nonetheless, we note that, because metastability of stratified equilibria is a nonlinear effect, it is not captured by Boussinesq-like equations that employ linear approximations of equilibrium gradients.

\subsection{Explosive instability}

The equation of motion of a small fluid parcel displaced by a distance $\delta z\sim \delta z_c $ that is much smaller than any scale height $H$ of the stratification is
\begin{equation}
    \frac{\dd^2 \delta z}{\dd t^2} = \left(-\delta z+\frac{\delta z^2}{\delta z_c}\right)\mathcal{L}.\label{Eqmotion}
\end{equation}It follows from~\eqref{Eqmotion} that, for ${0<\delta z_c \ll \delta z \ll H}$ (the case of $\delta z_c<0$ is analogous), the motion of a fluid parcel is explosive, viz.,
\begin{equation}
    \delta z \propto \frac{1}{(C-t)^2},
\end{equation}where the constant $C$ is determined by initial conditions. Explosive growth of $\delta z$ persists until either ${\delta z \sim H}$, whereupon~\eqref{Eqmotion} is no longer valid, the rising fluid element is shredded by Kelvin--Helmholtz instability or its speed approaches that of sound and thus~\eqref{F2pressure} no longer applies.

\subsection{Case of 2D MHD\label{sec:metastabilityMHD}}

In the remainder of this paper (with the exception of Appendix~\ref{app:moist}, where we consider moist hydrodynamics), we focus on the case of an equilibrium supported against gravity both by thermal pressure $p$ and by the magnetic pressure associated with a straight magnetic field with spatially dependent strength $B$. Thus, ${P = p + B^2/2}$. We restrict attention to 2D dynamics in the plane perpendicular to the magnetic field (i.e., to 2D interchanges of flux tubes). Then the specific magnetic flux
\begin{equation}
    \chi = \frac{B}{\rho}\label{specificmagflux}.
\end{equation}is conserved in a Lagrangian sense. For symmetry with this definition of $\chi$, we take advantage of the fact that any monotonically increasing function of specific entropy constitutes a good choice for the conserved quantity $s$, and thus let 
\begin{equation}
    s = \frac{p^{1/\gamma}}{\rho}\label{entropyfn},
\end{equation}
where $\gamma$ is the adiabatic index. To avoid confusion with the true specific entropy, which is proportional to $\exp(s/C_v)$ with $C_v$ the specific heat capacity, we hereafter refer to~$s$ as the ``entropy function'' (akin to potential temperature in atmospheric science, see Appendix~\ref{app:moist}). Thus, $\bQ = (s, \chi)$ and
\begin{equation}
    P(\rho,\bQ)=\rho^\gamma s^\gamma + \rho^2 \chi^2/2.
\end{equation}With these choices,~\eqref{linearstability} yields the linear-stability condition
\begin{equation}
    \mathcal{L}=  \frac{c_s^2}{c^2} \frac{\dd \ln s}{\dd z} +\frac{v_A^2}{c^2}\frac{\dd \ln \chi}{\dd z}>0,\label{Lmagnetic}
\end{equation}where $c_s\equiv\sqrt{\gamma p/\rho}$ is the sound speed, $v_A\equiv B/\sqrt{\rho}$ the Alfv\'{e}n speed and $c = \sqrt{c_s^2 + v_A^2}$ is the velocity of compressive waves [equation~\eqref{Lmagnetic} is sometimes called the modified Schwarzschild criterion]. The compressibility $\kappa$~\eqref{compressibility} is
\begin{equation}
    \kappa(P,s,\chi)= \frac{1+\beta}{2+\gamma \beta},\label{kappa_mhd}
\end{equation}where $\beta\equiv 2p/B^2$ is the plasma beta (the ratio of thermal to magnetic pressures) which is determined from $P$, $s$ and~$\chi$ via
\begin{equation}
    \frac{1}{\beta}\left(1+\frac{1}{\beta}\right)^{\frac{2}{\gamma}-1} = \frac{1}{2}\left(\frac{\chi}{s}\right)^2 P^{\frac{2}{\gamma}-1}.\label{betaP}
\end{equation}

Equation~\eqref{kappa_mhd} reveals that $\kappa$ increases monotonically with~$\beta$ for~$\gamma<2$, from $\kappa = 1/2$ at $\beta = 0$ to $\kappa \to 1/\gamma$ as ${\beta \to \infty}$. It follows that the nonlinear buoyancy response in~\eqref{metastability_compressibility} is destabilising when fluid with large $\beta$ moves through ambient fluid with smaller $\beta$. According to~\eqref{betaP}, the $\beta$ of a flux tube with given $s$ and $\chi$ depends on pressure and therefore is not constant during its motion.\footnote{According to~\eqref{betaP}, $\beta$ increases when a flux tube with fixed $s$ and $\chi$ rises while in total-pressure balance with its surroundings. For this reason, an equilibrium with $\beta(z)=\const$ is metastable to upwards perturbations if it is sufficiently close to marginal linear stability.} At any given pressure, however, $\beta$ is a monotonically increasing function of the ratio~$s/\chi$.
Therefore, an equilibrium that is sufficiently close to marginal linear stability is always nonlinearly unstable in the direction in which $s/\chi$ decreases [the same conclusion can be obtained from~\eqref{Hkappa} or by expanding~\eqref{F2pressure} to quadratic order in~$\delta z$ directly, see Appendix~\ref{quadratic}].

\subsection{Examples of metastable equilibria \label{sec:examples}}

Explicit examples of metastable equilibria may be obtained as follows. First, we change variables from height~$z$ to the total mass supported at height $z$,
\begin{equation}
    m = \int_z^{\infty} \rho(z')\dd z',\label{suppported_mass}
\end{equation}
so that the equilibrium condition becomes ${P = mg}$, which may be expressed as
\begin{equation}
    \rho^{\gamma}s^{\gamma}+\frac{1}{2}\rho^2 \chi^2 = mg,\label{equilibrium_rho}
\end{equation}or, using~\eqref{betaP}, as
\begin{equation}
    \frac{1}{\beta}\left(1+\frac{1}{\beta}\right)^{\frac{2}{\gamma}-1} = \frac{1}{2}\left( \frac{\chi}{s} \right)^2 (mg)^{\frac{2}{\gamma}-1}.\label{equilibrium}
\end{equation}Throughout this work, we shall find it convenient to use the supported mass $m$ as a proxy for height because the mass of a flux tube is preserved as it moves, while its cross-sectional area is not (note that $m$ decreases with increasing~$z$). 

We seek an equilibrium that is close to marginal linear stability [so that the second integral in~\eqref{metastability_compressibility} stands a chance of dominating the first]. We therefore demand that
\begin{equation}
    \mathcal{L}=\frac{\epsilon}{H_P},\label{L=epsonrho}
\end{equation}where $\mathcal{L}$ is given by~\eqref{Lmagnetic}, $H_P$ is the total-pressure scale height $[\dd \ln P /\dd z]^{-1}=m/\rho$ and $\epsilon \ll 1$ is a small number. Equation~\eqref{L=epsonrho} constitutes a differential equation involving $s$, $\chi$ and their gradients in $m$---to solve it, we specify a relationship between $s$, $\chi$ and $m$ that defines the particular equilibrium under consideration. We choose to specify as a function of $m$ the ratio $s/\chi$, which controls the compressibility of the fluid (Section~\ref{sec:metastabilityMHD}). We therefore recast~\eqref{L=epsonrho} [with $\mathcal{L}$ given by~\eqref{Lmagnetic}] as
\begin{equation}
    \frac{\dd \ln s}{\dd m} = \frac{2}{\gamma \beta + 2}\frac{\dd}{\dd m}\ln\left(\frac{s}{\chi}\right)-\frac{\epsilon}{m},\label{dlnsdm}
\end{equation}which we integrate numerically for~$s$ as a function of~$m$. We determine the dependence of all other quantities on $m$ (or $z$) via their definitions and~\eqref{suppported_mass} and~\eqref{equilibrium}.

\begin{figure*}
    \centering
    \includegraphics[width=\textwidth]{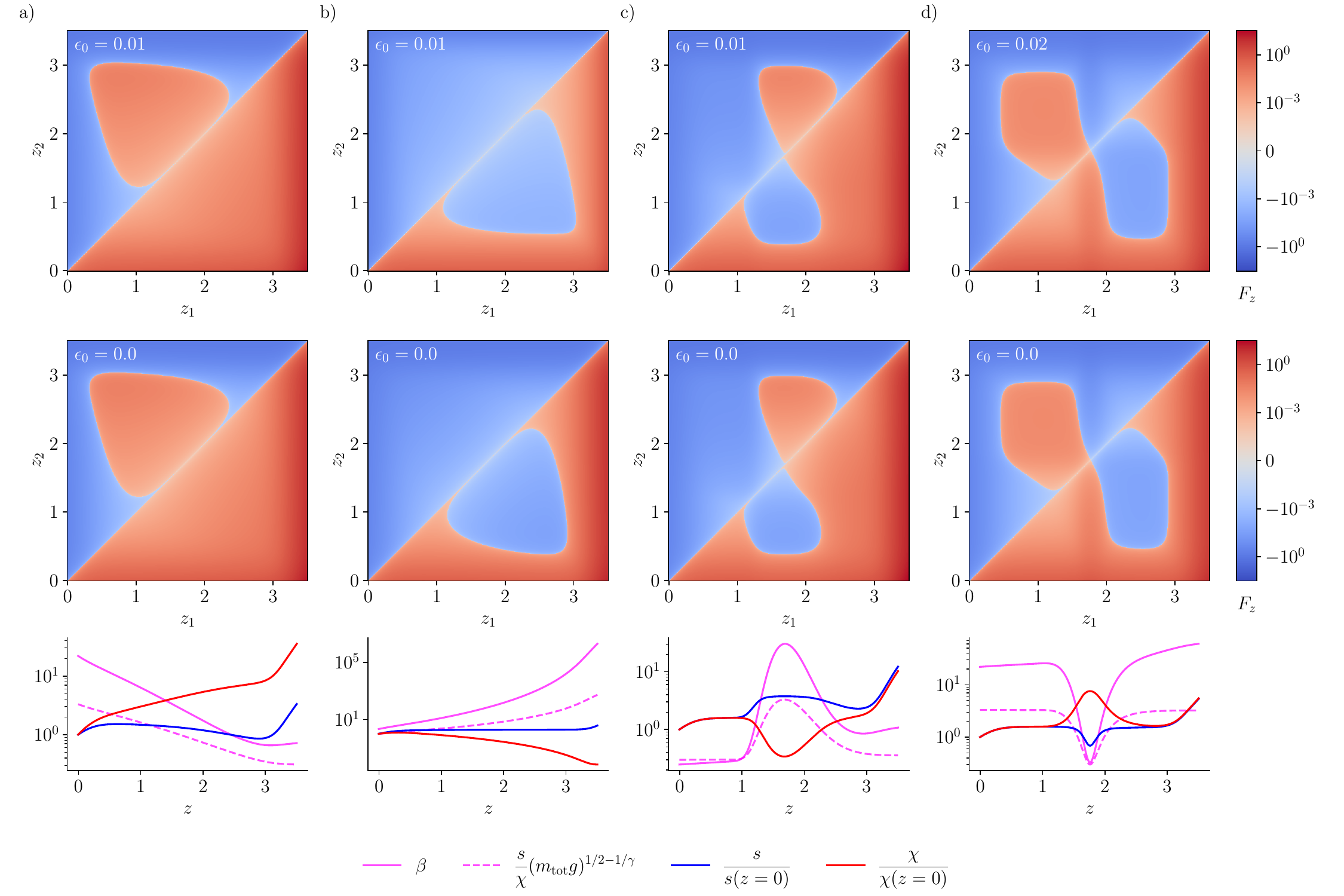}
    \caption{Top row: the upward force~\eqref{F2pressure} per unit mass on a small fluid parcel moved in pressure balance and without diffusion from height $z_1$ to $z_2$ for each of the profiles described in Section~\ref{sec:examples}. Values of $\epsilon_0$ in~\eqref{epsilon_buffer} are indicated on each panel. Middle row: the same as the top row, but with $\epsilon_0=0$ (marginal linear stability in the bulk). Bottom row: profiles of the entropy function $s=p^{1/\gamma}/\rho$, specific magnetic flux $\chi=B/\rho$, their ratio, and the plasma $\beta$~\eqref{equilibrium} as a function of height $z$ (these profiles correspond to the top row specifically, but the profiles that correspond to the middle row look essentially the same because the differences in $\epsilon_0$ are very small). Panel~(a) corresponds to~\eqref{m_profile}, (b) to~\eqref{invm}, (c) to~\eqref{bump} and (d) to~\eqref{dip}.}
    \label{fig:forcez1z2}
\end{figure*}

Some example cases are visualised in figure~\ref{fig:forcez1z2}, where we show $s$, $\chi$, $s/\chi$ and $\beta$ as functions of $z$ (lower row of panels) together with the force~\eqref{F2pressure} per unit mass on a small parcel of fluid moved from height $z_1$ to $z_2$ as a function of $z_1$ and $z_2$ (upper two rows of panels). Figure~\ref{fig:forcez1z2}(a) shows the case of
\begin{equation}
    \frac{s}{\chi} = (m_{\mathrm{tot}}g)^{1/\gamma - 1/2}  \left(3\frac{m}{m_\mathrm{tot}}+0.3\right),\label{m_profile}
\end{equation}for which $s/\chi$ increases with $m$. The most compressible material is therefore at the bottom of the atmosphere, and the equilibrium is unstable to upwards displacements (towards larger $z$). This is the equilibrium whose relaxation is visualised in figures~\ref{fig:fig1} and~\ref{fig:fig2}. Figure~\ref{fig:forcez1z2}(b) shows the case of
\begin{equation}
    \frac{s}{\chi} = (m_{\mathrm{tot}}g)^{1/\gamma - 1/2}  \left(3\frac{m}{m_\mathrm{tot}}+0.3\right)^{-1},\label{invm}
\end{equation}for which $s/\chi$ decreases with $m$, so produces an equilibrium unstable to downward displacements (towards smaller $z$). Figure~\ref{fig:forcez1z2}(c) shows the case of
\begin{equation}
    \frac{s}{\chi} = (m_{\mathrm{tot}}g)^{1/\gamma - 1/2}  \left[3\exp\left(-\frac{(m/m_{\mathrm{tot}}-0.2)^2}{0.1^2}\right)+0.3\right],\label{bump}
\end{equation}for which $s/\chi$ has a maximum at $m=0.2m_{\mathrm{tot}}$; the equilibrium is therefore unstable to both upwards and downwards displacements in the vicinity of the maximum. Finally, figure~\ref{fig:forcez1z2}(d) shows the case of
\begin{equation}
    \frac{s}{\chi} = (m_{\mathrm{tot}}g)^{1/\gamma - 1/2} \left[3\left(1-\exp\left(-\frac{(m/m_{\mathrm{tot}}-0.2)^2}{0.1^2}\right)\right)+0.3\right],\label{dip}
\end{equation}for which $s/\chi$ has a minimum at $m=0.2m_{\mathrm{tot}}$. The most compressible fluid is therefore situated at the top and bottom of the atmosphere, so it is unstable to both downwards and upwards displacements.

In figures~\ref{fig:fig1}-\ref{fig:forcez1z2} and in the rest of the paper, we choose units of mass and length such that $\rho=1$ and ${m = m_{\mathrm{tot}}=1}$ at the bottom of the atmosphere (${z=0}$), so the total-pressure scale height $H_P=1$ there. We choose units of time such that $g=1$; it follows that the time taken for a compressive wave to traverse a total-pressure scale height ${H_P/c\sim \sqrt{H_P/g}}$ is~$1$ at the bottom of the atmosphere. We limit the total height of the atmospheres to ${z_{\mathrm{max}}=3.5}$ in these units, where $\rho \lesssim 10^{-2}$ in each of the four cases described above. We stabilise the equilibria near $z=0$ and $z=z_{\mathrm{max}}$ by choosing
\begin{equation}
    \epsilon = \epsilon_0 + \tanh\left(\frac{z-z_u}{\Delta_u}\right) - \tanh\left(\frac{z-z_l}{\Delta_l}\right) + 2\label{epsilon_buffer}
\end{equation}in~\eqref{L=epsonrho}, with $\Delta_l=\Delta_u = 0.25$, $z_l = 0.25$, $z_u = 3.25$. This ensures that artificial boundary effects do not impact our numerical or theoretical analyses. We choose ${\epsilon_0 = 10^{-2}}$ for the simulations visualised in figures~\ref{fig:fig1} and~\ref{fig:fig2}. 

The top row of panels in figure~\ref{fig:forcez1z2} correspond to equilibria with small positive values of $\epsilon_0$; these are linearly stable, and exhibit the various sorts of metastability described above. Unless stated otherwise, we shall in the rest of the paper focus attention on the case of marginal linear stability, i.e., $\epsilon_0 = 0$ in~\eqref{epsilon_buffer}. This is because we expect nonlinear instability to be triggered for equilibria close to marginal linear instability in practice. The middle row of figure~\ref{fig:forcez1z2} corresponds to the $\epsilon_0 = 0$ case.

\section{Available potential energy\label{sec:ground_state_theory}}

In this section, we calculate the energy that can be liberated from a metastable MHD equilibrium by 2D rearrangement of flux tubes. 
Because the gravitational potential energy depends only on $z$, we shall assume a priori that the state with minimum energy is 1D: all quantities depend only on $z$ (or, equivalently, on $m$). It will turn out that this assumption can, and often does, fail, but the 2D minimum-energy states will be extractable from the solution to the 1D minimisation. 

The total potential energy per unit horizontal length of a 1D state (not necessarily in equilibrium) is
\begin{equation}
    E_{\mathrm{tot}} = \int_0^\infty \dd z \left(\frac{p}{\gamma -1}+\frac{B^2}{2}+\rho g z\right) = \int_0^{m_{\mathrm{tot}}} \dd m \left[\mcE(P, s, \chi)+\frac{mg - P}{\rho(P,s,
    \chi)}\right],\label{curlyE}
\end{equation}where $\rho(P,s,\chi)$ is determined implicitly from the definition of total pressure, ${P= \rho^{\gamma}s^{\gamma}+\rho^2\chi^2/2}$. In the second equality of~\eqref{curlyE} we have integrated by parts and introduced the specific enthalpy,
\begin{equation}
    \mcE\equiv \frac{1}{\rho}\left(\frac{p}{\gamma -1} +\frac{B^2}{2}+P\right) = \frac{\gamma}{\gamma - 1}\rho^{\gamma -1}s^{\gamma} + \rho \chi^2.\label{mcE_rho}
\end{equation}It is readily verified that
\begin{equation}
    \left(\frac{\p \mcE}{\p P}\right)_{s,\chi}=\frac{1}{\rho},\label{d_mcE_P=rho}
\end{equation}whence
\begin{equation}
    E_{\mathrm{tot}} = \int_0^{m_{\mathrm{tot}}} \dd m \Bigg\{\mcE(mg, s, \chi) +\frac{1}{2}\frac{P^2}{\rho^2 c^2}\left(\frac{\delta P}{mg}\right)^2 + \mathcal{O}\left[\left(\frac{\delta P}{mg}\right)^3\right]\Bigg\},\label{quadratic_energy}
\end{equation}where $\delta P \equiv P-mg$. Evidently, $E_{\mathrm{tot}}$ is minimal with respect to $P$ when ${P=mg}$. It follows that, when looking for the minimum-energy state, we can restrict attention to those states with $P=mg$, i.e., those in static equilibrium at all~$m$. Equation~\eqref{quadratic_energy} then reduces to
\begin{equation}
    E_{\mathrm{tot}}=\int_0^{m_{\mathrm{tot}}} \dd m\,\mcE(m g, s, \chi).\label{curlyEfinal}
\end{equation}

Following \citet{Lorenz55}, we discretise the integral~\eqref{curlyEfinal} by ``slicing'' the atmosphere into thin layers of equal mass~$\Delta m$ that we label by the index~$i$. We consider the 1D equilibria formed by rearranging the slices while conserving the entropy and flux in each slice. Each possible rearrangement is a permutation map $i\to j = \sigma (i)$, under which the slice that initially supports mass $m_i$ now supports mass $m_j$. The energy of the rearranged equilibrium is
\begin{equation}
    E_{\mathrm{tot}} \simeq \Delta m \sum_{i=0}^N \mcE(m_{\sigma(i)}g, s_{i}, \chi_{i}),\label{discrete}
\end{equation}where $N=m_{\mathrm{tot}}/\Delta m$, and $s_i$, $\chi_i$ are, respectively, the entropy function and specific magnetic flux of slice $i$. We seek the permutation $\sigma$ that gives the smallest possible value of~$E_{\mathrm{tot}}$, which we denote $E_{\mathrm{min}}$.

Because the energy associated with assigning slice $i$ to support a mass of $m_j$ only depends on $i$ and $j$ and not on the assignments of other slices, minimising~\eqref{discrete} over permutations $\sigma$ is a combinatorial optimisation problem known as linear sum assignment (LSA) [see, e.g., \citet{burkard2012}]. The LSA problem is canonically described as one of minimising the total cost associated with assigning a number of ``agents'' to the same number of ``tasks''---in our case, the ``agents'' are the slices of atmosphere with given $s$ and $\chi$, while their ``tasks'' are to occupy discrete positions in the atmosphere corresponding to each possible value of the discretised supported mass. The matrix of costs associated with assigning agent $i$ to task $j$ is~${\mcE(m_{j}g, s_{i}, \chi_{i})}$.

For economy of notation, we hereafter denote the cost matrix $\mcE(m_{j}g, s_{i}, \chi_{i})$ by ${{\mcE(m_{j}, m_{i})}}$: this is the energy cost associated with assigning the slice initially at~$m_i$ to~$m_j$ (in a minor abuse of notation, we use the same symbol, $\mcE$, for both functions). Despite our discretisation in $m$, ${\mcE(m_{j}, m_{i})}$ remains a continuous function of its arguments, and we shall frequently be required to integrate or take derivatives with respect to one or the other in what follows. We shall, therefore, introduce the continuous variable $\mu$ to denote the supported mass of a slice in the initial state, and denote the continuous form of ${\mcE(m_{j}, m_{i})}$ by $\mcE(m,\mu)\equiv \mcE(m g,s(\mu), \chi(\mu))$. Similarly, we shall write $\rho(m,\mu)$ as a shorthand for $\rho(mg, s(\mu), \chi(\mu))$. In this notation,~\eqref{d_mcE_P=rho} becomes 
\begin{equation}
    \frac{\p \mcE(m,\mu)}{\p m} = \frac{g}{\rho(m,\mu)}.\label{d_mcE=rho}
\end{equation}

\subsection{Estimating the available energy\label{sec:smallE}}

Before proceeding to solve the LSA problem, which can only be achieved numerically in most cases, let us try to estimate the outcome analytically: What is the typical available potential energy of a metastable atmosphere? This turns out to be a small fraction of the total potential energy, a fact we that we utilise in Section~\ref{sec:LyndenBell} to predict the relaxation of destabilised equilibria.

The change in potential energy $\delta E$ that results from moving a slice of atmosphere upwards from supported mass $m_a$ to new supported mass $m_b$, shuffling downwards the slices that it passes on the way, is
\begin{align}
    \frac{\delta E}{\Delta m} & = \mcE(m_b, m_a)-\mcE(m_a, m_a)+\sum_{i=b}^{a-1}\left[\mcE(m_{i+1},m_i)-\mcE(m_{i},m_i)\right] \nonumber\\&
    \simeq \int^{m_b}_{m_a}\dd m \left[\frac{g}{\rho(m,m_a)}-\frac{g}{\rho(m,m)}\right] \nonumber \\&
     = - g\int^{z_a}_{z_b}\dd z \left[\frac{\rho(m(z),m(z))}{\rho(m(z),m_a)}-1\right],\label{deltaE/dm}
\end{align}where $m(z_a) = m_a$ and $m(z_b) = m_b$. We have used~\eqref{d_mcE=rho} in moving from the first to the second line of~\eqref{deltaE/dm}. The integrand that appears in the last line of~\eqref{deltaE/dm} is straightforwardly recognised as the net buoyancy force~\eqref{F2pressure} per unit mass of fluid, so, sensibly, $\delta E$ is just the work done by this force on the moving slice\footnote{Work is done in moving the slice through its series of adjacent equilibria because interchanging two slices in practice involves the fluid in each of them passing through non-equilibrium states.}. Evidently, the energy that can be liberated by this process is maximal when the ratio that appears inside the integrand in the last line of~\eqref{deltaE/dm}---i.e., of the density of the ambient fluid to the density of the moving slice---is maximal. We can evaluate this ratio using~\eqref{metastability_compressibility}, which yields, in the most optimistic case of (\textit{i}) marginal linear stability, i.e., $\mathcal{L}=0$, (\textit{ii})~$\kappa=1/\gamma$ inside the slice (maximally compressible large-$\beta$ fluid) and (\textit{iii}) $\kappa = 1/2$ for the ambient fluid (minimally compressible small-$\beta$ fluid),
\begin{equation}
    \frac{\rho(m,m_a)}{\rho(m,m)} = \left(\frac{m}{m_a}\right)^{1/\gamma - 1/2}.\label{rho_fraction_optimistic}
\end{equation}Substituting this into~\eqref{deltaE/dm}, choosing $m_a=m_{\mathrm{tot}}$ (the slice originates from the bottom of the atmosphere) and evaluating integrals, we find that
\begin{equation}
    \frac{\delta E}{E_0} = \frac{3}{4}\frac{\Delta m}{m_{\mathrm{tot}}}\left[\frac{\gamma}{\gamma - 1}\left(\left(\frac{m_b}{m_{\mathrm{tot}}}\right)^{1-1/\gamma}-1\right)-2\left(\left(\frac{m_b}{m_{\mathrm{tot}}}\right)^{1/2}-1\right)\right],\label{Esmall}
\end{equation}where $E_0$ is the initial total energy of the atmosphere [assuming that it is mostly populated with small-$\beta$ fluid, as is consistent with assumption (\textit{iii}) above].
\begin{figure}
    \centering
    \includegraphics[width=0.6\columnwidth]{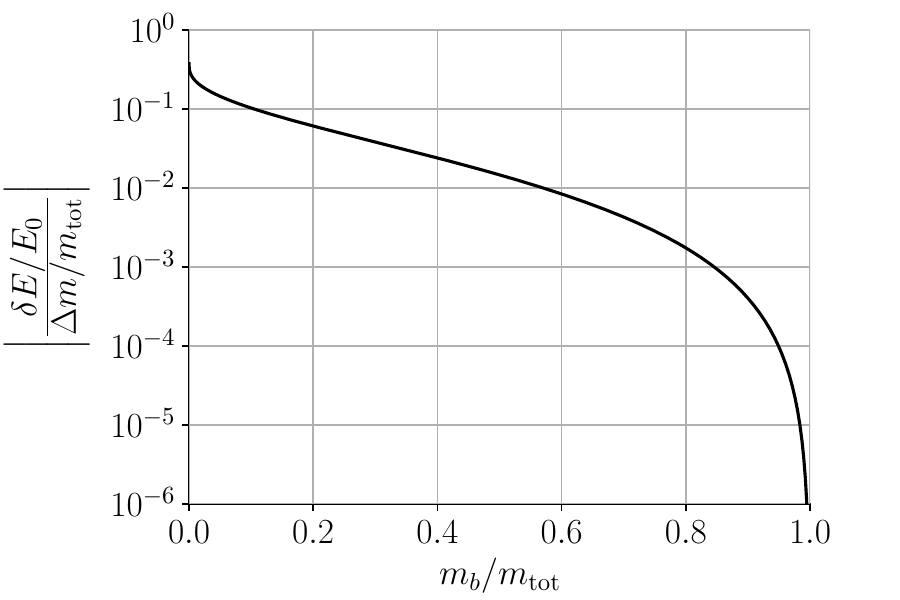}
    \caption{The fraction~\eqref{Esmall} of energy liberated when a slice of fluid with mass $\Delta m$ is moved from the bottom of an atmosphere at marginal linear stability to a new position where the supported mass is~$m_b$, under the most optimistic assumptions about the compressibility of the slice and that of the fluid through which it moves.}
    \label{fig:analytic_estimate}
\end{figure}

We plot~\eqref{Esmall} in figure~\ref{fig:analytic_estimate} for the case of $\gamma = 5/3$. We observe that the fraction of energy liberated per mass fraction of fluid moved decreases sharply as a function of $m_b/m_{\mathrm{tot}}$~(even though its limiting value of $3/8$ as $m_b/m_{\mathrm{tot}}\to 0$ is finite). This is significant as, because fluid parcels exclude each other (i.e., different slices cannot all be assigned to the same supported mass), the vast majority of slices that are interchanged in a global rearrangement experience an order-unity change in supported mass. Such reassignments are far less profitable than ones that take a slice all the way to the top of the atmosphere (i.e., $m_b/m_{\mathrm{tot}}\to 0$). For example, figure~\ref{fig:analytic_estimate} shows that the fraction of energy liberated per mass fraction of fluid moved is around $10^{-2}$ for $m_b = m_{\mathrm{tot}}/2$. The reason is that an order-unity change in pressure only results in a small change in density, owing to the smallness of the exponent $1/\gamma - 1/2 = 0.1$ that appears on the right-hand side of~\eqref{rho_fraction_optimistic}. A more detailed calculation (see Appendix~\ref{sec:anaytic_ground_state}) reveals that $10^{-2}$ is indeed a good estimate for the fractional available energy: the maximum fractional available energy of a (marginally) stable atmosphere consisting of small-$\beta$ fluid above a layer of large-$\beta$ fluid is $1.75\%$. 

The above estimates correspond to the most optimistic assumptions: we shall find that the fractional available energies of the equilibria described in Section~\ref{sec:examples} and represented in figure~\ref{fig:forcez1z2} are less than $1\%$ (by roughly an order of magnitude), because (\textit{i}) these equilibria have finite $\beta$, and thus do not have the extremal values of the fluid compressibility considered here; and (\textit{ii}) the stable buffer regions contribute to the potential energy of the equilibrium (and prevent the relaxing fluid from accessing large pressure differences) but do not participate in the reassignment.

\subsection{The Hungarian algorithm\label{sec:HungarianTheory}}

The discrete energy-minimisation problem \eqref{discrete} can be solved without recourse to numerical optimisation only in the limits $\beta\to\infty$ and $\beta\to0$, i.e., the cases where only one of thermal or magnetic pressure support the atmosphere against gravity. For ${\beta \to \infty}$, ${\mcE \to \gamma s(mg)^{1-1/\gamma}/(\gamma - 1)}$, which increases monotonically with both $m$ and $s$. The arrangement with least total energy is therefore the one for which the slice with largest $s$ has smallest $m$, the slice with the next largest $s$ has the next smallest $m$, and so on. It follows that the profile with the smallest energy is the unique rearrangement for which $s$ is a monotonically increasing function of height. Indeed, we already know this state to be nonlinearly stable by~\eqref{Schwarzschild}. A similar conclusion is obtained for ${\beta\to 0}$, for which ${\mcE \to 2(mg)^{1/2} \chi}$ and so the nonlinearly stable atmosphere is the one with $\chi$ increasing monotonically with height. Analogous simple constructions for the minimum-energy configuration do not exist in the finite-$\beta$ case: the optimal solution that balances the competing imperatives of ``entropy should increase upwards'' and ``flux should increase upwards'' is non-trivial. 

Solution of the LSA problem in general relies on the observation that the \textit{modified} cost matrix $\tilde{\mcE}(m_{j}, m_i)$, where
\begin{equation}
    \tilde{\mcE}(m_{j}, m_i) = \mcE(m_{j}, m_i) - a(m_{j}) - b(m_{i}),\label{normalform}
\end{equation}has the same optimal assignment of agents to tasks as does the original cost matrix ${\mcE(m_{j}, m_{i})}$. This is intuitive: if the cost of assigning a given agent (slice) to each task (position) decreases by the same amount, their optimal assignment will not change (although the total cost of the solution will decrease). Likewise, if a particular task becomes more expensive by the same amount for all agents, the optimal choice of agent for that task will remain the same. $\tilde{\mcE}$ is called the ``normal form'' of the cost matrix $\mcE$ if it satisfies the following properties~\citep{burkard2012}:
\begin{enumerate}[(a)]
    \item $\tilde{\mcE}(m_{j}, m_i)\geq0$,
    \item There exists at least one bijection $\sigma$ such that $\tilde{\mcE}(m_{\sigma(i)}, m_i)=0, \,\,\,\forall i$.
\end{enumerate}A proof that an $\tilde{\mcE}(m_j,m_i)$ with these properties can always be found with suitable choices of $a(m_j)$ and $b(m_i)$ is provided by the \textit{Hungarian algorithm}~\citep{Kuhn55,Munkres57}, which consists of a series of row and column operations on the cost matrix that are guaranteed to reduce it to normal form in polynomial (in $N$) time. The utility of the normal form~$\tilde{\mcE}(m_j,m_i)$ is that each of the bijections $\sigma$ constitutes an optimal assignment.\footnote{Typically, the optimal assignment is unique. This is true for each of the equilibria introduced in Section~\ref{sec:examples}. One can construct counter-examples, however: the optimal assignment is non-unique when multiple slices have the same values of $s$ and $\chi$, for example.}

\subsection{Energy minimisation: numerical results and interpretation\label{sec:HungarianResults}}

In this section, we present the outcomes of energy minimisation using the Hungarian algorithm (Section~\ref{sec:HungarianTheory}) for each of the equilibria described in Section~\ref{sec:examples} and represented in figure~\ref{fig:forcez1z2}.

\subsubsection{Metastable upwards,~\eqref{m_profile} \label{sec:selfsimilar}}

\begin{figure*}
    \centering
    \includegraphics[width=.8\textwidth]{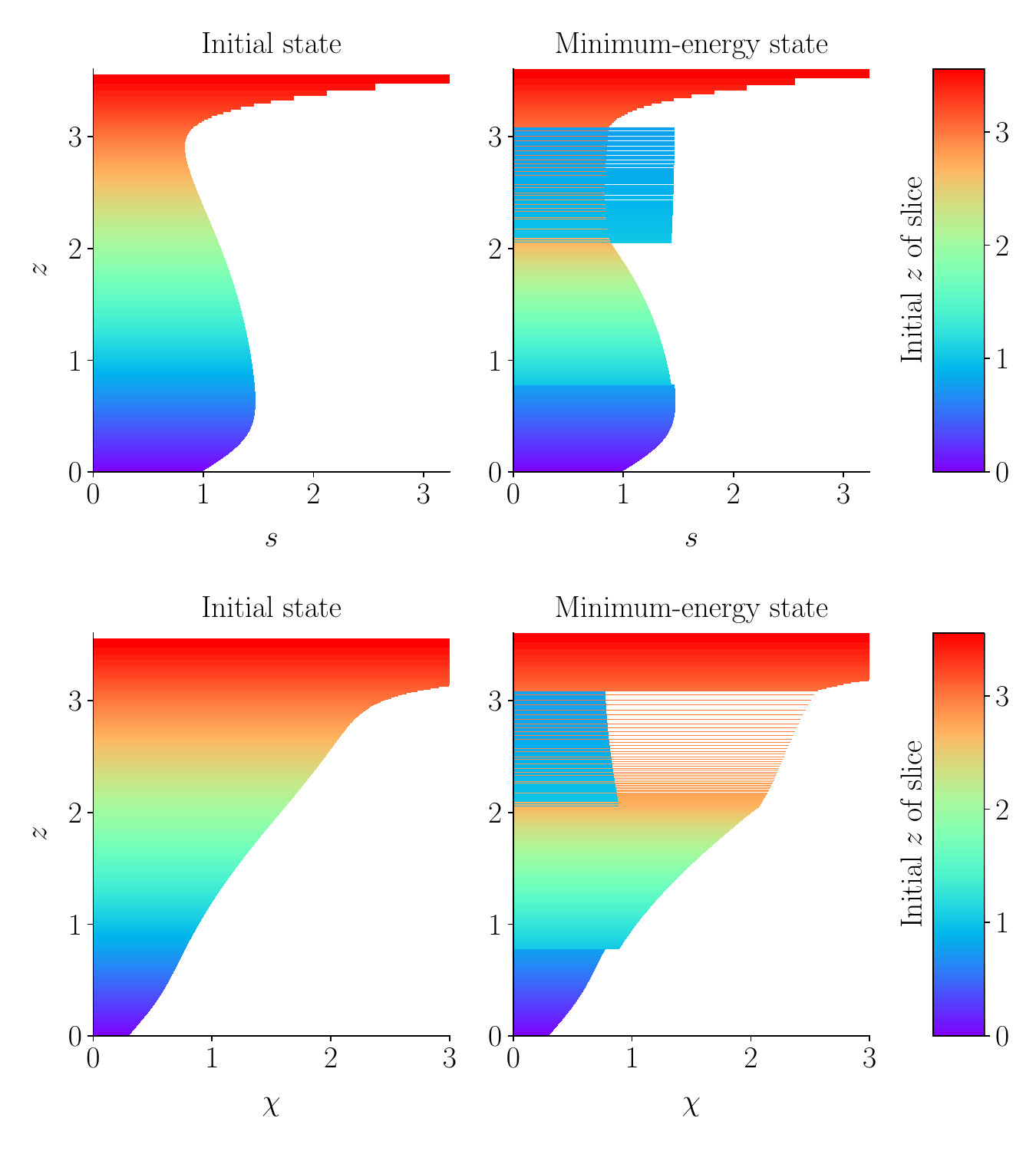}
    \caption{Visualisation of the assignment of 1D slices that minimises the total energy,~\eqref{discrete}, with $\Delta m = 5\times 10^{-4}m_{\mathrm{tot}}$, for the upwards-unstable profile defined by~\eqref{m_profile}. Panels on the left show the initial profiles of $s$ and $\chi$ as functions of height $z$, while panels on the right show the minimum-energy assignment. The slices are coloured by their height $z$ in the initial state to aid comparison. Blue slices from $0.8<z<1.0$ are moved to $2.1<z<3.1$, reversing order and foliating with red slices originally from $2.7<z<3.1$. Each slice has vertical extent $\Delta z = \Delta m/\rho$ with $\rho$ given by~\eqref{equilibrium_rho}.}
    \label{fig:selfsimilar_h}
\end{figure*}

We first consider the equilibrium defined by~\eqref{m_profile}, which is nonlinearly unstable to upward displacements [figure~\ref{fig:forcez1z2}(a)]. The solution of the LSA problem is visualised in figure~\ref{fig:selfsimilar_h}, which compares the initial and minimum-energy assignments of the slices. In the minimum-energy assignment, material initially from the bottom of the metastable part of the atmosphere [i.e., $z\gtrsim z_l$; see~\eqref{epsilon_buffer}] is reassigned to the top $z\lesssim z_u$, as is intuitive. The order in which the slices that are re-assigned are stacked also reverses. This happens because, as the material moves to smaller $m$, its $\beta$ increases [see~\eqref{equilibrium}], and therefore the contribution of~$s$ to the linear-stability criterion becomes more important relative to~$\chi$. As a consequence, the stacking order reverses so that $s$ increases upwards.

\begin{figure*}
    \centering
    \includegraphics[width=\textwidth]{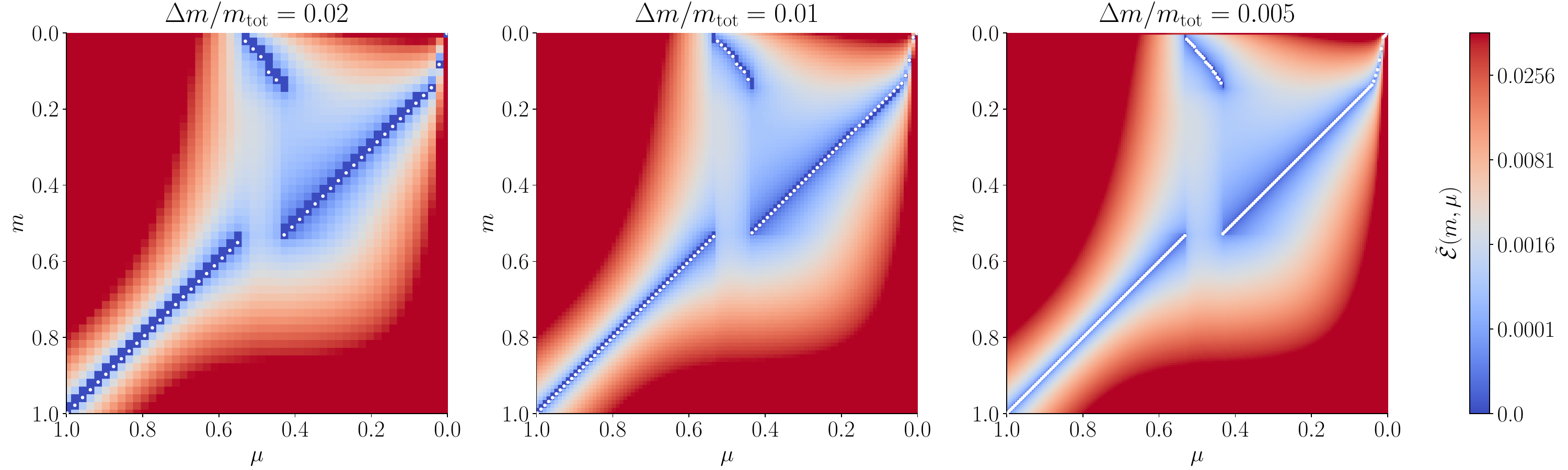}
    \caption{The normal form of the cost matrix $\tilde{\mcE}(m,\mu)$ [see~\eqref{normalform} for its definition] for the unstable-upwards profile defined by~\eqref{m_profile}, for three different choices of the discretisation scale $\Delta m$. White dots show the optimal assignment.} 
    \label{fig:selfsimilar_normalform}
\end{figure*}

A striking feature of the minimum-energy state is that slices that were adjacent at the bottom of the atmosphere do not remain adjacent in the minimum-energy state. Instead, slices from the bottom become foliated with those at the top. The scale of the foliation is set by $\Delta m$, and therefore is arbitrarily small as $\Delta m \to 0$. This is illustrated by figure~\ref{fig:selfsimilar_normalform}, which shows, for three different values of $\Delta m$, the normal-form cost matrix $\tilde{\mcE}(m_j,m_i)$ [defined by~\eqref{normalform}], with its zeros [which indicate the optimal assignment $j=\sigma(i)$] marked with white circles. Because the scale of foliation depends on~$\Delta m$, the optimal assignment of slices does not converge as $\Delta m \to 0$, although it does converge in a coarse-grained sense: the proportion of slices assigned to any small finite range of~$m$ that originated from any similarly small given range of~$\mu$ converges as $\Delta m \to 0$ (provided that $\tilde{\mcE}$ converges as $\Delta m\to 0$).\footnote{In the continuous limit, these proportions can be determined from the gradient of the locus of points for which $\tilde{\mcE}=0$. For example, denoting this locus, i.e., the curve to which the white circles in figure~\ref{fig:selfsimilar_normalform} converge as $\Delta m \to 0$, by $m_{\mathrm{opt}}(\mu)$, then, provided $m_{\mathrm{opt}}(\mu)$ is single valued (as is the case in figure~\ref{fig:selfsimilar_normalform}), the proportion of slices assigned to the vicinity of $m_2$ that originally had supported mass $m_1$ is $|\dd m_{\mathrm{opt}}/\dd \mu|$ evaluated at $m_1$ (by the conservation of mass). In the absence of foliation, ${|\dd m_{\mathrm{opt}}/\dd \mu|^{-1}=1}$, but, where there is foliation, it must be the case that ${|\dd m_{\mathrm{opt}}/\dd \mu|>1}$, so $m_{\mathrm{opt}}(\mu)$ steepens. Both cases may be observed in figure~\ref{fig:selfsimilar_normalform}. The generalisation to the case where $m_{\mathrm{opt}}$ is multi-valued is somewhat more complex, but it remains true in that case that the fractional assignments are determined by gradients of the optimal-solution curve (see Appendix~\ref{app:geometry}).}

Foliation occurs when the material properties of the fluid vary with $m$ at a different rate in the part of the atmosphere from which a slice originates than in the part to which it is reassigned. We demonstrate this fact by considering the motions of two slices from the bottom of the atmosphere to the top, each displacing downwards the slices through which they pass (as in Section~\ref{sec:smallE}). Let the first slice have initial assignment $m_a$ and new assignment $m_b$. Because the new assignment is optimal, the total energy has a local minimum as a function of displacement of this slice when its density in its new location, $\rho (m_b,m_a)$, is equal to the density $\rho (m_b,m_b)$ of the slices that surround it there (neutral buoyancy); this makes the integrand in the second line of~\eqref{deltaE} zero at $m=m_b$.\footnote{Intuitively, if the densities were different, the new equilibrium would be Rayleigh--Taylor unstable either at the upper or lower surface of the slice.} Now let us consider a second slice of fluid that initially neighbours the first, i.e., that originates from supported mass $m_a + \Delta m$. This slice reaches neutral buoyancy at a different supported mass $m_b+\delta m$. If the density of the background equilibrium changes more slowly with supported mass at~$m_b$ than at~$m_a$, then $\delta m > \Delta m$. Setting the density of the second slice, $\rho (m_b+\delta m,m_a+\Delta m)$, equal to that of the ambient fluid at supported mass $m_b+\delta m$, i.e., ${\rho (m_b+\delta m,m_b+\delta m)}$ (because we are concerned with the motion of only two slices of infinitesimal thickness, we neglect the fact that the reassignment of the first slice might have changed the identity of the fluid at $m_b$), we discover that
\begin{equation}
    \cfrac{\delta m}{\Delta m}={\cfrac{\p \rho(m_b,\mu)}{\p \mu}\bigg|_{\mu=m_a}\Bigg/\cfrac{\p \rho(m_b,\mu)}{\p \mu}\bigg|_{\mu=m_b}}.\label{delta_m1/delta_m2}
\end{equation}Thus, the adjacency of the slices if not preserved ($\delta m\neq \Delta m$) if the rate of change with supported mass of the density that a slice \textit{would} have if moved to $m_b$ is different for slices originating at $m_a$ than at $m_b$.

Although~\eqref{delta_m1/delta_m2} reveals the physical reason for foliation, it fails if the fraction on the right-hand side is small ($\delta m \lesssim \Delta m$ is not allowed for the discrete problem because slices exclude each other). It also does not apply in the case where a substantial mass of fluid is reassigned, in which case the background equilibrium through which the first slice moves is different from that through which the last slice does. A general treatment of such cases is as follows. Let us suppose that we are somehow given all the optimal assignments $\sigma(i)$, except those to some small range of $m$, i.e., those with $m_{\sigma(i)}=m_b+\delta m_{\sigma(i)}$, and seek the condition under which the optimal choice of the remaining assignments will be a foliated state. The contribution to the total energy of the slices that remain to be assigned is
\begin{align}
    \delta E & = \Delta m \sum_{i} \mcE(m_b+\delta m_{\sigma(i)}, m_i) \nonumber \\ & = \Delta m \sum_{i} \bigg[ \mcE(m_b, m_i)+ \frac{\delta m_{\sigma(i)}g}{ \rho(m_{b}, m_{i})}+\mathcal{O}(\delta m_{\sigma(i)}^2)\bigg],\label{deltaE}
\end{align}where sums are over all indices $i$ of the slices that remain to be assigned, and we have used~\eqref{d_mcE=rho} to obtain the second equality. The first term inside the square bracket in the second line of~\eqref{deltaE} is independent of the assignment $\sigma$. The second term takes the form of the differential supported mass $\delta m_{\sigma(i)}$ multiplied by a quantity that does not depend on $\sigma(i)$, viz.,~$1/{\rho(m_b,m_i)}$. This yields a local stacking rule: $\delta E$ is minimised when the slice with largest $\rho(m_{b}, m_i)$ is assigned to the largest supported mass, the next largest $\rho(m_b, m_i)$ is assigned to the next largest supported mass, and so on. This is physically intuitive: if more dense slices were situated above less dense ones, the equilibrium would be Rayleigh--Taylor unstable. We deduce that in order for slices from different initial locations to become foliated in the final state they must have
\begin{equation}
\rho(m_b,m_{i'})-\rho(m_b,m_i)=\mathcal{O}(\Delta m).\label{samerho}
\end{equation}As $\Delta m\to 0$, slices can be foliated in the vicinity of some $m$ only if they have the same density at that~$m$.

\begin{figure*}
    \centering
    \includegraphics[width=1.0\textwidth]{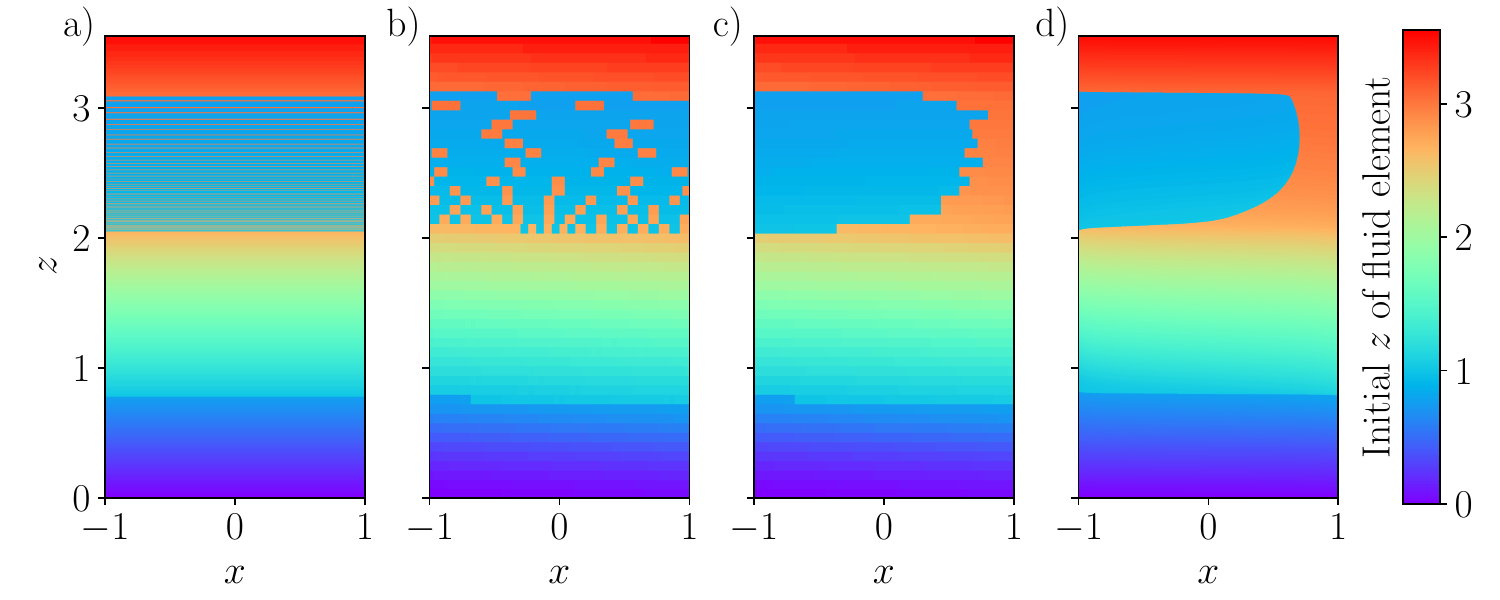}
    \caption{2D minimum-energy states. Panel~(a) shows the 1D stacking with minimum energy in the $xz$-plane. Panel~(b) shows a discrete approximation to the 2D ground state, obtained by arranging vertical slices sequentially into fixed vertical bins of fixed extent $\Delta z$, while preserving the $P$, $s$ and $\chi$ of each slice. Each slice occupies the same area as before (because its mass and density each remain constant) but now the slices are arranged horizontally within the bin, with left to right corresponding to increasing height in panel (a). Panel~(c) shows the state in panel (b) but sorted horizontally, which does not change the energy and removes the imprint of the foliation. Panel~(d) shows the expected equivalent of panel~(c) at very large resolution, but is obtained differently, by taking the small-thermodynamic-temperature limit of Lynden-Bell statistical mechanics (Section~\ref{sec:LyndenBellEquations}).}
    \label{fig:2Drestacking}
\end{figure*}

A foliated minimum-energy state may be considered a 1D representation of a minimum-energy state that is actually 2D, as follows. Because the scale of foliation is set by $\Delta m$, we can reconfigure the foliated state [figure~\ref{fig:2Drestacking}(a)] into a 2D state [figure~\ref{fig:2Drestacking}(b)] by rearranging fluid parcels locally in $z$, i.e., over a small vertical distance $\delta z \sim  \Delta m / \rho$, and then sorting globally in $x$ (with $P$, $s$ and $\chi$ fixed for each parcel) to remove small-scale variation [figure~\ref{fig:2Drestacking}(c) and~(d)]. Because the foliated state is a stable equilibrium with respect to local rearrangements in $z$, the force acting on each fluid parcel in the new state will be proportional to $\delta z$, and therefore vanishes as $\Delta m \to 0$. Thus, the 2D state is also an equilibrium state with the same energy as the foliated one.\footnote{More formally, the difference in energy between the two states under the operation described is $E_{\mathrm{2D}}-E_{\mathrm{1D}} =  \int \dd z \dd x\, \rho g\delta z$. Because $\delta z =\mathcal{O}( \Delta m)$ as $\Delta m \to 0$, we would have that ${E_{\mathrm{2D}}-E_{\mathrm{1D}}=\mathcal{O}(\Delta m)}$ if there existed a finite density difference between neighbouring slices. However, if the difference in density between neighbouring slices is $\mathcal{O}( \Delta m)$ [as required by~\eqref{samerho}], then $\delta z$ is the only rapidly varying function of $x$ and $z$ in the integral and therefore can be replaced by its coarse-grained average. This average is zero, because there is no net displacement of fluid parcels in each horizontal band. Thus $E_{\mathrm{2D}}-E_{\mathrm{1D}} = \mathcal{O}(\Delta m^2)$, so the difference in energy per fluid parcel vanishes.} In Section~\ref{viscous_case}, we shall show with direct numerical simulations that minimum-energy 2D states are the result of the nonlinear relaxation of a destabilised metastable state in certain regimes.

\begin{figure}
    \centering
    \includegraphics[width= .6\textwidth]{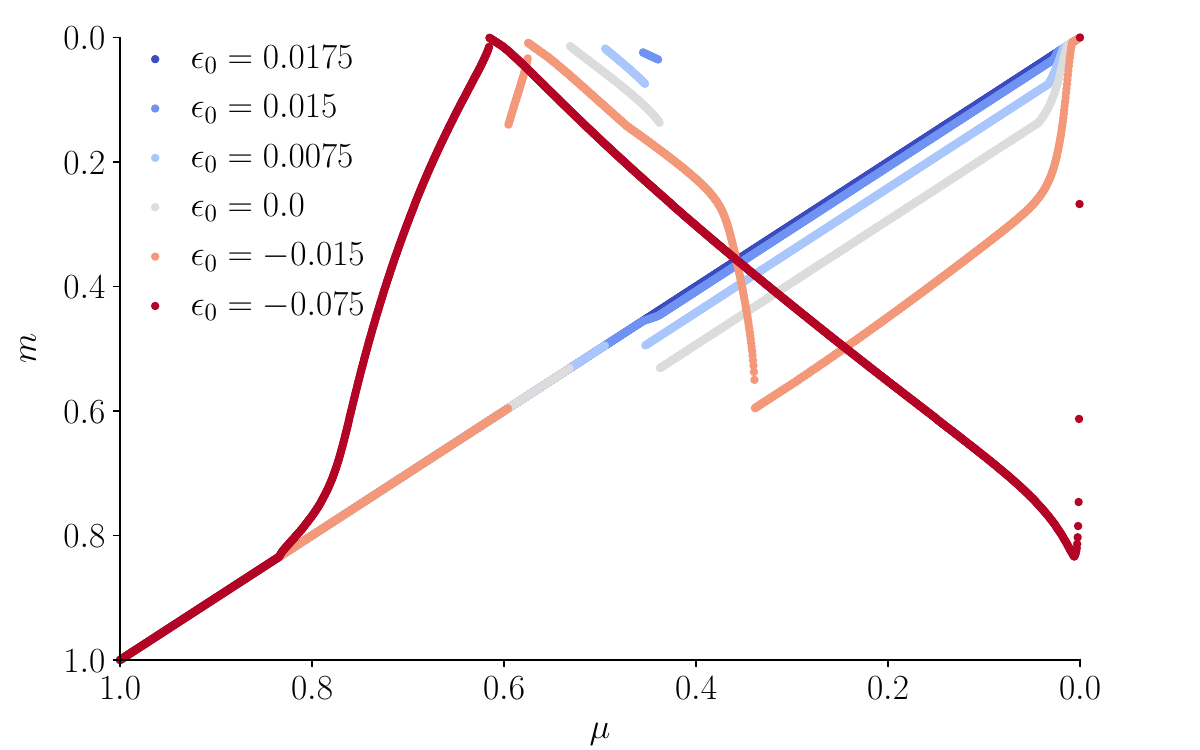}
    \caption{The energy-minimising assignments of slices from initial supported mass~$\mu$ to new supported mass~$m$ for the initial profile~\eqref{m_profile} with different values of the parameter $\epsilon_0$, which controls linear stability via~\eqref{epsilon_buffer}.}
    \label{fig:constbeta_epilons}
\end{figure}

To conclude our discussion of minimum-energy states of equilibria defined by~\eqref{m_profile}, we present in figure~\ref{fig:constbeta_epilons} a comparison of energy-minimising assignments for different values of the parameter $\epsilon_0$, which controls the distance from marginal stability [see~\eqref{epsilon_buffer}]. First, we note that $\epsilon_0 = 0$, i.e., marginal linear stability, is not a special point as far as the minimum-energy assignments are concerned: the assignments with $\epsilon_0 = 0.0075$ (linearly stable), $\epsilon_0 = 0.0$ (marginal) and $\epsilon_0 = -0.015$ (linearly unstable) are qualitatively similar, although, in the unstable cases of $\epsilon_0 = -0.015$ and $\epsilon_0 = -0.075$, foliation occurs over a much wider range of supported masses (and material from three different initial locations are foliated together near the top of the atmosphere). 

\begin{figure}
    \centering
    \includegraphics[width= .6\textwidth]{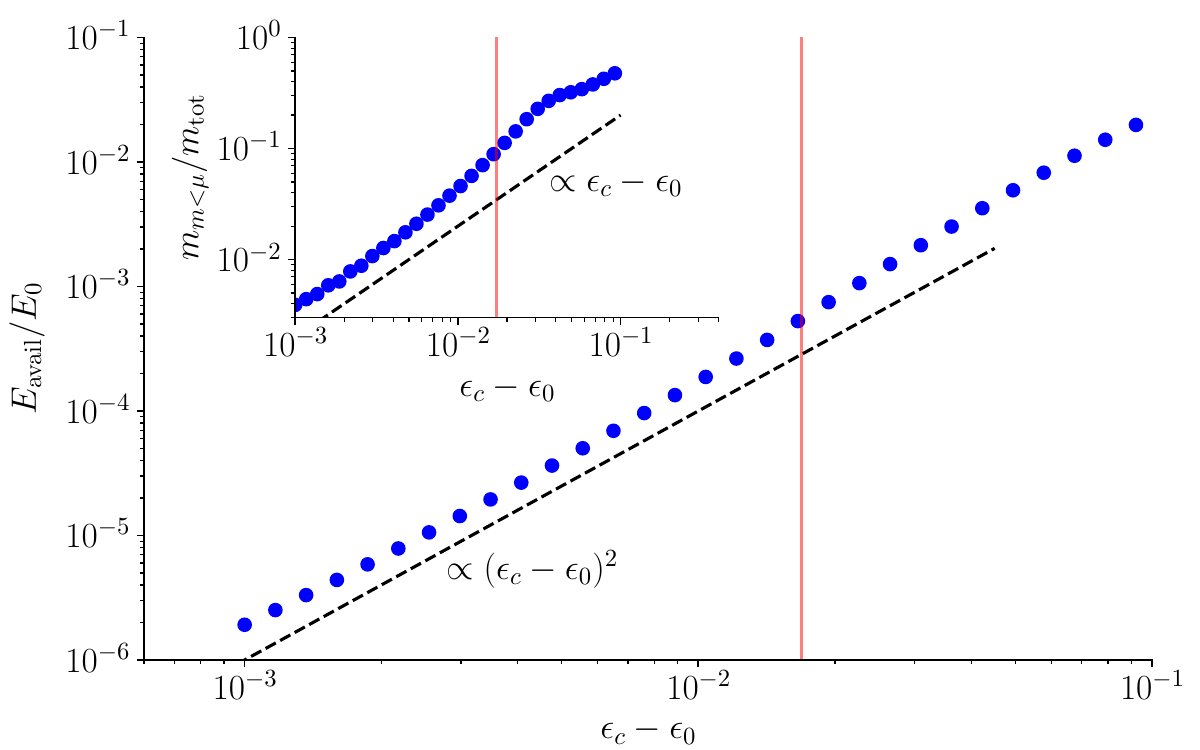}
    \caption{The available energy $E_{\mathrm{avail}}=E_0-E_{\mathrm{min}}$ as a fraction of the initial potential energy $E_0$, plotted as a function of $\epsilon_c-\epsilon_0$, where $\epsilon_c\simeq 1.7\times 10^{-2}$ is the largest value of $\epsilon_0$ in~\eqref{epsilon_buffer} for which the initial state is metastable. The inset shows the fraction of fluid that is assigned to a smaller supported mass than its initial one under optimal reassignment. Red lines correspond to $\epsilon_0=0$.}
    \label{fig:constbeta_deltaE}
\end{figure}

Figure~\ref{fig:constbeta_deltaE} shows the ratio of the available energy ${E_{\mathrm{avail}}=E_0-E_{\mathrm{min}}}$ to the original potential energy $E_0$ as a function of $\epsilon_c-\epsilon_0$, where $\epsilon_c\simeq 1.7\times 10^{-2}$ is the largest value of $\epsilon_0$ for which the atmosphere has a restacking with smaller energy. We see that $E_{\mathrm{avail}}/ E_0$ is small: it is around $10^{-3}$ for $\epsilon_0 = 0$, which is the value that corresponds to figures \ref{fig:selfsimilar_h} and \ref{fig:selfsimilar_normalform}. This is despite the fact that the minimum-energy assignment involves significant rearrangement of the atmosphere (around $10\%$ by mass of the atmosphere is reassigned upwards for $\epsilon_0 = 0$, see inset to figure~\ref{fig:constbeta_deltaE}). As explained in Section~\ref{sec:smallE}, the reason for the smallness of $E_{\mathrm{avail}}/ E_0$ is the fact that fluid slices exclude each other and so only very few of them can experience a significant change in total pressure as a result of reassignment. 

We observe from figure~\ref{fig:constbeta_deltaE} that 
\begin{equation}
    \frac{E_{\mathrm{avail}}}{E_0}\propto (\epsilon_c-\epsilon_0)^2 \,\,\, \mathrm{as}\,\,\epsilon_c-\epsilon_0\to 0;\;\label{dE/E}
\end{equation}this scaling is readily interpreted as a result of the fact that both (\textit{i}) the typical amount of energy liberated when a slice is reassigned from the bottom of the atmosphere to the top, and (\textit{ii}) the number of slices that are reassigned in this way, are proportional to $\epsilon_c - \epsilon_0$ when the latter is small [see inset to figure~\ref{fig:constbeta_deltaE}; $s$, $\chi$ and, therefore, the buoyancy force on a displaced fluid element,~\eqref{F2pressure}, depend linearly on $\epsilon_0$, by~\eqref{dlnsdm}]. 
This argument being independent of the particular profile under consideration, we expect a quadratic dependence of $E_{\mathrm{avail}}/E_0$ on $\epsilon_c - \epsilon_0$ for any metastable profile equilibrium with  $\epsilon_0$ sufficiently close to $\epsilon_c$. We shall find in Section~\ref{sec:invm} that a quadratic scaling is indeed reproduced for the profile represented by~\eqref{invm}.

\subsubsection{Metastable downwards,~\eqref{invm} \label{sec:invm}}

\begin{figure}
    \centering
    \includegraphics[width=0.8\columnwidth]{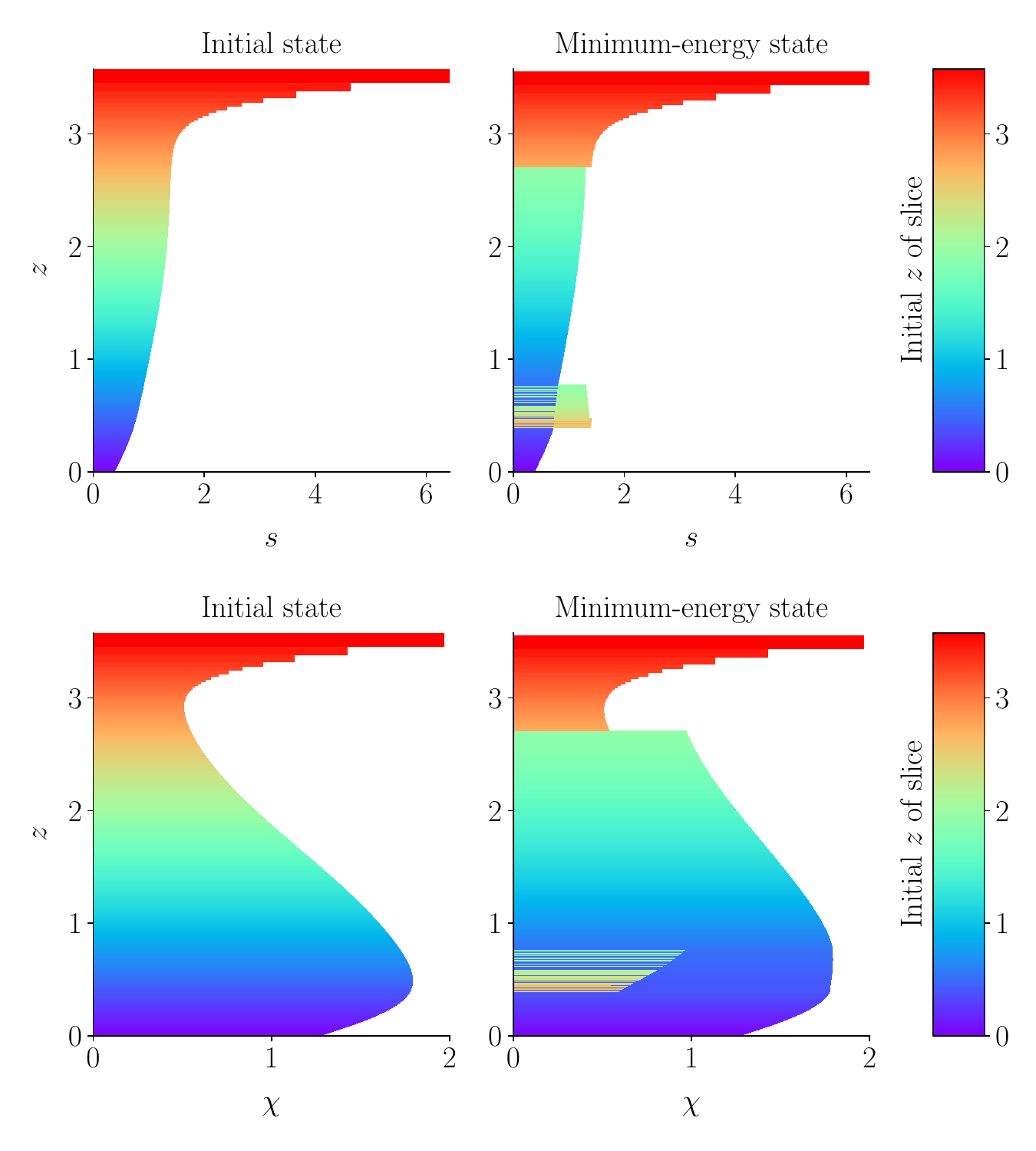}
    \caption{Visualisation of the minimum-energy assignment for the equilibrium defined by~\eqref{invm}, with $\epsilon_0=0$ in~\eqref{epsilon_buffer}. Details are the same as for figure~\ref{fig:selfsimilar_h}.}
    \label{fig:invm_h}
\end{figure}

We now turn to the second example case introduced in Section~\ref{sec:metastability_theory},~\eqref{invm}, which describes an initial state that is metastable to downwards displacements. The 1D minimum-energy state associated with this profile [with $\epsilon_0=0$ in~\eqref{epsilon_buffer}] is shown in figure~\ref{fig:invm_h}, which is the analogue for~\eqref{invm} of figure~\ref{fig:selfsimilar_h}. The minimum-energy assignment is similar qualitatively to the one examined in Section~\ref{sec:selfsimilar}: in this case, material from the top of the atmosphere moves to the bottom, reverses stacking order, and becomes foliated with the material already there (in fact, material from three different initial heights becomes foliated). 

Figure~\ref{fig:invm_assignments} shows the dependence of the optimal assignment on the value of $\epsilon_0$ in~\eqref{epsilon_buffer}. A qualitatively new feature appears in the cases of ${\epsilon_0=-0.025}$ and ${\epsilon_0=-0.069}$: for these unstable equilibria, there exists a range of the initial supported mass coordinate (in the vicinity of $\mu\simeq 0.25$ for the former case and $\mu\simeq 0.4$ for the latter) for which slices that are neighbouring in the initial state are alternately assigned to two different final locations. Like foliation, this phenomenon can be interpreted as a consequence of our seeking a 1D optimisation when, in fact, the true optimal assignment is higher-dimensional in the continuous limit $\Delta m \to 0$. In this case, the optimal assignment involves splitting the fluid at given height in a horizontal sense, and reassigning it to multiple new locations. In Appendix~\ref{app:onetomany}, we derive a necessary condition for this kind of one-to-many assignment and prove that two is, in fact, the largest possible value of ``many''.

\begin{figure}
    \centering
    \includegraphics[width=.6\textwidth]{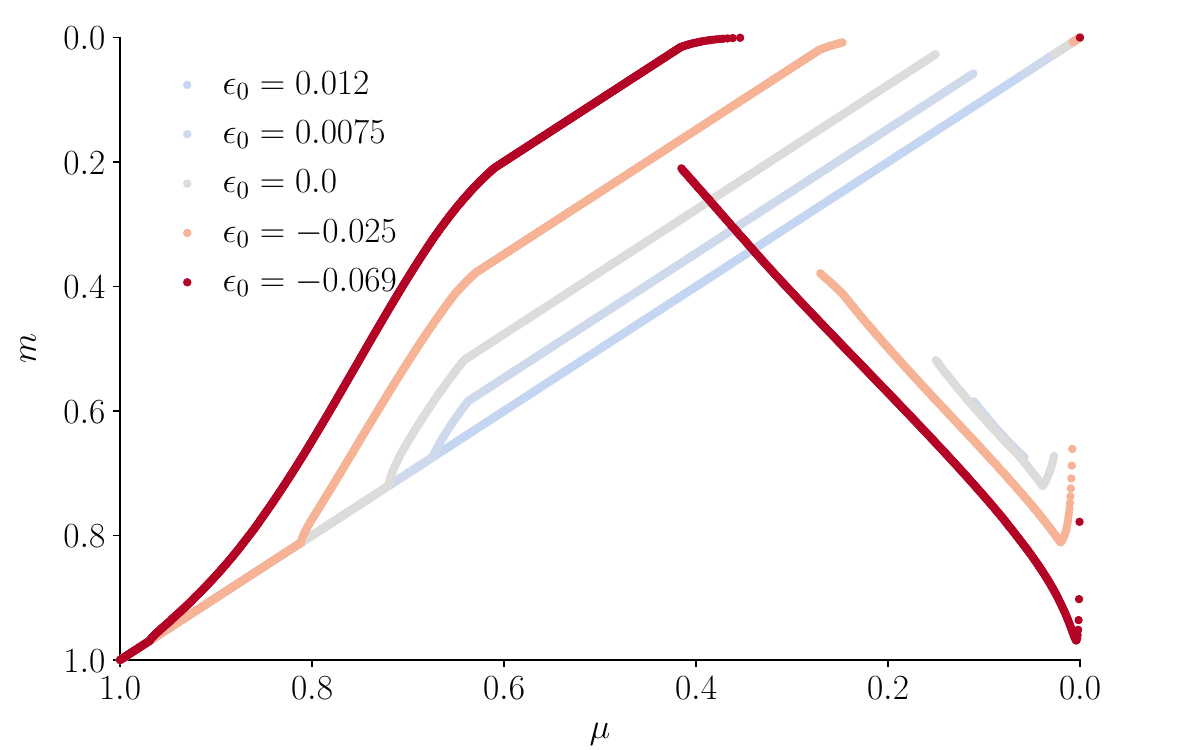}
    \caption{Minimum-energy assignments for the profile~\eqref{invm}. We observe that the optimal assignment is one to two over certain ranges of $m_1$ in the cases with $\epsilon_0<0$.}
    \label{fig:downwards_epilons}
\end{figure}

\begin{figure}
    \centering
    \includegraphics[width=.6\textwidth]{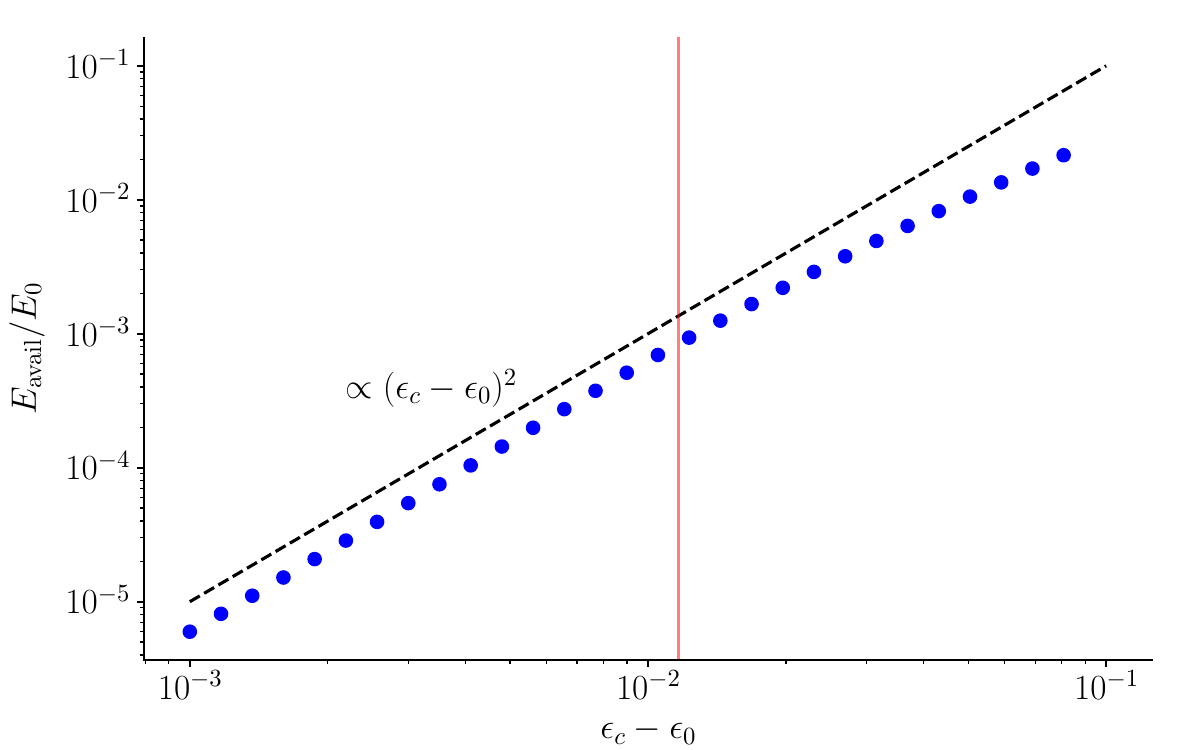}
    \caption{Fractional available energy as a function of $\epsilon_c - \epsilon_0$ for the profile~\eqref{invm}, where $\epsilon_c=1.2\times10^{-2}$ is the largest value of~$\epsilon_0$ in~\eqref{epsilon_buffer} for which the equilibrium is metastable.}
    \label{fig:invm_assignments}
\end{figure}

\begin{figure}
    \centering
    \includegraphics[width=0.8\columnwidth]{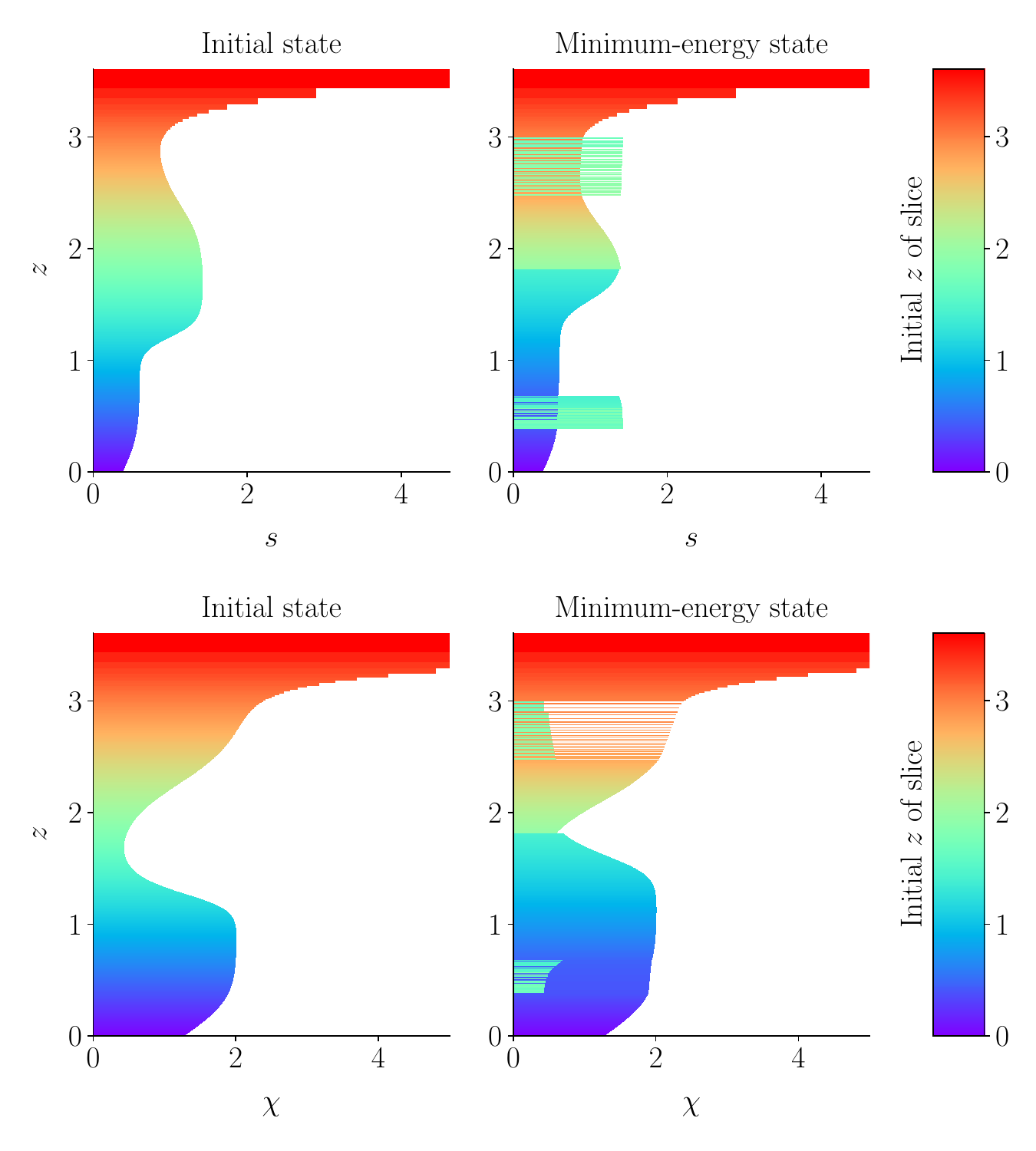}
    \caption{Visualisation of the minimum-energy assignment for the equilibrium defined by~\eqref{bump}, with $\epsilon_0=0$ in~\eqref{epsilon_buffer}. Details are the same as for figure~\ref{fig:selfsimilar_h}.\label{fig:bump_h}}
\end{figure}

\begin{figure}
    \centering
    \includegraphics[width= .6\textwidth]{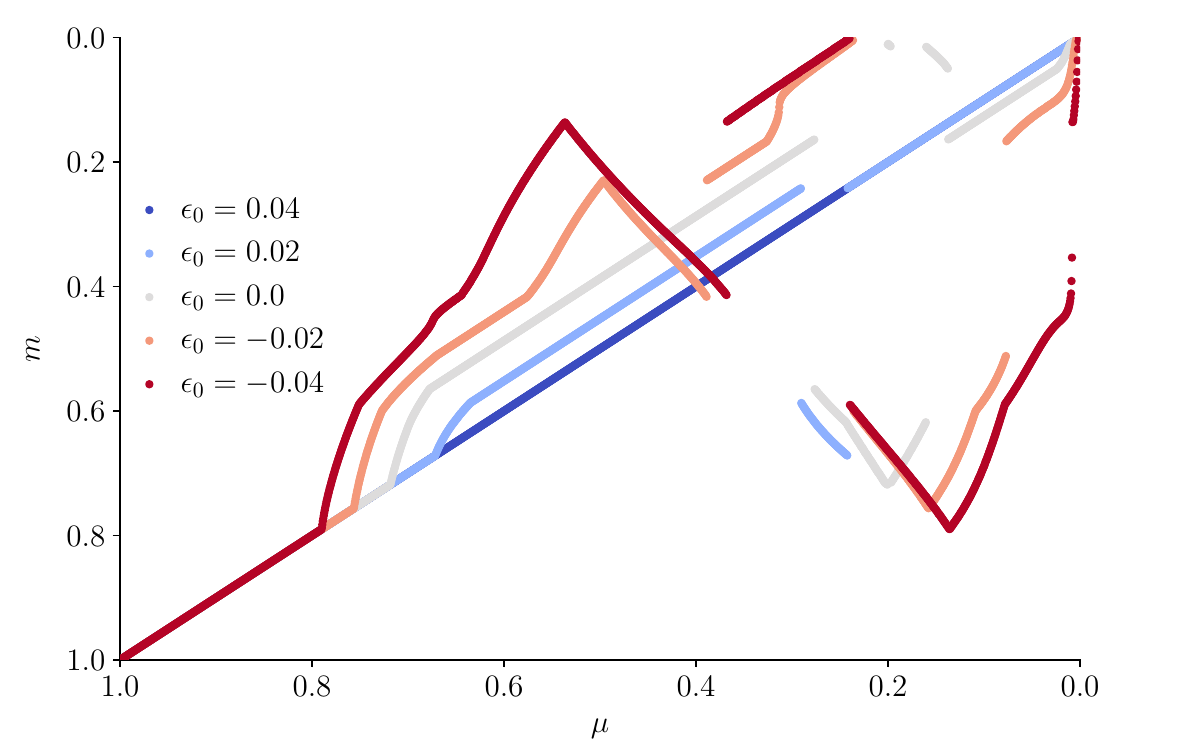}
    \caption{Minimum-energy assignments for the profile~\eqref{bump} for different values of $\epsilon_0$ in~\eqref{epsilon_buffer}.}
    \label{fig:bump_assignments}
\end{figure}

\subsubsection{Bi-directional metastability,~\eqref{bump}\label{sec:bump}}

We visualise in figure~\ref{fig:bump_h} the optimal assignment for the profile defined by~\eqref{bump}, which has a local maximum in its profile of $s/\chi$ at $z\simeq 1.75$ [see figure~\ref{fig:forcez1z2}(c)] and therefore the fluid is metastable both to upward and downward displacements there. It is intuitive, therefore, that the minimum-energy state should be obtained by reassignment of slices from the vicinity of $z\simeq 1.75$ to both the top and the bottom of the region that is at marginal linear stability. This is indeed the case in figure~\ref{fig:bump_h} (with foliation between moved and ambient fluid).

Figure~\ref{fig:bump_assignments} shows the optimal assignments for different values of $\epsilon_0$. The available energy associated with this profile is a fraction $6\times 10^{-4}$ of the initial total energy for $\epsilon_0 = 0$, which is comparable to the equilibria considered in sections~\ref{sec:selfsimilar} and~\ref{sec:invm}.

\subsubsection{Overturning metastability,~\eqref{dip}\label{sec:dip}}

\begin{figure}
    \centering
    \includegraphics[width=0.8\columnwidth]{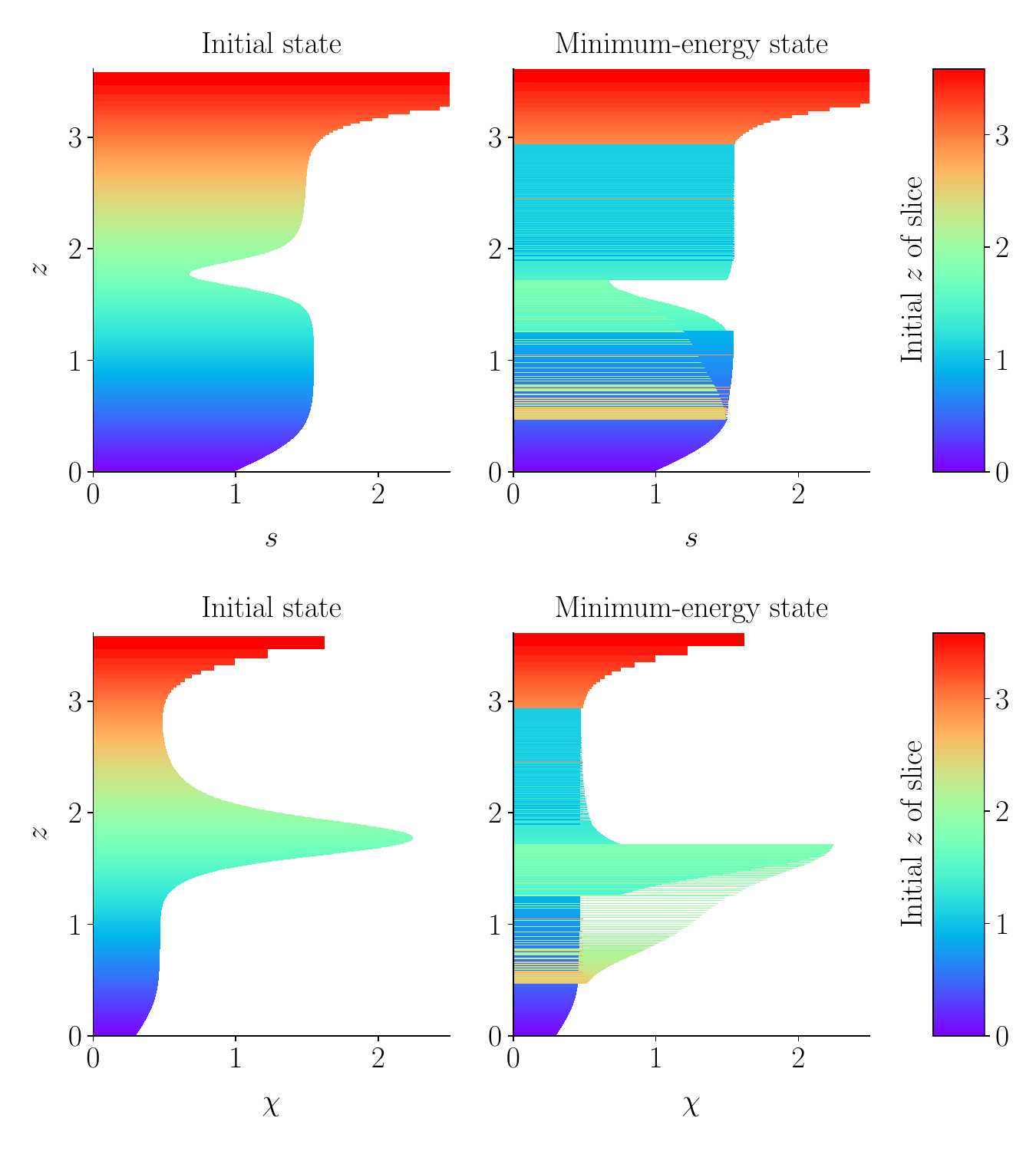}
    \caption{Visualisation of the minimum-energy assignment for the equilibrium defined by~\eqref{dip}, with $\epsilon_0=0$ in~\eqref{epsilon_buffer}. Details are the same as for figure~\ref{fig:selfsimilar_h}.\label{fig:dip_h}}
\end{figure}

\begin{figure}
    \centering
    \includegraphics[width= .6\textwidth]{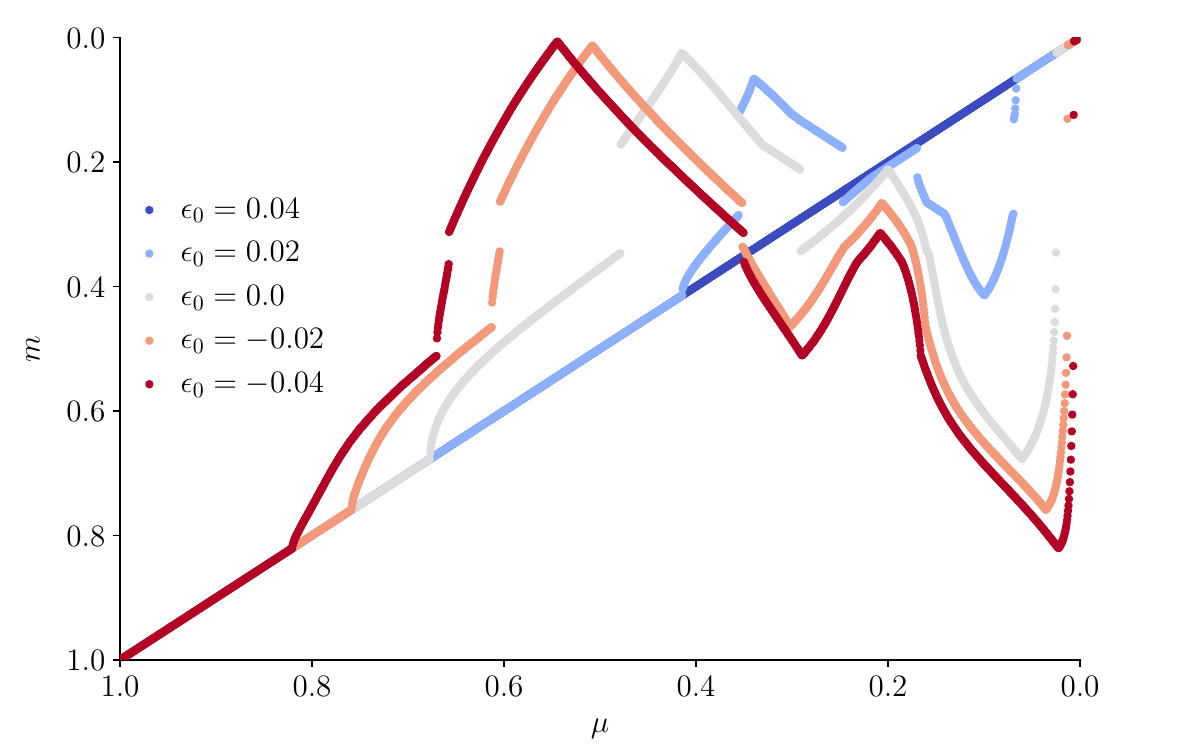}
    \caption{Minimum-energy assignments for the profile~\eqref{dip} for different values of $\epsilon_0$ in~\eqref{epsilon_buffer}.}
    \label{fig:dip_assignments}
\end{figure}

In figure~\ref{fig:dip_h}, we visualise the optimal assignment for the profile defined by~\eqref{dip}, which, in contrast to~\eqref{bump}, has a local minimum in its profile of $s/\chi$ at $z\simeq 1.75$ [see figure~\ref{fig:forcez1z2}(d)]. The fluid at the top of the atmosphere is therefore nonlinearly unstable to downwards motions, while the fluid at the bottom is nonlinearly unstable to upwards motions---we expect therefore that the minimum-energy state will be reached by an ``overturning'' of the atmosphere. This is roughly what we observe in figure~\ref{fig:dip_h}, although the precise optimal assignment is remarkably complex [see figure~\eqref{fig:dip_assignments} for a visualisation of the optimal assignments for different values of $\epsilon_0$].  The available energy associated with this profile at $\epsilon_0 = 0$ is $2\times 10^{-3}$ of the initial total, which is somewhat more than that of the equilibria considered in Sections~\ref{sec:selfsimilar},~\ref{sec:invm} and~\ref{sec:bump}.

\section{Relaxation without diffusion\label{viscous_case}}

In the remainder of this paper, we consider the problem of predicting the state to which a metastable equilibrium relaxes when destabilised. We shall focus on the case where relaxation is \textit{complete}, i.e., the destabilisation of the initial equilibrium is sufficiently violent to liberate the system from its metastable equilibrium completely. We assume that the system thereafter explores its configuration space freely, only subject to the constraints imposed by conservation laws. We shall assess with numerical simulations whether this is indeed a good assumption in Sections~\ref{sec:LowReNumerics} and~\ref{sec:numerics2}.

The relevant conservation laws are those of total energy and, to the extent that non-ideal processes (i.e., thermal conduction and resistive and viscous heating; see Appendix~\ref{app:numerical} for a statement of the MHD equations including these effects) can be neglected, of $s$ and $\chi$ for each fluid element. We shall assume in what follows that viscous heating is negligible because the kinetic energy that develops during relaxation is limited by the available energy of the initial equilibrium, which is small compared with the internal energy (Section~\ref{sec:smallE}). It follows that the ultimate deposition of kinetic energy as heat does not change the internal energy by very much (even locally). On the other hand, diffusion of $s$ and $\chi$ by thermal conduction and resistivity in a well-mixed state may change their values significantly, since $s$ and $\chi$ can vary by an order-unity fraction between fluid elements. Two qualitatively different types of relaxation may therefore be distinguished: the one in which $s$ and $\chi$ do not diffuse during relaxation and the one in which they do. In this section, we consider the former case, which is realised when turbulent mixing is suppressed by viscosity. We shall argue that relaxation is then to the minimum-energy state calculated in Section~\ref{sec:ground_state_theory}. We address the case of fully turbulent relaxation in Section~\ref{sec:LyndenBell}.

\subsection{Conditions for diffusion to be absent\label{sec:conditions}}

In order for diffusion of $s$ and $\chi$ to be negligible during the whole period of relaxation, the flow generated during relaxation must decay before it is able to mix $s$ and $\chi$ to scales $l$ such that their diffusion timescales ($l^2/K$ and $l^2/\eta$, respectively, where $K$ and $\eta$ are the thermal and magnetic diffusivities) become comparable to the flow's decay timescale. The latter is $H^2/\nu$ ($\nu$ is the kinematic viscosity) independently of whether the flow is laminar or turbulent, because 2D turbulence does not cascade energy to smaller scales. A simple regime in which diffusion may be negligible is the one in which the flow is laminar. This requires a Reynolds number $\mathrm{Re}\equiv UH/\nu \lesssim 1$, where $U$ is the characteristic velocity developed during relaxation and $H$ the stratification height, which we take to be the characteristic outer scale of the flow (this being necessary for complete relaxation). By~\eqref{deltaE/dm}, $\delta \rho \sim \rho E_{\mathrm{avail}}/E_0$, so the net gravitational force on a fluid element is $g\delta \rho \sim \rho c^2 E_{\mathrm{avail}}/HE_0$, where $c$ is the characteristic speed of compressive waves (assumed constant here for simplicity). Balancing this with the viscous force $\rho \nu U/H^2$ to estimate $U\sim Hc^2 E_{\mathrm{avail}}/\nu E_0$, we find that relaxation is laminar if
\begin{equation}
    \nu \gtrsim Hc\sqrt{\frac{E_{\mathrm{avail}}}{E_0}}.\label{viscous_condition}
\end{equation}If~\eqref{viscous_condition} is satisfied, the flow turns over at most once before it decays, so $s$ and $\chi$ are not mixed to smaller scales. They are therefore well conserved provided that
\begin{equation}
    \mathrm{Pr}_t,\,\,\mathrm{Pr}_m\gg 1,\label{Pm_viscous_condition}
\end{equation}where $\mathrm{Pr}_{t}\equiv \nu/K$ and $\mathrm{Pr}_{m}\equiv \nu/\eta$ are the thermal and magnetic Prandtl numbers, respectively.

For $\mathrm{Re}\gg 1$, the outer-scale flow has velocity ${U\sim  (E_{\mathrm{avail}}/E_0)^{1/2}c}$ and is turbulent. It turns over $\mathrm{Re}$ times before decaying (in 2D), so $s$ and $\chi$ are mixed to the scale ${l\sim H\exp(-\mathrm{Re})}$. Diffusion at this scale can be neglected if its timescale is longer than the decay time of the turbulence, i.e., if
\begin{equation}
    \ln \mathrm{Pr}_{t},\,\,\ln \mathrm{Pr}_{m} \gg \mathrm{Re}\sim \frac{cH}{\nu}\sqrt{\frac{E_{\mathrm{avail}}}{E_0}}.\label{inviscid_nodiffusion_condition}
\end{equation}

In addition to mixing by the flow at scale~$H$, Rayleigh--Taylor instability at interfaces of fluid with different densities may generate small-scale vortices that mix the fluid (this effect was evident in the ${t=70}$ panel of figure~\ref{fig:fig2}). The fastest-growing Rayleigh--Taylor mode, which develops at scale $L_{\mathrm{RT}}$, is limited by viscosity: its growth rate is ${\gamma_{\mathrm{RT}}\sim(gL_{\mathrm{RT}}\delta \rho/\rho)^{1/2}\sim \nu/ L_{\mathrm{RT}}^2}$, whence ${L_{\mathrm{RT}}\sim\mathrm{Re}^{-2/3} H}$ and ${\gamma_{\mathrm{RT}}\sim\mathrm{Re}^{1/3} U/H}$. Taking the nonlinear turnover rate of the developed vortex to be equal to $\gamma_{\mathrm{RT}}$, this mode turns over $\mathrm{Re}^{4/3}$ times before the outer-scale turbulence decays (at which point we assume that no Rayleigh--Taylor-unstable interfaces remain). An argument similar to the one that led to~\eqref{viscous_condition} indicates that the Rayleigh--Taylor vortices will not establish diffusion-scale structure in $s$ and $\chi$ provided that
\begin{equation}
    \ln \mathrm{Pr}_{t},\,\,\ln \mathrm{Pr}_{m} \gg \mathrm{Re}^{4/3}\sim \left(\frac{cH}{\nu}\right)^{4/3}\left(\frac{E_{\mathrm{avail}}}{E_0}\right)^{2/3}.\label{viscous_condition_RT}
\end{equation}Equation~\eqref{viscous_condition_RT} is a stricter criterion than~\eqref{inviscid_nodiffusion_condition}; because of their faster turnover rate than the flow at scale $H$, the Rayleigh--Taylor-generated vortices are more effective at mixing.

If~\eqref{viscous_condition_RT} is satisfied [or if~\eqref{viscous_condition} and~\eqref{Pm_viscous_condition} are, for the case of $\mathrm{Re}\lesssim 1$], the relaxation flow decays before $s$ and $\chi$ diffuse via thermal conduction or ohmic heating. Provided that the initial destabilisation was sufficiently thorough (so that the system does not become trapped in a new metastable state), we expect the final static state of the system to be the one with smallest potential energy subject to fluid-element-wise conservation of~$s$ and~$\chi$. Because the amount of heat per unit mass generated by viscosity is $U^2\sim c^2 E_{\mathrm{avail}}/E_0\ll c^2$, the fractional change in $s$ of each fluid element due to viscous heating during relaxation is small. Thus, the final state reached by the system ought to be, to first approximation, the states of minimum energy calculated in Section~\ref{sec:ground_state_theory}. In the next section, we verify with numerical simulations that this is indeed the case.

\subsection{Numerical results\label{sec:LowReNumerics}}

\begin{figure*}
    \centering
    \includegraphics[width=\textwidth]{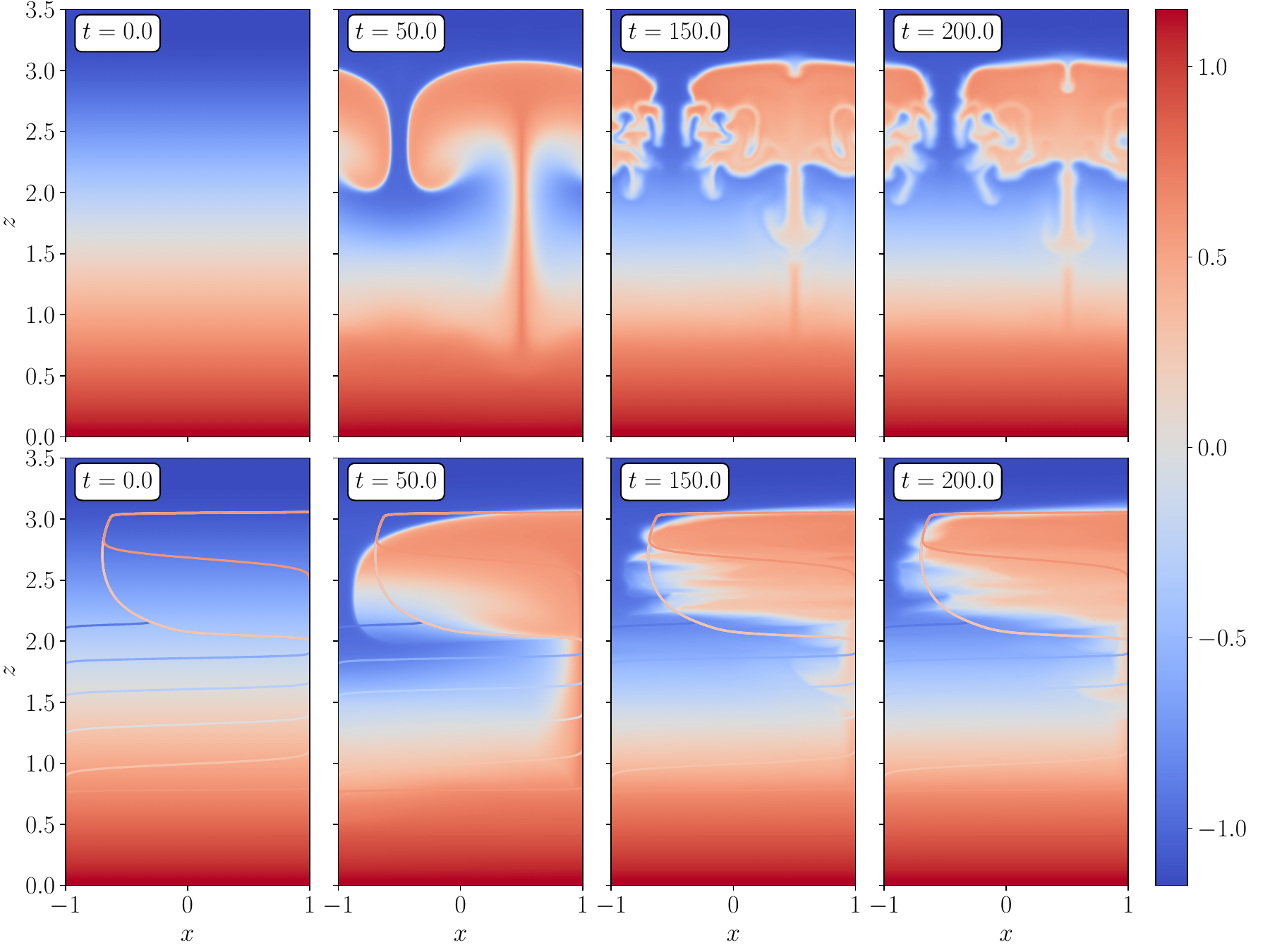}
    \caption{The relaxation of the equilibrium defined by~\eqref{m_profile} at $\mathrm{Re}\sim 10^2$. The initial velocity field is given by~\eqref{u0}. Upper panels show the evolution of $\ln(s/\chi)$ in $x$-$z$ space. Lower panels show the same quantity but sorted horizontally at each $z$, with contours of the theoretical minimum-energy state [figure~\ref{fig:2Drestacking}(d)] overlaid.}
    \label{fig:upwards_2D_sim}
\end{figure*}

\begin{figure*}
    \centering
    \includegraphics[width=.85\textwidth]{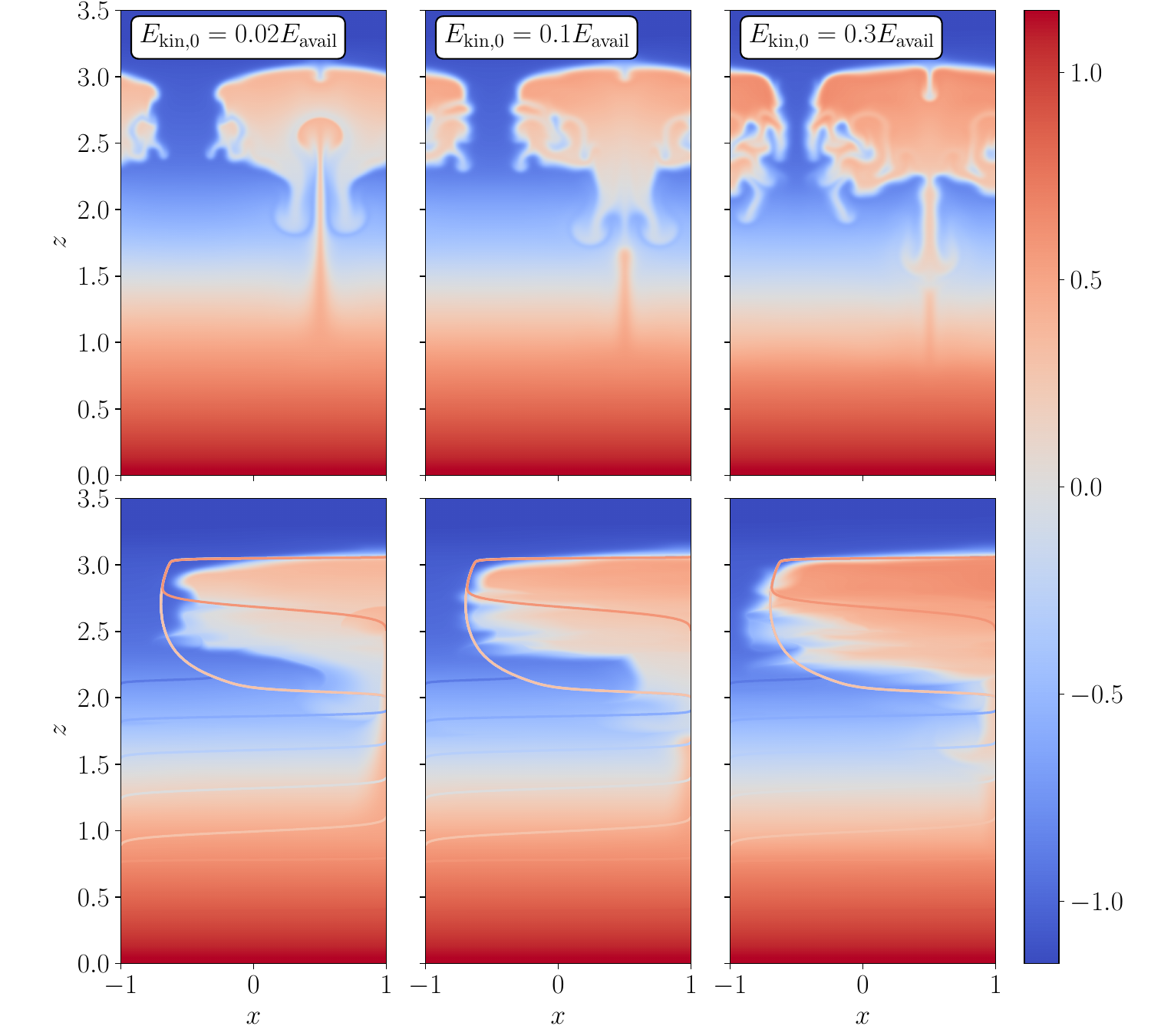}
    \caption{The distribution of $\ln(s/\chi)$ at $t=300$ for simulations analogous to the one visualised in figure~\ref{fig:upwards_2D_sim} but for three different values of $u_0$. As in figure~\ref{fig:upwards_2D_sim}, upper panels visualise the state of the simulation, while lower panels are sorted horizontally and overlaid with the 2D minimum-energy state [figure~\ref{fig:2Drestacking}(d)].}
    \label{fig:upwards_2D_v0_comparison}
\end{figure*}

Figure~\ref{fig:upwards_2D_sim} visualises the relaxation of the equilibrium defined by~\eqref{m_profile} after the application of an impulsive force that accelerates the fluid to a velocity
\begin{equation}
    \bu = u_0 \boldsymbol{\hat{z}}\sin\left(\frac{2\pi x}{L_x}\right)\exp\left(-\frac{(z-z_0)^2}{\Delta z^2}\right),\label{u0}
\end{equation}where $L_x = 2$ is the size of the simulation domain in the~$x$ direction, $u_0=0.1$, $z_0=1.0$ and $\Delta z = 0.5$ (see Section~\ref{sec:examples} for an explanation of our system of units). This corresponds to an initial kinetic energy $E_{\mathrm{kin,}0} \simeq 0.3 E_{\mathrm{avail}}$, where the available energy $E_{\mathrm{avail}}\sim 10^{-3}E_0$. The kinematic viscosity is $1.6\times 10^{-3}$ in these units, so the Reynolds number of the flow at the initial time is~$\mathrm{Re}\sim u_0 L_x/\nu\sim 10^{2}$. The magnetic and thermal Prandtl numbers are $\mathrm{Pr}_m = 400$ and $\mathrm{Pr}_t = 670$, respectively. We provide further details of the numerical set-up in Appendix~\ref{app:numerical}.

Although $\mathrm{Re}\sim 100>1$, figure~\ref{fig:upwards_2D_sim} shows that $\mathrm{Re}
$ is insufficiently large for turbulence to develop, either at the outer scale or driven by Rayleigh--Taylor instability (which does nonetheless lead to the development of structure at scales smaller than $L_x\sim H$). Thus, relaxation takes place without significant diffusion of $s$ and $\chi$ because the Prandtl numbers are large [see~\eqref{Pm_viscous_condition}]. We observe that the upwards plume generated by the initial impulse forms a long-lived 2D state (upper panels of figure~\ref{fig:upwards_2D_sim}), which is indeed consistent with the minimum-energy state obtained in Section~\ref{sec:selfsimilar} (see lower panels of figure~\ref{fig:upwards_2D_sim}).

In order to assess the sensitivity of the final state to the initial perturbation, we visualise in figure~\ref{fig:upwards_2D_v0_comparison} the late-time state developed by simulations identical to the one shown in figure~\ref{fig:upwards_2D_sim}, but with different values of the initial kinetic energy. Specifically, we choose $u_0=0.05$ (centre panel) and $u_0=0.025$ (left panel), so that $E_{\mathrm{kin,0}}\simeq 0.1 E_{\mathrm{avail}}$ and $E_{\mathrm{kin,0}}\simeq 0.02 E_{\mathrm{avail}}$, respectively. We observe that there is some sensitivity to the amplitude of the initial perturbation: somewhat more material is displaced upwards at $E_{\mathrm{kin,0}}\simeq 0.3 E_{\mathrm{avail}}$ than $E_{\mathrm{kin,0}}\simeq 0.02 E_{\mathrm{avail}}$. This weak, but measurable, dependence of the final state on initial conditions is despite the fact that the equilibrium is initially at marginal linear stability, so there is no potential barrier to be overcome in order to trigger instability. However, partial relaxation stabilises the atmosphere (see Appendix~\ref{sec:anaytic_ground_state}), so that, while the first magnetic-flux tube to move upwards experiences no potential barrier, later ones do.

In Appendix~\ref{sec:2D_sims_down}, we present analogous simulations of viscous relaxation for the equilibrium defined by~\eqref{invm} (i.e., metastable to downwards perturbations). The results are qualitatively similar to those presented in this section.

\section{Statistical theory of relaxation at large Reynolds number\label{sec:LyndenBell}}

In this section, we consider relaxation for which the Reynolds number is sufficiently large that inequality~\eqref{viscous_condition_RT} no longer holds. In this case, turbulent mixing generates sufficiently fine-scale structure in $s$ and $\chi$ to enable diffusion. Consequently, the equilibrium reached once the velocity field has decayed cannot be obtained by an ideal rearrangement of the initial state and so will, in general, differ from the minimum-energy states discussed in Section~\ref{sec:ground_state_theory}. 

Given that diffusion is to occur, the key question is \textit{which fluid parcels are brought into contact by the turbulent flow?} If we could identify in advance which fluid parcels were to diffuse with which others and thus become locally homogenised, then we could predict their new values of the ideal Lagrangian invariants $s$ and $\chi$ from the conservation of net magnetic flux and enthalpy during diffusion. The relaxed state would then be the one with minimum energy subject to ideal rearrangements of this post-diffusion state.

Motivated by the chaotic nature of turbulent mixing, we shall treat this problem probabilistically, i.e., with statistical mechanics. We assume that the timescale for thorough mixing is much shorter than that for the onset of diffusion, so that these processes can be treated separately. Specifically, we assume that, \textit{before diffusion acts}, turbulent mixing causes the system to explore all possible two-dimensional distributions of $s$ and $\chi$ (microstates) consistent with its potential energy.\footnote{For simplicity, we take the potential energy to be constant and equal to the total initial energy of the system (where we make comparison with numerical simulations in Section~\ref{sec:numerics2}, this includes the kinetic energy of the perturbation).} We seek the probability-distribution function (the macrostate) for fluid at a given spatial position to have originated from a different given position. We obtain it by maximising the number of microstates for which it gives correct coarse-grained statistics, i.e., by maximising mixing entropy. Finally, we shall determine the diffused states by taking moments of this distribution, as we  explain in Section~\ref{sec:diffusion}.

\subsection{Equivalence of 1D and 2D microstates\label{sec:1dmicrostates}}

\begin{figure*}
    \centering
    \includegraphics[width=\textwidth]{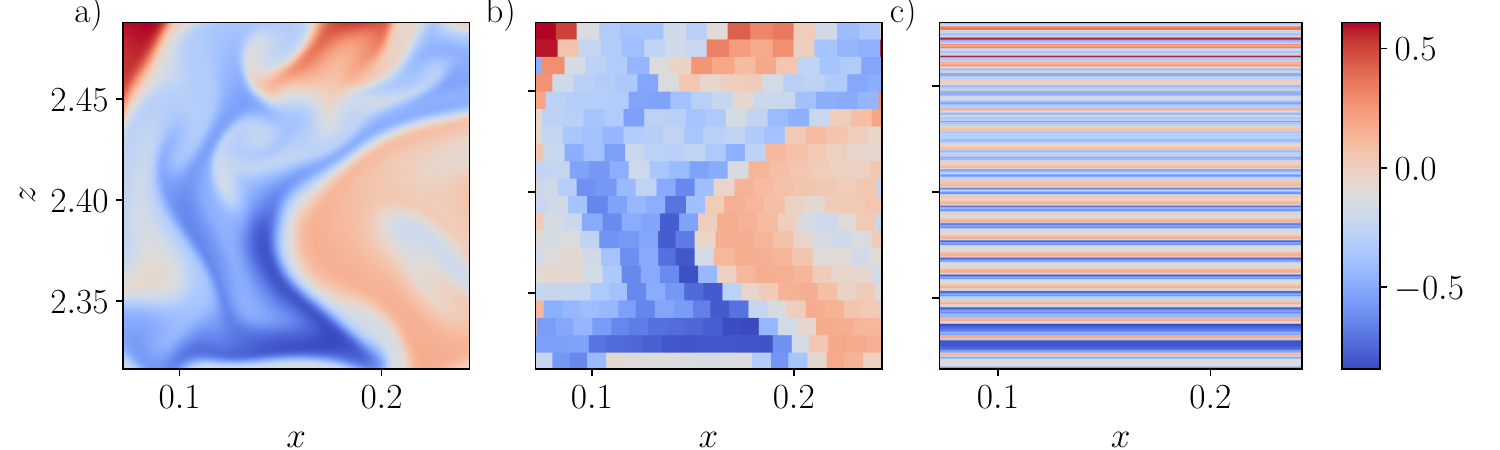}
    \caption{At $\mathrm{Re}\gg 1$, turbulence mixes the advected scalars $s$ and $\chi$ [panel~(a); this is a subsection of the state visualised in the $t=70$ panel of figure~\ref{fig:turb_simulation_upwards}]. Panel~(b) is a copy of panel~(a) but with the fluid discretised into parcels of equal mass. The 2D non-equilibrium states obtainable by shuffling these parcels while maintaining horizontal pressure balance constitute microstates in our theory. As explained in the main text, there exists a bijection between such 2D states and 1D static equilibria with the same energy [panel~(c)]. We may therefore take microstates to be 1D equilibria.}
    \label{fig:2d_to_1d_restack}
\end{figure*}

Even through the relaxing system explores two-dimensional non-equilibrium microstates, maximising the number of them that are consistent with a given macrostate (i.e., the multiplicity of the macrostate) is formally equivalent to doing so for the equilibria obtainable by 1D rearrangements (similar to those we used to find minimum-energy states in Section~\ref{sec:ground_state_theory}). This is because, for every 2D non-equilibrium state in horizontal pressure balance, there exists a distinct 1D equilibrium state with the same total energy and mass of fluid with each value of $s$ and $\chi$, as we now demonstrate.

We define the set of 2D microstates as follows. We partition the initial equilibrium into horizontal slices of fixed width $\Delta z$, which we further subdivide into parcels of equal mass~$\Delta m$. The number of parcels in each horizontal slice is not fixed, and parcels may ``spill'' over into adjacent slices if the total mass in the slice is not an integer multiple of~$\Delta m$, although the fraction of parcels that do vanishes as $\Delta m \to 0$. We consider labelling each parcel by an index that increases along each slice from left to right, starting from the lowest slice and then moving upwards. Then, the full set of microstates is the set of possible permutations of these indices, i.e., the set of 1D shuffles of parcels with fixed $s$ and $\chi$, assembled in 2D space as described [see panels (a) and (b) of figure~\ref{fig:2d_to_1d_restack}].

A given permutation of parcel indices is not a complete description of the microstate, as the pressure~$P$ in each parcel remains to be specified. As noted above, the 2D microstates are not equilibria: horizontal density variations induce baroclinic torques. On the other hand, we do not expect significant horizontal variation in pressure: because ${E_{\mathrm{avail}}/E_0\ll 1}$, the flow that develops during relaxation is subsonic, i.e., $U/c\ll 1$. Fluctuations~$\delta P$ of the total pressure $P$ about its horizontal mean~$P_0$ are therefore small, ${\delta P/P_0\sim U^2/c^2 \ll 1}$. We can evaluate $P_0(z)$ by integrating the $z$-component of the momentum equation~\eqref{momentum} over $x$ and from the given $z$ to $z=\infty$, neglecting inertial terms (which correspond to the pressure fluctuations). This yields $P_0= m g$, where $m$ is the total mass of fluid above height $z$ per unit horizontal length, given by
\begin{equation}
    m = \frac{1}{L_x}\int_0^{L_x} \dd x \int^{\infty}_z\dd z' \rho(x,z'),\label{m2d}
\end{equation}where $L_x$ is the horizontal extent of the system.

To leading order in $\Delta z$, \eqref{m2d} evaluated at a given parcel is the total mass of parcels that have greater indices. Consequently, if we were to rearrange the parcels to be stacked vertically in order of their indices (preserving the $P$, $s$ and $\chi$ from the 2D state), $P_0$ is the pressure they would have in equilibrium. Restacking in this way produces no change in the total energy as $\Delta z, \Delta m \to 0$, so the total energy of both states can be evaluated using~\eqref{quadratic_energy}. The leading-order term in $\delta P$ in the integrand is ${\sim c^2 \delta P^2/P_0^2 \sim (U^2/c^2)U^2\ll U^2}$, so the contribution of $\delta P$ to the energy of the state can be neglected. Thus, there exists a correspondence between 2D non-equilibrium (but horizontally pressure-balanced) states and 1D equilibria: for every 2D state, we can find a 1D equilibrium state with the same energy, by the process described above [visualised in figure~\ref{fig:2d_to_1d_restack}(c)]. Thus, in what follows, we consider microstates to be 1D equilibrium states, with energy given by~\eqref{discrete}.

\subsection{Lynden-Bell statistical mechanics of MHD atmospheres\label{sec:LyndenBellEquations}}

As explained in Section~\ref{sec:1dmicrostates}, we may construct our statistical mechanics on the space of 1D equilibria, this being fully equivalent to doing so on the space of 2D non-equilibria in horizontal pressure balance. We follow the formulation of Lynden-Bell's~(\citeyear{LyndenBell67}) statistical mechanics of distinguishable particles with an exclusion principle---originally derived for collisionless stellar systems and plasma---by \citet{Chavanis03}. 

We introduce $\mathcal{P}(m, \mu)\dd \mu$ as the probability of finding the material with initial supported mass in the range ${[\mu, \mu+\dd \mu]}$ to have a supported mass of $m$ in the final state. We obtain $\mathcal{P}(m, \mu)$ by maximising the number of microstates with which it is consistent after coarse graining. This corresponds to maximising the mixing entropy~\citep{RobertSommeria91}
\begin{equation}
    S = - \int \dd m \int \dd \mu \,\mathcal{P}(m, \mu) \ln \mathcal{P}(m, \mu).\label{entropy}
\end{equation}We maximise $S$ subject to the constraints of fixed total probability (i.e., the normalisation of $P$),
\begin{equation}
    \int \dd \mu \, \mathcal{P}(m, \mu) = 1,\quad\forall m;\label{lambda}
\end{equation}fixed potential energy $E_{\mathrm{pot}}$,
\begin{equation}
    \int \dd m \int \dd \mu \, \mcE(m,\mu) \mathcal{P}(m, \mu) = E_{\mathrm{pot}};\label{betaT}
\end{equation}and fixed mass of fluid with each value of $\mu$,
\begin{equation}
    \int \dd m \mathcal{P}(m, \mu) = 1, \quad\forall \mu.\label{mu}
\end{equation}This constrained maximisation of $S$ is equivalent to unconstrained maximisation of
\begin{align}
    & -\int \dd m \int \dd \mu \mathcal{P}(m, \mu) \ln \mathcal{P}(m, \mu) - \beta_T \int\dd m\lambda(m) \left[\int \dd \mu \, \mathcal{P}(m, \mu) - 1\right] \nonumber\\
    & - \beta_T \left[\int \dd m\int \dd \mu \, \mcE(m,\mu) \mathcal{P}(m, \mu) - E_{\mathrm{pot}}\right] - \beta_T \int \dd \mu \psi(\mu) \left[\int \dd m \, \mathcal{P}(m, \mu) - 1\right]\label{entropy-lagrangeMs}
\end{align}over the probability $\mathcal{P}(m,\mu)$ and the Lagrange multipliers $\beta_{T}$, $\lambda(m)$ and $\psi(\mu)$. The solution is
\begin{equation}
    \mathcal{P}(m, \mu) = \displaystyle e^{\displaystyle-\beta_T [\mathcal{E}(m, \mu)-\psi(\mu)-\lambda(m)]}\label{P}
\end{equation}where the Lagrange multipliers $\beta_T$ (the thermodynamic beta, to be identified with the inverse of the statistical mechanical temperature), $\psi (\mu)$ and $\lambda(m)$ are determined from the constraints~\eqref{lambda},~\eqref{betaT} and~\eqref{mu}.\footnote{Let us make explicit the analogy between these formulae and their equivalents in the Lynden-Bell theory of collisionless stellar systems and plasma. In those contexts, $\mathcal{P}(m,\mu)$ is replaced by $\mathcal{P}(\bx,\bv,\eta)$, where position~$\bx$ and velocity~$\bv$ are the phase space coordinates (analogous to $m$) and $\eta$ is the phase-space density, which, analogously to $\mu$, is conserved under rearrangements of phase space (Liouville's theorem). The energy density $\mcE(m,\mu)$ is replaced by the energy associated with $\eta$ particles occupying the~$(\bx,\bv)$ coordinates of phase space, $\eta[v^2/2 + \Phi(\bx)]$ (or appropriate generalisations), where $\Phi$ is potential energy. Finally, on the right-hand side of the constraint~\eqref{mu}, $1$ is replaced by a function of $\eta$, sometimes called the ``waterbag content'', which gives the total volume of phase space with density $\eta$. The equivalent object is a constant in our formalism because the mass of fluid in the range $[\mu,\mu+\dd\mu]$ is $\dd \mu$, independently of $\mu$. In Appendix~\ref{sec:alternative} we present an alternative formulation of our statistical mechanics, with $\mu$ replaced by $s$ and $\chi$, which are also preserved under rearrangement. In that formulation, a function $M(s,\chi)$ appears on the right-hand side of the equation analogous to~\eqref{mu}; this is the total mass of fluid with given $s$ and $\chi$.}

\begin{figure*}
    \centering
    \includegraphics[width=1.0\textwidth]{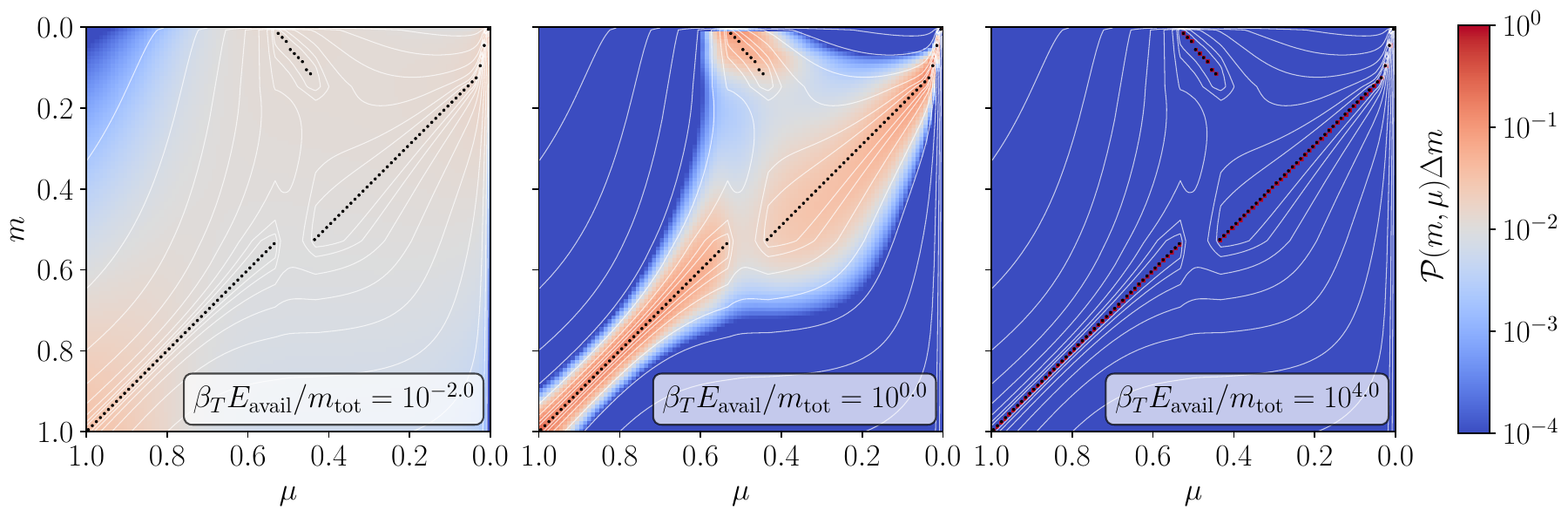}
    \caption{Convergence of $\mathcal{P}(m,\mu)$ [equation~\eqref{P}] to the solution of the LSA problem [black circles; see Section~\ref{sec:HungarianTheory}] as $\beta_T E_{\mathrm{avail}}\to\infty$ for the case of the unstable-upwards profile~\eqref{m_profile} with all integrals discretised at scale $\Delta m = 0.01$. Contours of the normal form of the cost matrix~$\tilde{\mcE}_{ij}$ (as visualised in figure~\ref{fig:selfsimilar_normalform}) are plotted in white. }
    \label{fig:LB_probs}
\end{figure*}

In Appendix~\ref{sec:LyndenBelltoLSA}, we prove that, as $\beta_T\to\infty$, $\mathcal{P}(m,\mu)$ becomes increasingly sharply peaked around the solution to the LSA problem of Section~\ref{sec:ground_state_theory} [see figure~\ref{fig:LB_probs}]. Thus, the minimum-energy states are the $\beta_T\to\infty$ limit of our statistical mechanics. Mathematically, this happens because the exponent in~\eqref{P} has the form of a modified cost matrix [see~\eqref{normalform}] multiplied by $\beta_T$; as $\beta_T\to \infty$, $\mathcal{P}(m,\mu)$ vanishes except in the vicinity of the zeros of the normal form of the cost matrix [see Section~\ref{sec:HungarianTheory}], which represent the optimal assignment.

The algorithm we use to evaluate $\mathcal{P}(m,\mu)$ numerically for given $E_{\mathrm{pot}}$ is analogous to the one proposed by~\citet{Ewart23}. A brief summary is as follows. First, we choose a trial value of $\beta_T$ and, with suitable discretisation for $m$ and $\mu$, calculate
\begin{equation}
    \mathcal{P}(m, \mu) = \frac{\displaystyle e^{\displaystyle-\beta_T [\mathcal{E}(m, \mu)-\psi(\mu)]}}{\displaystyle\int \dd \mu' e^{\displaystyle-\beta_T [\mathcal{E}(m, \mu')-\psi(\mu')]} },\label{Pfull}
\end{equation}where we obtain $\psi(\mu)$ from the iterative formula
\begin{equation}
    e^{\displaystyle-\beta_T \psi_{n+1} (\mu)} = \int \dd m \frac{\displaystyle e^{\displaystyle-\beta_T \mathcal{E}(m, \mu)}}{\displaystyle\int \dd \mu' e^{\displaystyle-\beta_T [\mathcal{E}(m, \mu')-\psi_n(\mu')]} }.\label{iterative}
\end{equation}Equation~\eqref{iterative} is the result of integrating~\eqref{Pfull} over $m$ and using~\eqref{mu}. Equation~\eqref{Pfull} satisfies the constraints~\eqref{mu} and~\eqref{lambda}, but does not necessarily correspond to the correct energy $E_{\mathrm{pot}}$; in this case, we increment $\beta_T$ and repeat the procedure described until the desired energy is obtained.

\subsection{Diffusion\label{sec:diffusion}}

The function $\mathcal{P}(m,\mu)$ gives the fractional abundances of parcels from supported mass $\mu$ at new supported mass $m$ in the ``most mixed'' state accessible by ideal rearrangements. We shall use these abundances to determine the result of diffusion: when sufficiently small scales are developed by mixing, diffusion homogenises nearby fluid parcels. Thus, diffused states may be obtained by taking suitable moments of $\mathcal{P}(m,\mu)$.

The standard method for deriving predictions from Lynden-Bell probability distribution functions is to argue that, due to the presumed stochastic nature of the underlying microstate, physically measurable quantities correspond to expectation values. In our case, these are
\begin{align}
    \langle s \rangle & \equiv \int \dd \mu\, s(\mu) \mathcal{P}(m, \mu), \label{expectations_s}\\ \langle  \chi \rangle & \equiv \int \dd \mu\, \chi(\mu) \mathcal{P}(m, \mu).\label{expectations_chi}
\end{align}
Unlike in the traditional contexts, however, expectation values---in particular,~\eqref{expectations_s}---are unsuitable as a model of the state that develops after diffusion acts. This is because diffusive processes do not preserve the mean value of $s$; instead, thermal conduction and ohmic heating increase thermal entropy.\footnote{The relevant analogue of~\eqref{expectations_s} and~\eqref{expectations_chi} for the collisionless-relaxation problem, i.e., ${\langle \eta \rangle = \int \dd \eta\, \eta P(\bx,\bv, \eta)}$,
\textit{does} constitute a plausible prediction for the particle-distribution function in the presence of small collision frequency (collisions act as diffusion in velocity space, smoothing the stochastic variation in the local value of $\eta$ to its mean). This is because collisions preserve the contribution of each patch of phase space to the total energy, momentum and number of particles, as these quantities are each linear in $\eta$. In the language of \citet{Chavanis03}, total energy, momentum and number of particles are ``robust integrals'', being the same for both the coarse- and fine-grained distributions of $\eta$ [although, see the discussion in Section~\ref{sec:discussion}].} It is readily verified that energy is not conserved under a collapse of the distributions of $s$ and $\chi$ onto their expectation values:
\begin{equation}
    \int \dd m \,\mcE(mg, \langle s\rangle, \langle \chi \rangle) \neq E_{\mathrm{pot}},\label{Ecg_not_E}
\end{equation}because $\mcE(m,\mu)$ is nonlinear in $s$ and $\chi$. In the terminology of \citet{Chavanis03}, energy is a ``fragile integral''---its value is different depending on whether it is computed using the coarse- or fine-grained distributions of $s$ and $\chi$. The difference between the left- and right-hand sides of~\eqref{Ecg_not_E} is typically much greater than the available energy of the original equilibrium---around ten times greater in the case of the unstable-upwards profile defined by~\eqref{m_profile}. 

\begin{figure*}
    \centering
    \includegraphics[width=.9\textwidth]{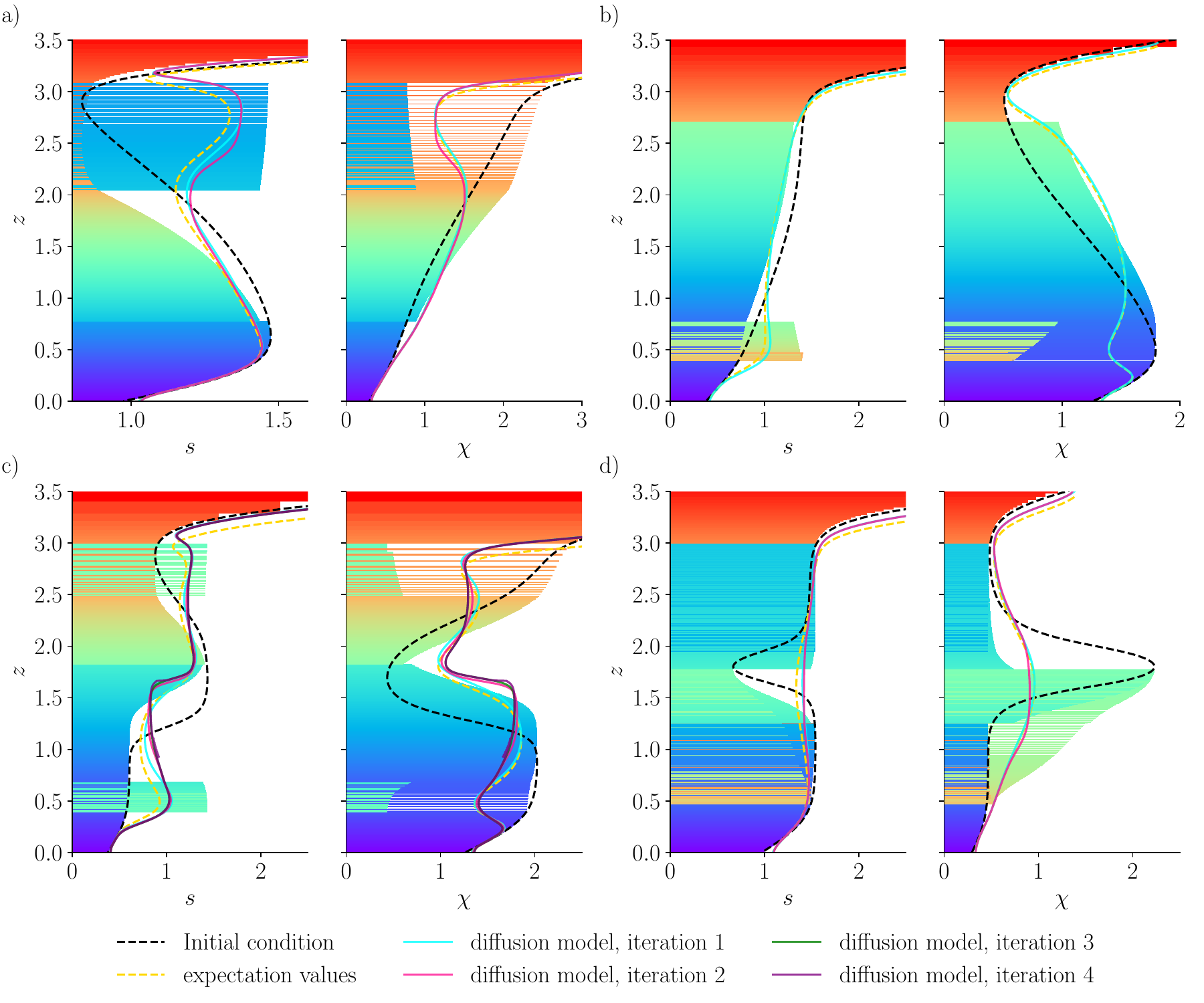}
    \caption{The predictions of the Lynden-Bell statistical mechanics (Section~\ref{sec:LyndenBellEquations}) for each of the profiles described in Section~\ref{sec:examples}. Panel~(a) corresponds to~\eqref{m_profile}, (b) to~\eqref{invm}, (c) to~\eqref{bump} and (d) to~\eqref{dip}. In each case, the black dashed line corresponds to the initial profile, the gold dashed line to $\langle s\rangle$ or $\langle \chi\rangle$  [see~\eqref{expectations_s} and~\eqref{expectations_chi}] and the cyan solid line to the predictions $\bar{s}$ and $\bar{\chi}$ for the result of diffusion [see~\eqref{energycons_mixing} and~\eqref{chi_mixing}]. Other coloured lines correspond to iterations of the statistical mechanical calculation, as described in Section~\ref{sec:secondaryrelaxation}.}
    \label{fig:LB_diffusion}
\end{figure*}

Equation~\eqref{expectations_s} being unsuitable, we instead determine the entropy function after diffusion,~$\bar{s}$, from the conservation of energy, i.e., from
\begin{equation}
\mcE(mg,\bar{s},\bar{\chi}) = \int \dd \mu\, \mcE(mg,s(\mu),\chi(\mu)) \mathcal{P}(m,\mu),\label{energycons_mixing}
\end{equation}with the diffused magnetic flux $\bar{\chi}$ given by
\begin{equation}
\bar{\chi}=\langle\chi\rangle\label{chi_mixing}
\end{equation}(the diffusion of straight magnetic field lines \textit{does} conserve magnetic flux). We plot $\bar{s}$ and $\bar{\chi}$ against $z$ [with density~${\rho(m,\bar{s} ,\bar{\chi} )}$] with a solid cyan line in figure~\ref{fig:LB_diffusion}. For comparison, we plot $\langle s \rangle $ and $\langle \chi \rangle $ against~$z$ [with density~${\rho(m,\langle s \rangle,\langle \chi \rangle )}$] with a gold dashed line---these profiles are appreciably different, and we will find in Section~\ref{sec:numerics2} that the former is indeed a better predictor of numerical simulations.

\begin{figure*}
    \centering
    \includegraphics[width=.8\textwidth]{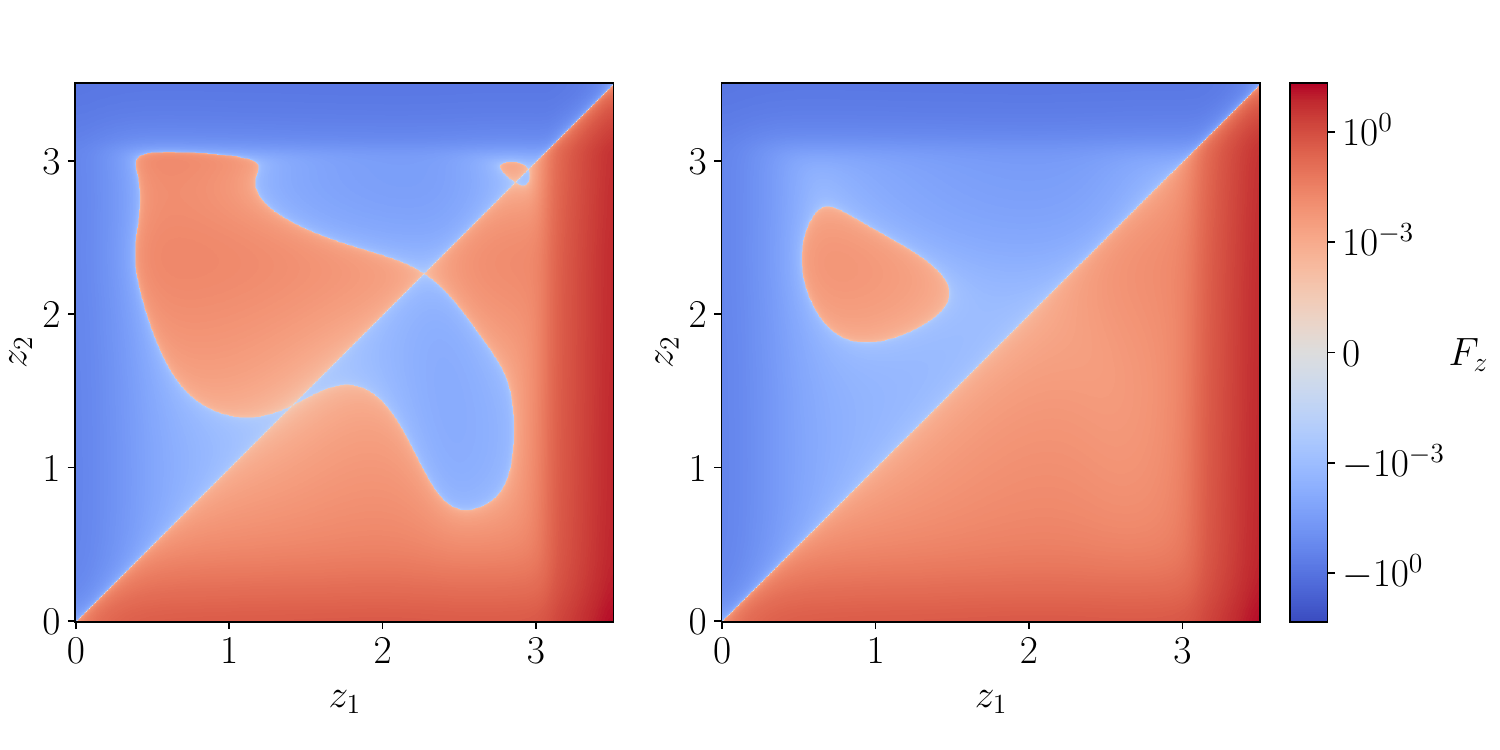}
    \caption{The force~\eqref{F2pressure} per unit mass of a small fluid parcel moved in pressure balance and without diffusion from height $z_1$ to $z_2$ for: left panel, the state that corresponds to the cyan lines in figure~\ref{fig:LB_diffusion}(a); and, right panel, the analogue of this profile for $E=E_0+0.3 E_{\mathrm{avail}}$. The left panel exhibits linear instability, the right panel nonlinear instability (metastability).}
    \label{fig:force_plots_relaxing}
\end{figure*}

\subsection{Secondary relaxation\label{sec:secondaryrelaxation}}

Equations~\eqref{energycons_mixing} and~\eqref{chi_mixing} need not, in general, represent the final state reached by relaxation, because the turbulent mixing flow remains present even after diffusion. This flow can cause further reorganisation if the post-diffusion state of the system has more than one accessible microstate, i.e., if it is not a nonlinearly stable minimum-energy state.\footnote{Because we assume that the energy of the flow is small, we neglect the possibility of it exciting the system into states with greater potential energy.} Remarkably, this often turns out to be the case: diffusion [in the sense of~\eqref{energycons_mixing} and~\eqref{chi_mixing}] of the state predicted by Lynden-Bell statistical mechanics tends to produce new states that are unstable to further (ideal) dynamics. This phenomenon has no analogue in the relaxation of collisionless stellar systems and plasma, for which the coarse-grained probability-distribution function is always a state of minimum energy with respect to rearrangements of phase space. That it is possible for MHD atmospheres is a consequence of the fact that diffusion produces changes in buoyancy. A well-known example of this is the phenomenon of ``buoyancy reversal'' in the terrestrial atmosphere: the nonlinear dependence of density on the advected Lagrangian invariants (in that context, the mixing ratio of water and potential temperature; see Appendix~\ref{app:moist}) means that a buoyant parcel of fluid that rises and mixes with denser ambient fluid can become denser than the ambient fluid, and sink as a result [see, e.g.,~\citet{Stevens05}]. 

The 1D equilibrium states given by~\eqref{energycons_mixing} and~\eqref{chi_mixing} may be linearly or nonlinearly unstable (metastable). Examples of each case are illustrated in figure~\ref{fig:force_plots_relaxing}. Panel~(a) plots the force~\eqref{F2pressure} per unit mass on a small parcel of fluid displaced from height $z_1$ to $z_2$ for the post-diffusion state of the metastable-upwards equilibrium~\eqref{m_profile}. This state is linearly unstable between $z \simeq 1.4$ and $z \simeq 2.3$, and also close to $z=3$. Panel~(b) is analogous, but for a slightly larger energy of $E = E_0+0.3 E_{\mathrm{avail}}$ in~\eqref{betaT} (this is the initial energy of the numerical simulation to be presented in figures~\ref{fig:turb_simulation_upwards} and~\ref{fig:upLB_fig0.1} in Section~\ref{sec:turb_simulations_up}). In this case, the new profile is linearly stable, but is unstable nonlinearly (metastable), and turns out to have states with lower energy (as can be confirmed by solving its LSA problem). Likewise, the post-diffusion states corresponding to the ``bi-directional''~\eqref{bump} and ``overturning''~\eqref{dip} profiles are not minimum-energy states. On the other hand, it turns out that diffusion \textit{does} yield a minimum-energy state in the case of the unstable-downwards profile~\eqref{invm}.

As noted above, in cases where the new state given by~\eqref{energycons_mixing} and~\eqref{chi_mixing} has more than one accessible microstate with the same energy, continued turbulent mixing can reshuffle fluid parcels and produce further diffusion until a minimum-energy state is reached. A simple model of this process is to apply the procedure described in Sections~\ref{sec:LyndenBellEquations} and~\ref{sec:diffusion} iteratively. This corresponds to the diffused system exploring the full space of states that are energetically accessible under ideal arrangements before diffusing again. We show in figure~\ref{fig:LB_diffusion} the profiles of $\bar{s}$ and $\bar{\chi}$ at the second, third and fourth iterations, plotted in pink, green and purple, respectively. The procedure terminates (i.e., reaches a minimum-energy state) after two iterations in the unstable-upwards~\eqref{m_profile} and ``overturning''~\eqref{bump} cases [see panels (a) and (d)], and after four iterations in the unstable-``bi-directional'' case~\eqref{dip}. [Note that, in the ``overturning'' case, we find at the fourth iteration that the state is sufficiently close to the ground state to become sensitive to the discretisation that we employ to compute the integrals in~\eqref{expectations_chi} and~\eqref{energycons_mixing}: see the jagged structure of the purple line in panel (c) of figure~\ref{fig:LB_diffusion}. At the fifth iteration, $\beta_T$ becomes so large as to preclude accurate computation of the relevant integrals, so we terminate the process at the fourth iteration.]

The difference in the profiles of $\bar{s}$ and $\bar{\chi}$ between the first and last stages of the iterative procedure turn out to be slight. Therefore, if, as in reality, secondary relaxations are incomplete or diffusion occurs concurrently with them, the final state reached ought not to be very different. We shall see in Section~\ref{sec:numerics2} that accounting for these diffusive rearrangements is, in practice, a precision overkill---greater discrepancies between numerical experiment and the theoretical prediction arise which appear to be a result of the tendency of relaxing profiles to become ``stuck'' in other metastable states. Nonetheless, it is interesting to note that, on a qualitative level, the chief outcome of the iterative procedure is the formation of plateaus (corresponding to thorough mixing and diffusion in regions where the first relaxed state is close to marginal linear stability). In the case of the unstable-upwards profile, for example, we see from figure~\eqref{fig:LB_diffusion}(a) that, between the first and second iterations, material at $z\lesssim 1.75$ moves upwards to settle in the range $1.75 \lesssim z \lesssim 2.75$, producing a flatter region between $2.5 \lesssim z \lesssim 3.0$ and in the vicinity of $z\simeq 2$. Similar plateaus are observed in panels (c) and (d), with panel (c) resembling a staircase. The intriguing possibility of modelling the formation of staircases (which are observed in myriad diffusing systems in geo- and astro-physical contexts) statistical mechanically is a topic to which we shall return in future work [see Section~\ref{sec:discussion}].

\section{Numerical simulations of relaxation at large Reynolds number\label{sec:numerics2}}

In this section, we present a comparison of the theoretical predictions obtained in Section~\ref{sec:diffusion} with the results of numerical simulations with $\mathrm{Re}\gg 1$. The numerical set-up is the same as in Section~\ref{sec:LowReNumerics}, but with the kinematic viscosity $\nu$ smaller by a factor of~$400$, such that the Reynolds number based on the initial velocity field is $u_0 L_x / \nu\sim 10^{5}$.

\subsection{Upwards metastability~\eqref{m_profile}\label{sec:turb_simulations_up}}

\begin{figure*}
    \centering
    \includegraphics[width=\textwidth]{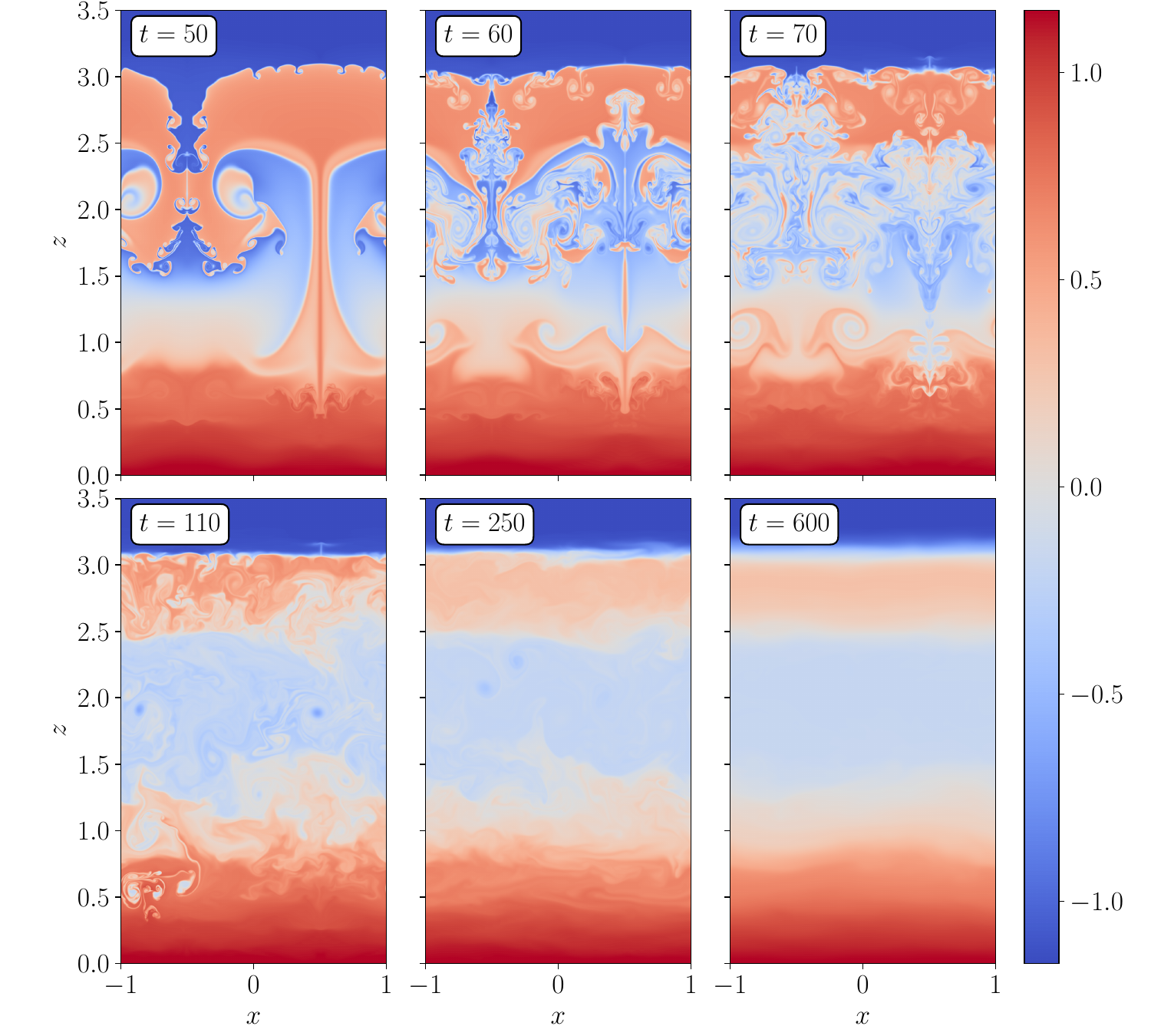}
    \caption{Numerical simulation of the relaxation of the equilibrium defined by~\eqref{m_profile} at $\mathrm{Re}\sim 10^5$. The quantity plotted is~$\ln(s/\chi)$. The initial velocity field is given by~\eqref{u0} with $u_0 = 0.1$, $z_0 = 1.0$ and $\Delta z = 0.5$, which corresponds to~${E_{\mathrm{kin},0}\simeq 0.3 E_{\mathrm{avail}}}$. A movie version of this figure is available in the published version of this article.}
    \label{fig:turb_simulation_upwards}
\end{figure*}

\begin{figure*}
    \centering
    \includegraphics[width=.6\textwidth]{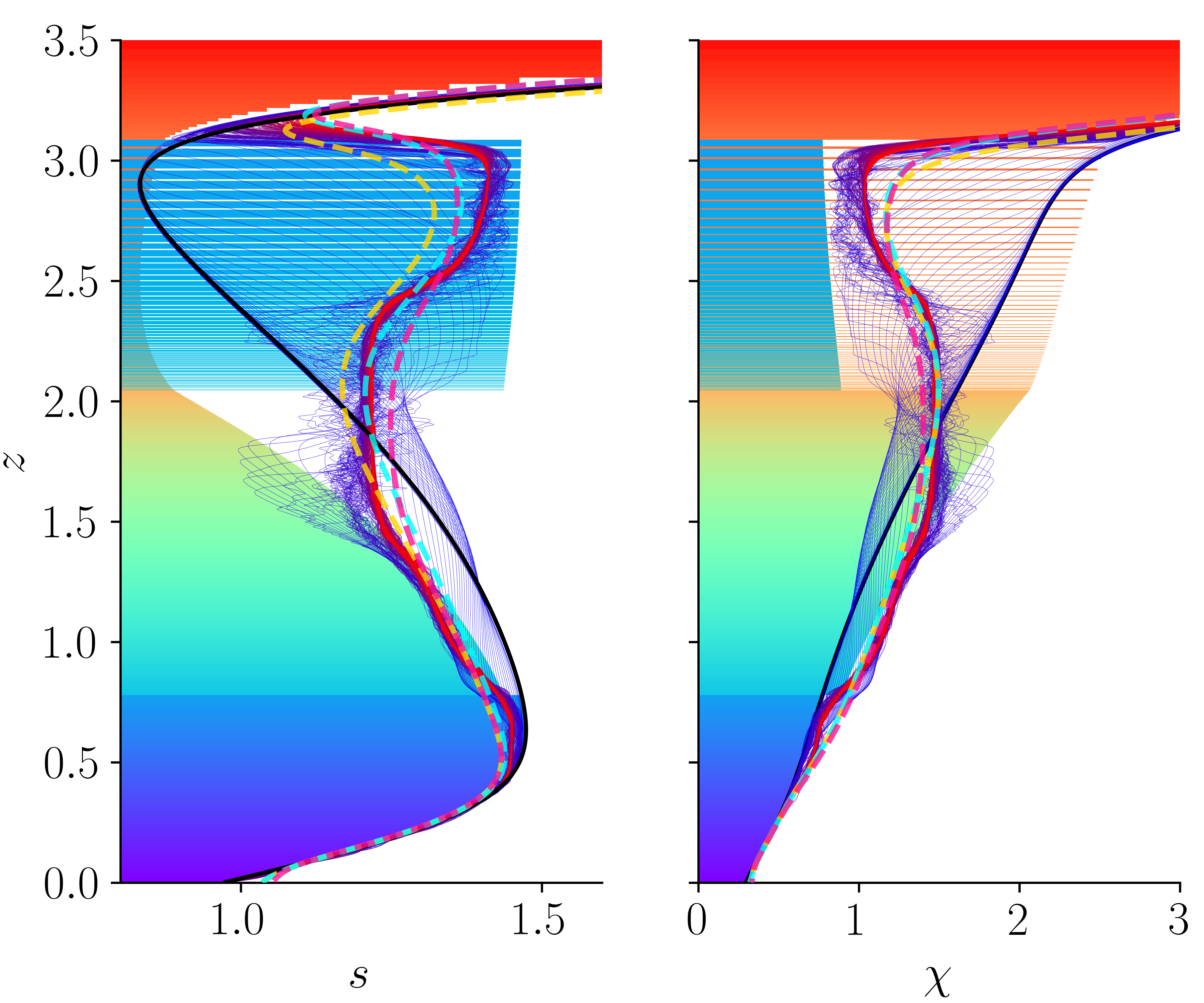}
    \caption{Horizontally averaged profiles of $s$ and $\chi$ plotted at intervals of $1$ code time unit, between $t=0$ (blue) to $t=600$ (red), for the simulation visualised in figure~\ref{fig:turb_simulation_upwards}. Also plotted are the profiles that correspond to expectation values of $\mathcal{P}(m,\mu)$ (gold dashed line; which we claim is not a suitable model of diffusion), from~\eqref{energycons_mixing} and~\eqref{chi_mixing} (cyan dashed line; these model energy- and flux-conserving diffusion), and after iterating the statistical mechanical prediction from the profile based on the cyan dashed line (pink dashed line). For reference, we also plot the minimum-energy state: this is as shown in figure~\ref{fig:selfsimilar_h}.}
    \label{fig:upLB_fig0.1}
\end{figure*}

We first consider the case of the unstable-upwards profile defined by~\eqref{m_profile}. Figure~\ref{fig:turb_simulation_upwards} visualises the distribution of $s/\chi$ (the advected scalar that controls the fluid compressibility---see Section~\ref{sec:metastability_theory}) during the relaxation that follows perturbation by a velocity field given by~\eqref{u0} with $u_0 = 0.1$ and $z_0=1.0$ (this corresponds to an initial kinetic energy $E_{\mathrm{kin},0}\simeq 0.3 E_{\mathrm{avail}}$). We observe that a substantial plume of material rises until reaching the stable region at the top of the simulation domain, where it overturns and develops Rayleigh--Taylor instabilities (upper left panel). Under advection by the increasingly chaotic flow, small-scale structures are developed (upper middle and right panels). These structures diffuse, ultimately leading to a one-dimensional but non-homogeneous state with no fine-scale structure (lower three panels).

Figure~\ref{fig:upLB_fig0.1} compares the horizontally averaged instantaneous profiles of $s$ and $\chi$ developed in the simulation with both $\langle s\rangle$ and $\langle \chi\rangle$ [as defined in~\eqref{expectations_s} and~\eqref{expectations_chi}] and~$\bar{s}$ and~$\bar{\chi}$ [as defined in~\eqref{energycons_mixing} and~\eqref{chi_mixing}]. The quantities are computed from $\mathcal{P}(m,\mu)$ calculated with $E_{\mathrm{pot}}=E_0+E_{\mathrm{kin},0}$ in~\eqref{betaT}.\footnote{The Mach number of the simulation, $\mathrm{Ma}\equiv u_0/\sqrt{v_A^2 + c_s^2}\sim 0.1$ is small, so that the compressible part of the initial velocity field rapidly propagates away as compressive waves. The energy associated with these waves (i.e., the initial energy of the non-solenoidal part of $\bu$) is, we assume, irrelevant to the otherwise quasi-incompressible dynamics, so we exclude it from the energy we use for $E_{\mathrm{pot}}$.} 

We make the following observations. First, the late-time profile of $s$ is almost everywhere larger than $\langle s \rangle$ (gold dashed line in figure~\ref{fig:upLB_fig0.1}). This validates the reasoning we used to reject~\eqref{expectations_s} as a predictor of the final state---evidently, dissipation causes the entropy of the fluid to grow. Secondly, $\bar{s}$ and $\bar{\chi}$ computed from~\eqref{energycons_mixing} and~\eqref{chi_mixing} (cyan dashed line in figure~\ref{fig:upLB_fig0.1}) constitute a very reasonable prediction of the relaxed state. The chief discrepancies are in the range $2.5 \lesssim z \lesssim 3.0$, where $\bar{s}$ and $\bar{\chi}$ are, respectively, somewhat smaller and larger than in the late-time profiles developed by the simulation. We interpret this as a consequence of the system not ``exploring'' the full surface of constant energy in configuration space, owing to the fluid that rises from the bottom of the equilibrium becoming trapped in a metastable state at the top. In support of this interpretation, we note that, in the range $0.9 \lesssim z \lesssim 1.5$, the cyan lines somewhat over- and under-predict the simulation result, respectively, indicating that ``too much'' material rose in the initial plume (and became stuck). 

A second discrepancy between the cyan line and the simulation result is that the latter exhibits a clear plateau in the range $1.5 \lesssim z \lesssim 2.3$. On the other hand, the cyan line is nonlinearly unstable [see figure~\ref{fig:forcez1z2}(b) for its force diagram]---the dashed pink line in figure~\ref{fig:upLB_fig0.1} shows the result of taking it as the initial state for a secondary relaxation [see Section~\ref{sec:secondaryrelaxation}]. The pink line features a plateau over roughly the same range of $z$ as the one that forms in the simulation, although the predicted plateau has slightly larger $s$ (smaller $\chi$). This might be interpreted as a consequence of the fact that the large-$s$ fluid that would have risen upwards under the secondary relaxation to form the plateau was, in the simulation, already displaced to the top of the atmosphere. As a consequence, the plateau forms with somewhat smaller $s$ than it would otherwise have had.

\begin{figure}
    \centering
    \includegraphics[width=0.6\columnwidth]{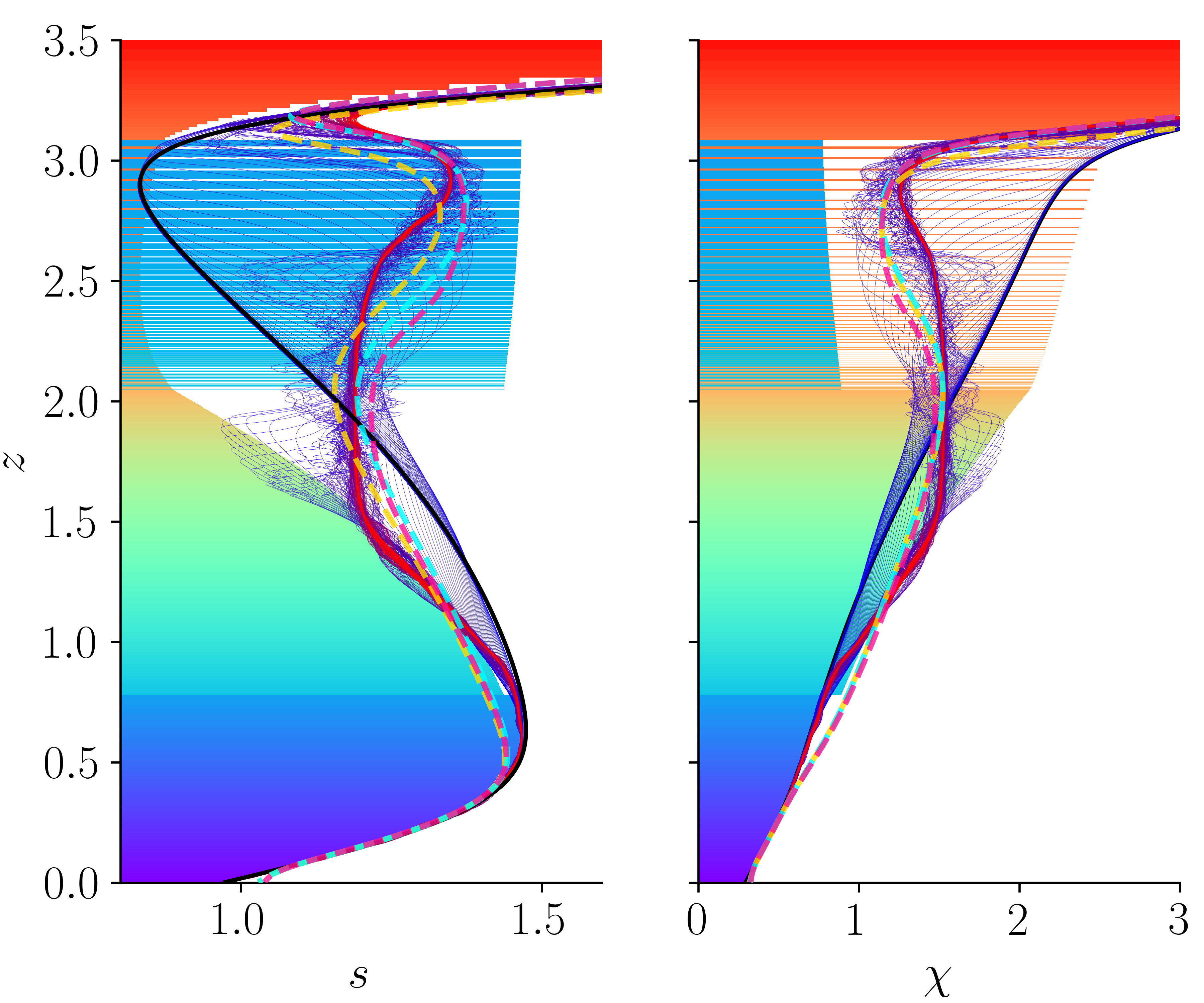}
    \caption{As in figure~\ref{fig:upLB_fig0.1}, but for a simulation initialised with $u_0=0.05$ in~\eqref{u0} ($E_{\mathrm{kin},0}\simeq 0.1 E_{\mathrm{avail}}$).}
    \label{fig:upLB_fig0.05}
\end{figure}
\begin{figure}
    \centering
    \includegraphics[width=0.6\columnwidth]{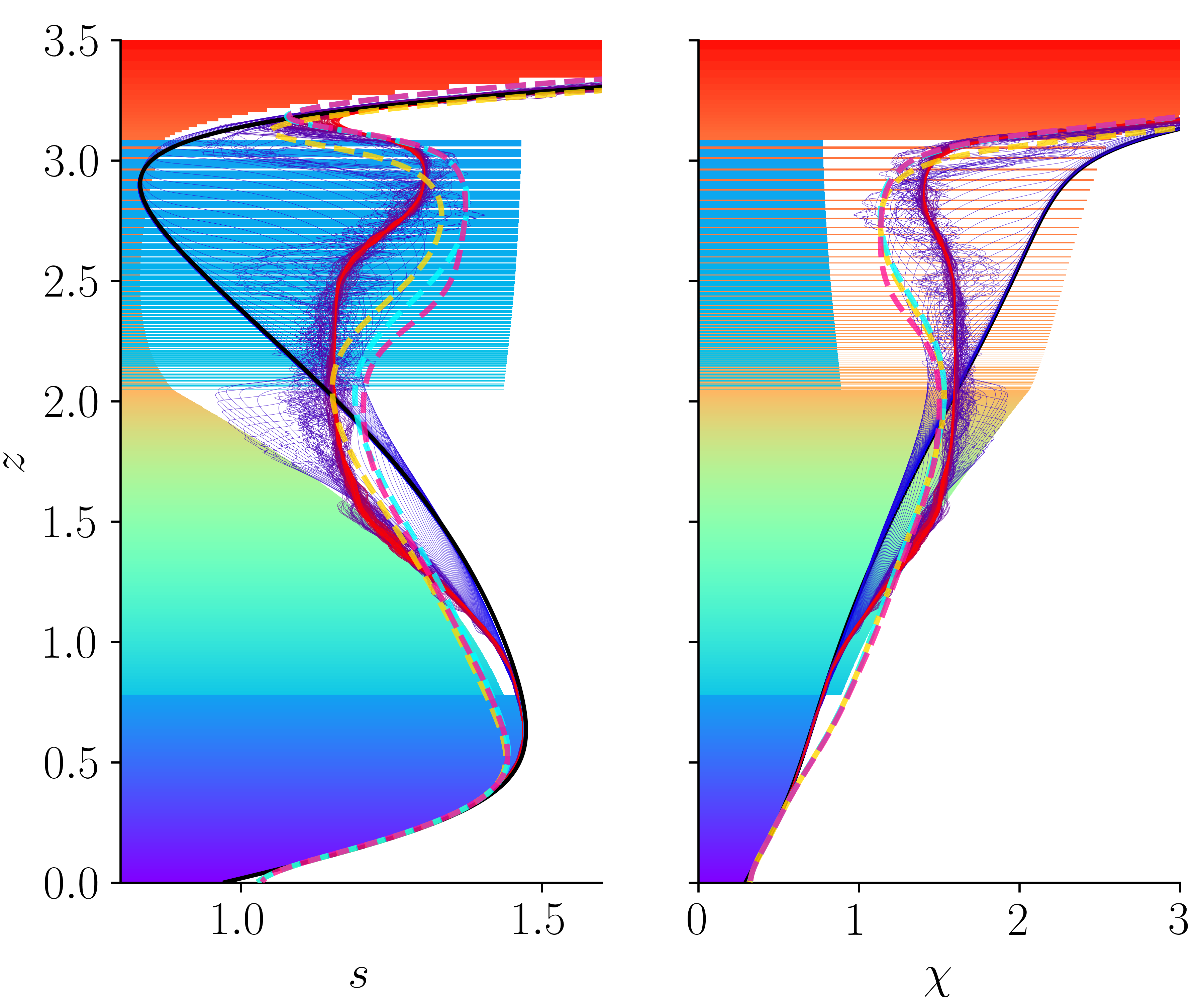}
    \caption{As in figure~\ref{fig:upLB_fig0.1}, but for a simulation initialised with $u_0=0.025$ in~\eqref{u0} ($E_{\mathrm{kin},0}\simeq 0.02 E_{\mathrm{avail}}$).}
    \label{fig:upLB_fig0.025}
\end{figure}

\begin{figure}
    \centering
    \includegraphics[width=0.6\columnwidth]{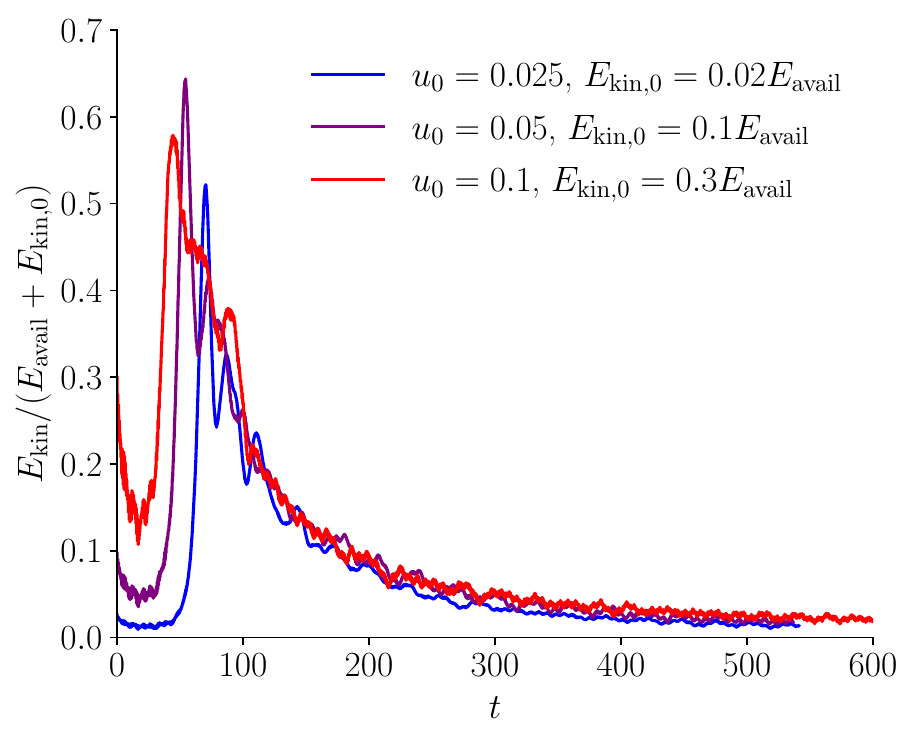}
    \caption{Evolution of the kinetic energy as a fraction of the total available energy, which is the kinetic energy plus the available potential energy of the initial state, for the simulations visualised in figures~\ref{fig:upLB_fig0.1},~\ref{fig:upLB_fig0.05} and~\ref{fig:upLB_fig0.025}.}
    \label{fig:energy_time_up}
\end{figure}

Figures~\ref{fig:upLB_fig0.05} and~\ref{fig:upLB_fig0.025} are analogous to figure~\ref{fig:upLB_fig0.1}, but for simulations with ${u_0 = 0.05}$ ($E_{\mathrm{kin},0}\simeq 0.1 E_{\mathrm{avail}}$) and ${u_0 = 0.025}$ ($E_{\mathrm{kin},0}\simeq 0.02 E_{\mathrm{avail}}$), respectively. In these cases, the quantitative agreement between simulation and theory is less good than in figure~\ref{fig:upLB_fig0.1}: with a smaller initial impulse, relaxation is incomplete. Figure~\ref{fig:energy_time_up}, shows that, at peak, between $50\%$ and $60\%$ of the available kinetic plus potential energy is in the form of kinetic energy for all three simulations, showing that the liberation of available potential energy during relaxation is fairly efficient.

\subsection{Downwards metastability~\eqref{invm}\label{sec:turb_simulations_down}}

We now report the results of analogous simulations to those in Section~\ref{sec:turb_simulations_down}, but for the unstable-downwards profile~\eqref{invm} and with $z_0 = 2.25$ in~\eqref{u0}.

\begin{figure*}
    \centering
    \includegraphics[width=\textwidth]{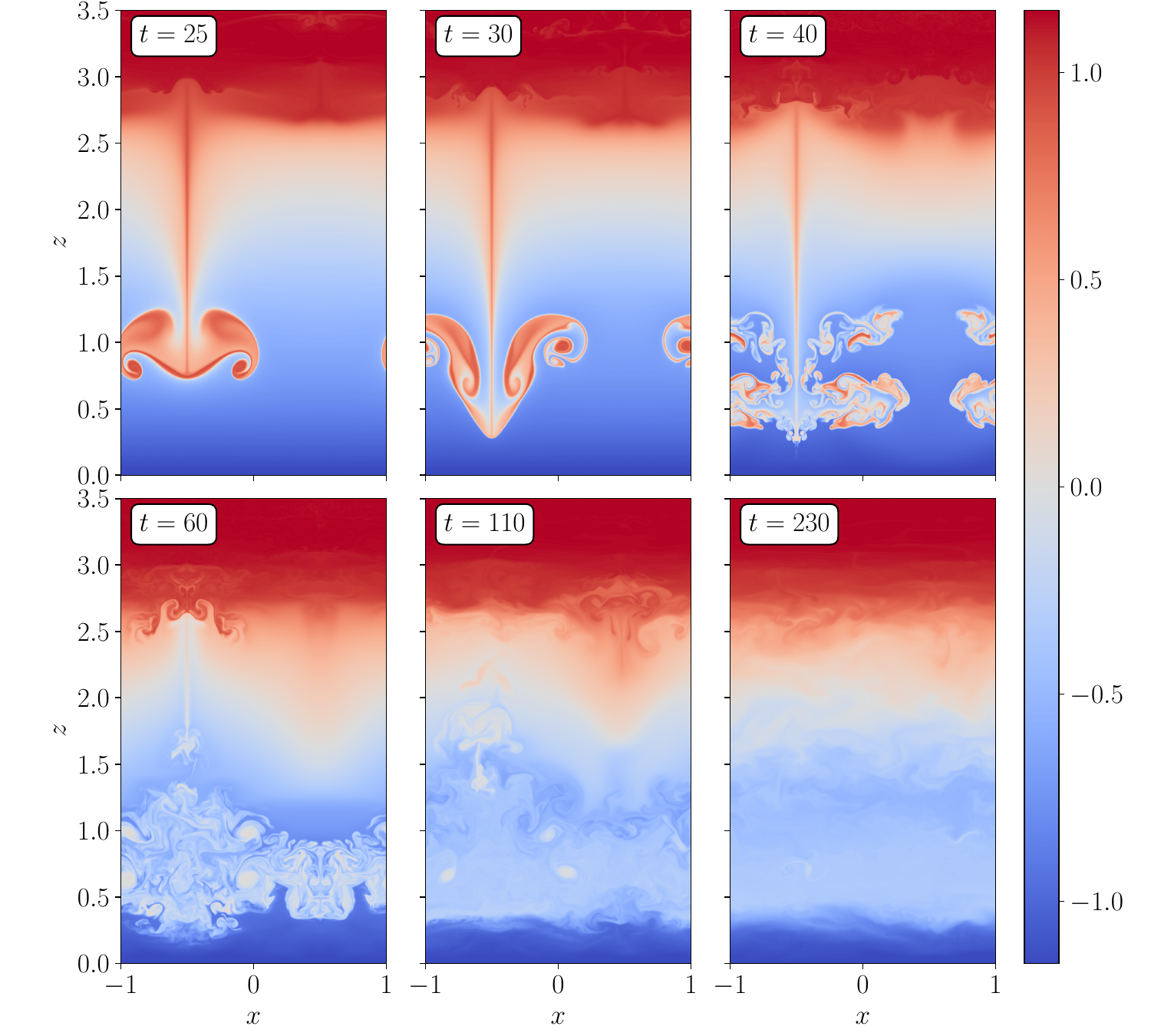}
    \caption{Numerical simulation of the relaxation of the equilibrium defined by~\eqref{invm} at $\mathrm{Re}\sim 10^5$. The quantity plotted is $\ln(s/\chi)$. The initial velocity field is given by~\eqref{u0} with $u_0 = 0.14$, $z_0 = 2.25$ and $\Delta z = 0.5$; this corresponds to an initial kinetic energy $\simeq 0.2$ times the available potential energy. A movie version of this figure is available in the published version of this article.}
    \label{fig:turb_simulation_downwards}
\end{figure*}

\begin{figure*}
    \centering
    \includegraphics[width=.6\textwidth]{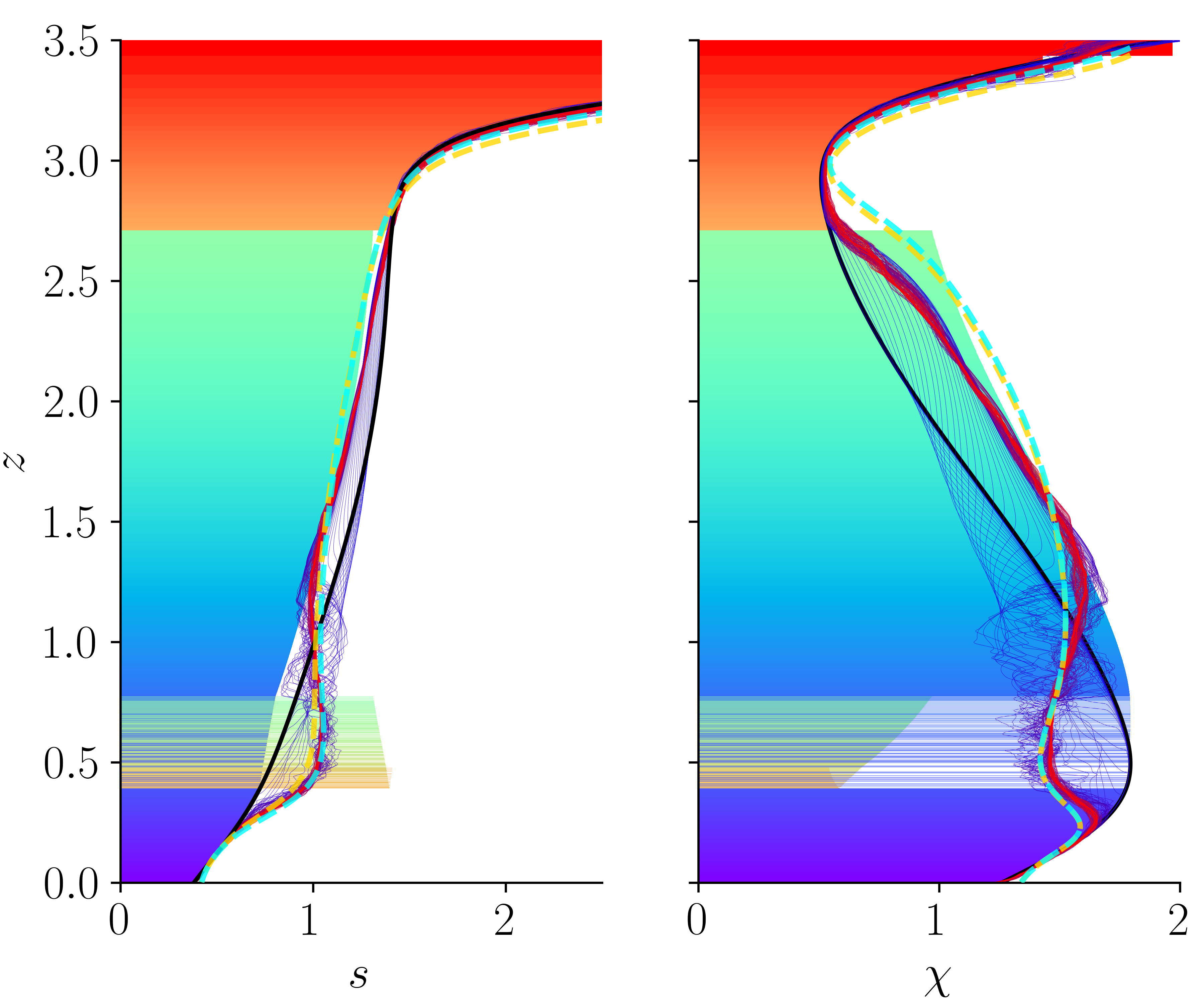}
    \caption{Horizontally averaged profiles of $s$ and $\chi$ plotted at intervals of $1$ code time unit, between $t=0$ (blue) to $t=600$ (red), for the simulation visualised in figure~\ref{fig:turb_simulation_downwards}. Also plotted are the profiles obtained by taking expectation values of $\mathcal{P}(m,\mu)$ (gold dashed line) and from~\eqref{energycons_mixing} and~\eqref{chi_mixing} (cyan dashed line).}
    \label{fig:downLB_fig0.14}
\end{figure*}

\begin{figure}
    \centering
    \includegraphics[width=.6\columnwidth]{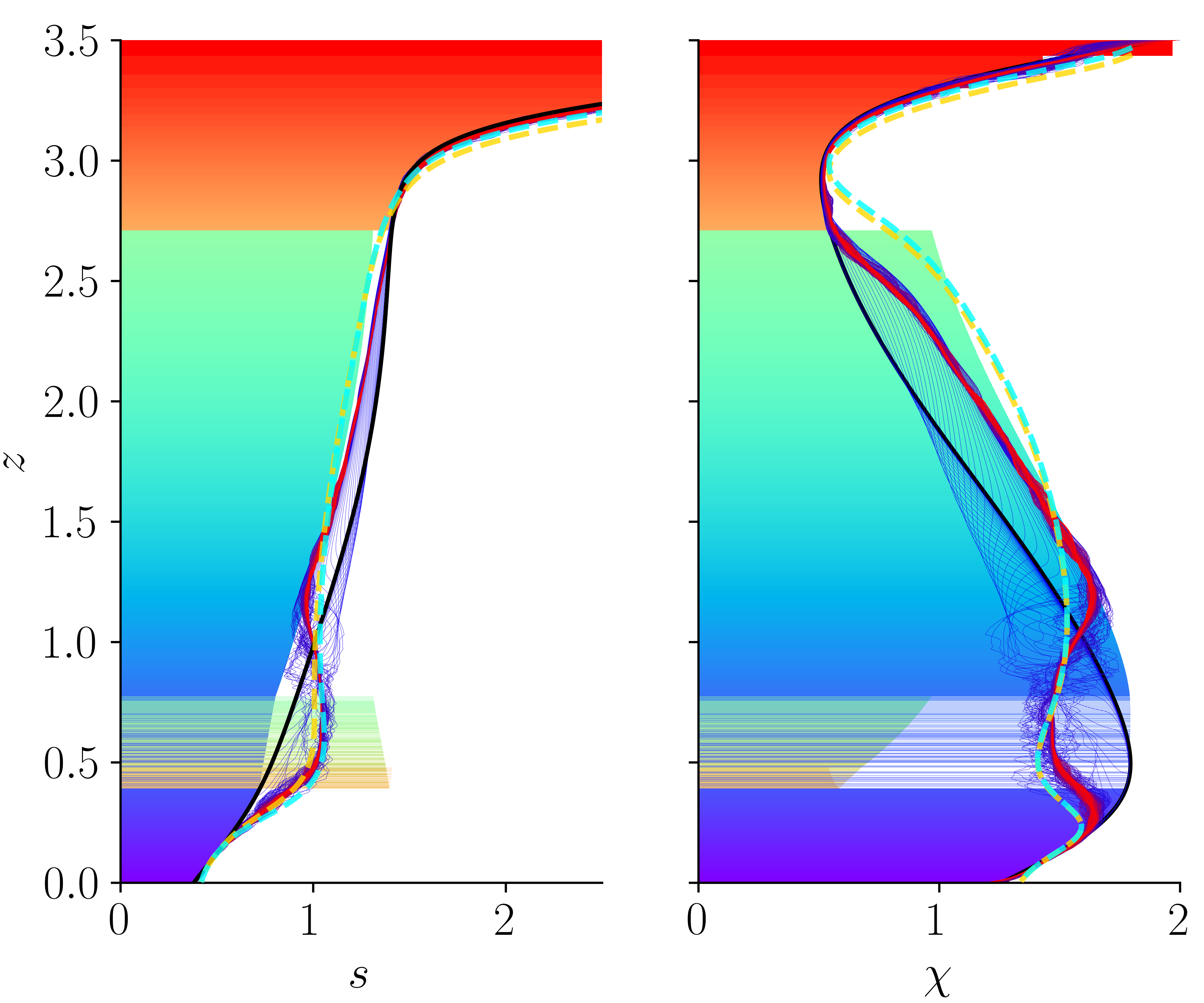}
    \caption{As in figure~\ref{fig:downLB_fig0.14}, but for a simulation initialised with $u_0=0.1$ in~\eqref{u0}, which corresponds to ${E_{\mathrm{kin},0}\simeq 0.1 E_{\mathrm{avail}}}$.}
    \label{fig:downLB_fig0.1}
\end{figure}

\begin{figure}
    \centering
    \includegraphics[width=.6\columnwidth]{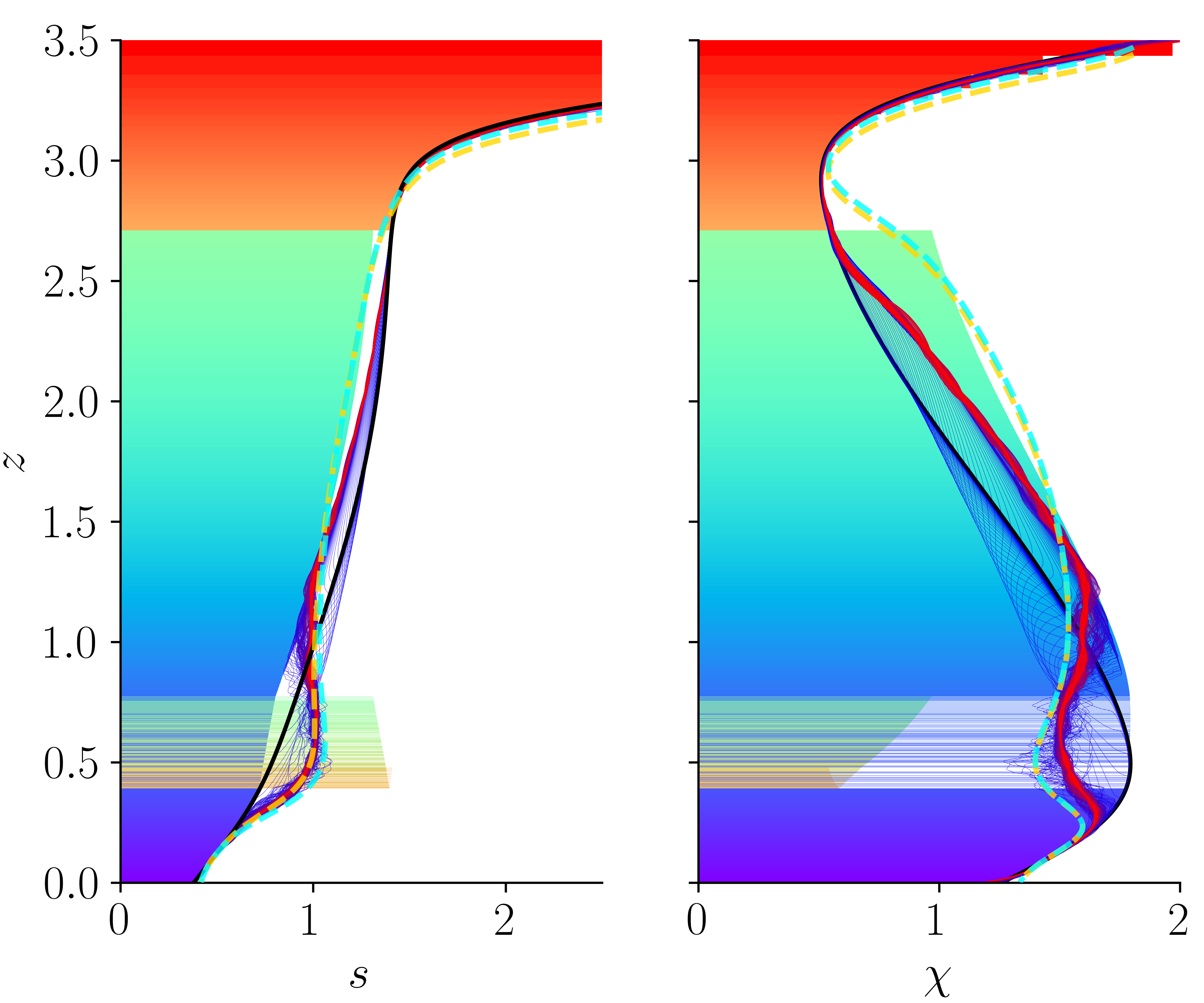}
    \caption{As in figure~\ref{fig:downLB_fig0.14}, but for a simulation initialised with $u_0=0.05$ in~\eqref{u0}, which corresponds to ${E_{\mathrm{kin},0}\simeq 0.03 E_{\mathrm{avail}}}$.}
    \label{fig:downLB_fig0.05}
\end{figure}

\begin{figure}
    \centering
    \includegraphics[width=.6\columnwidth]{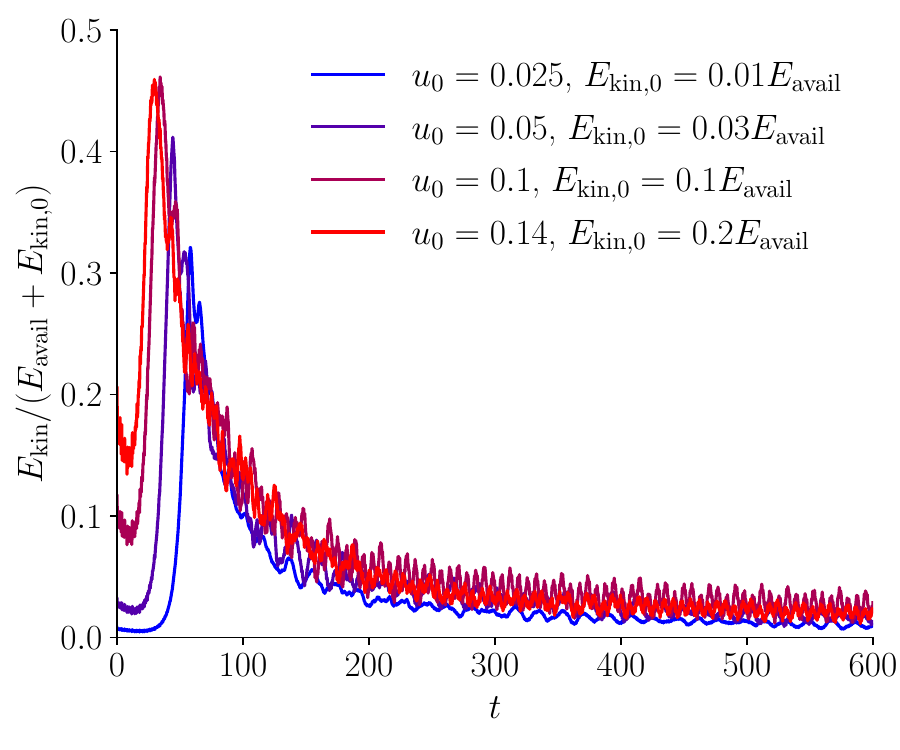}
    \caption{Evolution of the kinetic energy as a fraction of the total available energy, which is the kinetic energy plus the available potential energy of the initial state, for the simulations visualised in figures~\ref{fig:downLB_fig0.14},~\ref{fig:downLB_fig0.1} and~\ref{fig:downLB_fig0.05}, as well as for one with~$u_0 = 0.025$ in~\eqref{u0}.}
    \label{fig:energy_time_down}
\end{figure}

Figure~\ref{fig:turb_simulation_downwards} visualises the evolution in $x$-$z$ space for ${u_0=0.14}$ ($E_{\mathrm{kin},0}\simeq0.2E_{\mathrm{avail}}$), following the tracer $s/\chi$. Similarly to figure~\ref{fig:turb_simulation_upwards}, we observe that a descending plume reaches the stable buffer region at the bottom of the simulation domain, develops small-scale structure due to Rayleigh--Taylor instability and advection by chaotic motions, and ultimately diffuses. The evolution of the horizontally averaged profiles is displayed in figures~\ref{fig:downLB_fig0.14},~\ref{fig:downLB_fig0.1} and~\ref{fig:downLB_fig0.05} for the cases of $u_0 = 0.14$ ($E_{\mathrm{kin},0}\simeq0.2E_{\mathrm{avail}}$), $u_0 = 0.1$ ($E_{\mathrm{kin},0}\simeq0.1E_{\mathrm{avail}}$), and $u_0 = 0.05$ ($E_{\mathrm{kin},0}\simeq0.03 E_{\mathrm{avail}}$), respectively. Again, we observe reasonable agreement between the statistical mechanical prediction and the late-time limit of the simulations, although the predictions do somewhat under-predict $s$ and over-predict $\chi$ for $z\gtrsim 1.5$. Again, the reason appears to be partial relaxation: the degree of inaccuracy is greater in the simulations with smaller initial velocity fields. We also note that the theory fails to predict the plateau that forms in the vicinity of $z\simeq 0.5$ (no secondary relaxation is possible from the diffused state that corresponds to the cyan line, as it is nonlinearly stable, see Section~\ref{sec:secondaryrelaxation}). Nonetheless, the theoretical predictions agree reasonably well in this region with all three simulation profiles, the flatness notwithstanding. Finally, we show the evolution of the kinetic energy as a fraction of the total available energy in figure~\ref{fig:energy_time_down}. As in figure~\ref{fig:energy_time_up}, the relaxation is fairly efficient at liberating available potential energy. Large initial perturbations are not required to do so: over a factor $\simeq 30$ difference in the initial kinetic energy, the peak ratio of kinetic energy to available energy varies only between around $30\%$ and $45\%$.

\section{Conclusion\label{sec:conclusion}}

\subsection{Summary}

In this work, we have demonstrated that MHD equilibria with straight magnetic field lines can be metastable to 2D interchange-type motions in the plane perpendicular to the magnetic field~(Section~\ref{sec:metastability_theory}). This phenomenon occurs because fluid with large plasma~$\beta$ is more compressible than fluid with small~$\beta$. As a result, when displaced upwards (for example) by a large distance and thus exposed to a large change in pressure, the density of large-$\beta$ fluid can be less than that of the ambient fluid at the new location, even if this would not be the case for sufficiently local (i.e., linear) displacements.

The existence of metastability in 2D is particularly interesting because ideal (i.e., dissipationless) relaxation in 2D constitutes rearrangement of flux tubes subject to the Lagrangian invariance of entropy and magnetic flux. This enables the use of combinatorial and statistical techniques that are unavailable for the 3D problem (although see Section~\ref{sec:discussion}). We have determined the minimum-energy state consistent with 2D rearrangements by solving a linear sum assignment (LSA) problem in four different illustrative cases (Section~\ref{sec:ground_state_theory}). We find that the available potential energy is generally small, a result that can be traced back to the fact that fluid parcels exclude each other, so that the mass of fluid that can experience a significant change in pressure is limited (Section~\ref{sec:smallE}). An interesting finding is that the minimum-energy states are two-dimensional, despite the initial profiles being one dimensional (Section~\ref{sec:selfsimilar}). We show that the relaxed state that develops in numerical simulations at small Reynolds number approximates this 2D minimum-energy state (Section~\ref{viscous_case}), as the potential energy liberated during relaxation is dissipated by viscosity (changing the entropy of the fluid only slightly, owing to the small energy scales involved).

The two-dimensionality of the minimum-energy states may be interpreted as a consequence of the fact that the 1D states with least potential energy are, in general, Rayleigh--Taylor unstable. Provided that it is not suppressed by viscosity, this instability (and the potential energy liberated during relaxation) drives a turbulent flow that generates small-scale structure, enabling diffusion to act and violate the fluid-element-wise conservation of entropy and magnetic flux (Section~\ref{sec:LyndenBell}). We have proposed that such cases can be modelled theoretically by postulating (non-rigorously, but, apparently, usefully) a separation of timescales between the one on which the flow is mixed (i.e., the ideal dynamical timescale) and the one on which the advected invariants diffuse. We suggest that the final state of the former relaxation is the equilibrium state that maximises the Boltzmann mixing entropy subject to fixed energy (Section~\ref{sec:LyndenBellEquations}). This is analogous to the Lynden-Bell theory of ``violent relaxation'' of collisionless stellar systems and plasma, and to the Robert-Sommeria-Miller (RSM) theory of 2D vortex turbulence. The generalisation of these theories to more than one conserved quantity (thermal entropy and magnetic flux) turns out to be uncomplicated because the theory can be recast as a maximisation of the mixing entropy associated with rearrangements of 1D slices (Section~\ref{sec:1dmicrostates}). 

We take the latter, diffusive part of the relaxation to be a homogenisation of the stochastic small-scale structure present in a statistical-mechanical microstate. We obtain a prediction for the post-diffusion state by collapsing the statistical mechanical probability function onto a value determined by the local conservation of enthalpy and magnetic flux (Section~\ref{sec:diffusion}). Because the enthalpy density is a nonlinear function of the ideal Lagrangian invariants of the fluid, the states that one derives in this manner are not necessarily (or, indeed, usually) stable to ideal dynamics (Section~\ref{sec:secondaryrelaxation}). We propose that a natural scheme for dealing with this, which is consistent with the philosophy of timescale separation between ideal and non-ideal effects, is to iterate the procedure described above for the new profiles---i.e., to seek the mixing-entropy-maximising ideal rearrangement and then allow diffusion to act upon it---and to continue iterating until a stable profile is reached. The difference between the profiles predicted at the first and final iterations is not typically very large; the qualitative outcome of the subsequent iterations is to produce plateaus in the profiles of $s$ and $\chi$---see Section~\ref{sec:discussion} for further discussion.

We compare the theoretical predictions described above with the results of 2D numerical simulations at large Reynolds number in Section~\ref{sec:numerics2}. Provided that the equilibrium is perturbed sufficiently strongly to generate thorough mixing, we find remarkably good agreement between the late-time state of the numerical simulations and the predictions of our statistical mechanical theory. We also observe in the simulations the formation of well-mixed plateaus whose properties are consistent with the idea that they are formed by further mixing of diffused fluid. For weak initial perturbations, the agreement between the late-time states of the numerical simulations and our theory is less good. We interpret this as a consequence of the metastability phenomenon itself---when the relaxing system becomes ``stuck'' in a new metastable state, it cannot explore the full constant-energy surface in its configuration space.

\subsection{Discussion\label{sec:discussion}}

Our study suggests a number of questions for future investigation. One concerns the role of metastability in driven systems. The present study has been motivated by the fact that, unlike equilibria that are very unstable linearly, equilibria that are strongly metastable (in the sense of having a large amount of available energy) are realisable in real physical systems by an evolution of the potential-energy landscape that preserves the local minimum within which the system resides. It would be illuminating to understand whether, and under what conditions, the development of a metastable state actually happens in a dynamically evolving system, such as one driven towards convective instability by external heating (and/or cooling). If metastable states do develop, do they relax periodically via sporadic eruptions? How frequent are those eruptions? And are the corresponding relaxations ``complete'', in the sense of taking the system far from the linear-stability limit, or does the state of the system always remain close to this limit? Such questions are particularly pertinent in light of the observation that edge-localised modes in tokamaks---which are believed to be manifestations of metastable dynamics in driven systems [see the Introduction and \citet{CowleyArtun97, Hurricane97,WilsonCowley04, CowleyCowley15, Ham18}]---are observed to leave the plasma much below the threshold for linear instability post-eruption~[see, e.g.,~\citet{Kirk04, Kirk06}].

A natural question is whether the methods employed within this work can be adapted to fully 3D dynamics. In 3D, $B/\rho$ is not an ideal invariant: the magnetic flux through any material surface is conserved, but the density of the fluid in that surface can change due to motions perpendicular to the surface as well as parallel to it. Thus, relaxation in 3D is not simply a rearrangement of fluid parcels with associated invariants. Non-ideal processes too are significantly more complex in the 3D problem than the simple diffusion of scalar quantities---with magnetic field now a vector, magnetic topology imposes constraints [some, but not all, of which can be broken by magnetic reconnection in the subsequent relaxation~\citep{Taylor74,Taylor86,Zhou19,Bhat21,HoskingSchekochihin20decay}]. Nonetheless, we note that the Hungarian algorithm (Section~\ref{sec:HungarianTheory}) could still be employed to calculate a rigorous lower bound on the available energy of an equilibrium with initially straight magnetic field: the smallest potential energy that the field can have under 2D interchanges of field lines evidently constitutes such a bound. Furthermore, it seems plausible that the final state of 3D relaxation would, in fact, be 2D: bent field lines would tend to reconnect and straighten out under magnetic tension. Speculatively, if the development of such a final state were to constitute an effective series of interchanges, the statistical methods developed in this paper might well be applied usefully.

Another intriguing question for future work is whether combinatorial and statistical theories of relaxation can predict the nonlinear saturation of double-diffusive instabilities, for example, the fingering instability that occurs if a quantity whose stratification is stabilising is diffusive, so that its stabilisation is not felt on sufficiently small scales [see, for example,~\citet{HughesBrummell21}, and references therein]. Such systems saturate with staircase distributions of their compositional properties: thermohaline staircases in the ocean, which exhibit steps in their temperature and salinity profiles, are a prominent example. These staircases are apparently extremely stable---measurements of thermohaline staircases have revealed structure that persists over $\sim 100 \mathrm{km}$ and for a timescale of years [see~\citet{Merryfield00} for a review]. Metastability to diffusive modes has been mooted as a possible explanation for the staircases~\citep{Merryfield00}. Although diffusion is not naturally incorporated into the LSA problem of finding minimum-energy states (Section~\ref{sec:ground_state_theory}), one could imagine incorporating different rates of diffusion into the iterative model for post-diffusive relaxation described in Sections~\ref{sec:diffusion} and~\ref{sec:secondaryrelaxation}. Whether such a scheme would reproduce staircases, or offer qualitative insights into the mechanisms by which they form, remains to be seen. Some cautious optimism can be derived from results like those shown in figure~\ref{fig:LB_diffusion}(c), which constitutes a theoretical prediction of a staircase-like structure (although not as a result of double-diffusive instability).

In closing, let us consider whether the results of this work might be valuable for “violent relaxation” in other contexts. A straightforward yet useful result is that the statistical mechanical theory can work well in predicting relaxed states, provided the system is perturbed sufficiently strongly for it to become well mixed. A second result is that the theory appears to work well in systems that possess more than one invariant. \citet{Ewart23} have speculated about the possibility of predicting the relaxation of a collisionless magnetised plasma by enforcing the conservation of two invariants: the phase-space density $\eta$ and the magnetic moment ${\mu_b = mv_{\perp}^2/2B}$ (here, $m$ is the mass of a particle, and~$v_{\perp}$ is the magnitude of the velocity of the particle in the direction perpendicular to the magnetic field~$B$). It is interesting to note that, under such a scheme, the energy $\mcE$ associated with a volume element of phase space is a nonlinear function of conserved quantities, viz., $\mcE = \eta (m v_{\parallel}^2/2+\mu_b B)$. It follows that such a theory would need to address questions similar to the ones we considered in Section~\ref{sec:diffusion}, about how one extracts predictions from the Lynden-Bell probability distribution: the distribution function corresponding to its expectation values will, in general, not have the correct energy. We suggest that the fix might be, as in this work, to acknowledge that diffusion (in this case, particle collisions) does not conserve $\eta$ and $\mu_b$, but, rather, $\eta$ and~$\mcE$, and, therefore, to evaluate the distribution function according to a scheme analogous to~\eqref{energycons_mixing} and~\eqref{chi_mixing}. Whether or not the resulting distribution functions are unstable (as for the profiles that develop due to diffusion in our study), and, if so, the prediction of their further evolution, would, we suggest, be an interesting topic for future exploration.

\section*{Acknowledgements}

We thank Joshua Brown, Robbie Ewart, Thomas Foster, Alan Kerstein, Matt Kunz, Henrik Latter, Ewan McCulloch, Michael Nastac, Gordon Ogilvie, Eliot Quataert, Alex Schekochihin, Nicole Shibley and Anatoly Spitkovsky for helpful discussions and suggestions. In particular, we are indebted to Robbie Ewart and Alex Schekochihin for their suggestion that Lynden-Bell's violent-relaxation formalism might be applicable to our developing work on metastability in 2D MHD. We are grateful to the two anonymous referees, whose recommendations have improved this paper.

\section*{Funding}

This research received no specific grant from any funding agency, commercial or not-for-profit sectors.

\section*{Declaration of interests}

The authors report no conflict of interest.

\appendix

\section{Details of the numerical simulations\label{app:numerical}}

The numerical simulations presented in this work were conducted with the finite-difference MHD code Pencil~\citep{Pencil21}. The code uses sixth-order finite differences and a third-order-accurate time-stepping scheme to solve the equations of 2D MHD with a constant gravitational field, i.e.,
\begin{equation}
    \frac{\p \rho}{\p t}+\bu\bcdot \bnabla \rho = - \rho \bnabla \bcdot \bu,
\end{equation}
\begin{equation}
    \rho \left( \frac{\p \bu}{\p t} + \bu \bcdot \bnabla \bu \right ) = - \bnabla \left( p + \frac{B^2}{2} \right ) - \rho g \hat{\boldsymbol{z}} + \rho \nu \left( \nabla^2 \bu + \frac{1}{3} \bnabla \left( \bnabla \bcdot \bu \right ) \right ),
\end{equation}
\begin{multline}
    \frac{\p p}{\p t} + \bu \bcdot \bnabla p = - \gamma p \bnabla \bcdot \bu \\+ (\gamma - 1) \left[ \rho \nu \left( 2 e_{ij} e_{ij} - \frac{2}{3} \left( \bnabla \bcdot \bu \right )^2 \right ) + \eta \left| \bnabla \times (B \hat{\boldsymbol{y}}) \right|^2 + \rho K \nabla^2 \left( \frac{p}{\rho} \right )\right ],
\end{multline}where
\begin{equation}
    e_{ij} = \frac{1}{2} \left( \frac{\p u_i}{\p x_j} + \frac{\p u_j}{\p x_i} \right )
\end{equation}is the rate-of-strain tensor, and
\begin{equation}
    \frac{\p B}{\p t}+\bu\bcdot \bnabla B = - B \bnabla \bcdot \bu+\eta \nabla^2 B.
\end{equation}

In all simulations, we use reflecting boundary conditions [i.e., anti-symmetry for the component of velocity field in the direction normal to the boundary ($u_z$) and symmetry for the component in the other direction ($u_x$)] at the boundaries in $z$, while enforcing anti-symmetry relative to the value on the boundary (i.e., vanishing second derivative) for $B$, $p$ and $\rho$. For the simulations reported in Section~\ref{sec:introduction}, we simulate directly only the region with $0\leq x \leq 1$ and $0\leq z \leq 3.5$, with reflecting boundary conditions in the $x$ direction (and symmetric boundary conditions for $B$, $p$ and $\rho$)---we construct the visualisations in figures~\ref{fig:fig1} and~\ref{fig:fig2} by reflecting the simulation domain in the line $x=0$. For all other simulations, we employ periodic boundary conditions in the $x$ direction (and simulate the full range $-1 \leq x \leq 1$). The resolutions of the simulations are $1166\times 4080$ for those reported in Section~\ref{sec:introduction}, $1166\times 2040$ for those reported in Section~\ref{sec:LowReNumerics}, and $2332\times 4080$ for those reported in Section~\ref{sec:numerics2}.

In all simulations, we choose the adiabatic index ${\gamma = 5/3}$, magnetic diffusivity ${\eta = 4\times 10^{-6}}$ and thermal diffusivity $K=6\times 10^{-6} $. This means that ${\eta/K = 2/3 = \gamma -1}$, which is the critical ratio for stability to double-diffusive instabilities at any $\nu$~\citep{Hughes85}. The kinematic viscosity is ${\nu = 4\times 10^{-6}}$ for the simulations in Sections~\ref{sec:introduction} and~\ref{sec:numerics2} and ${\nu = 1.6\times 10^{-3}}$ for the simulations in Section~\ref{sec:LowReNumerics}. The details of the equilibrium states in which we initialise the simulations are explained in Section~\ref{sec:examples}.

\section{Comparison with moist hydrodynamics\label{app:moist}}

In this appendix, we apply the general formalism of Section~\ref{sec:generalEqState} to moist hydrodynamics, thus elucidating the analogy between metastability in 2D magnetohydrodynamics and in the terrestrial atmosphere. The appendix is mostly a review of standard results, but derived economically via the linear- and nonlinear-stability criteria~\eqref{linearstability} and~\eqref{metastability_compressibility}. 

\subsection{Overview}

Moist hydrodynamics is the fluid dynamics of a mixture of dry air, water vapour and liquid water in local thermodynamic equilibrium. In this appendix, we restrict attention to the case where the liquid water is suspended in the air, as in clouds and fog, and does not precipitate out. As for 2D MHD [see~\eqref{specificmagflux} and~\eqref{entropyfn}], moist hydrodynamics has two quantities that are conserved in a Lagrangian sense in the absence of diffusion. These are the specific entropy $S$ and the water mixing ratio $w$, which is the ratio of the mass of water (both liquid and vapour) to the mass of dry air. As with our use of the ``entropy function'' $s$ in the main text [see~\eqref{entropyfn}], it is more convenient to work with a quantity derived from $S$---in this case, potential temperature, $\theta$---than with $S$ directly. We therefore take the vector of conserved quantities~\eqref{adiabatic} for moist hydrodynamics to be~$\bQ = (\theta,w)$. 

In the following sections, we consider the linear and nonlinear stability of unsaturated air (composed of dry air and water vapour) and saturated air (in which water is in both vapour and liquid states) in turn. For unsaturated air, we review the definition of specific entropy~$S$ in Section~\ref{sec:specificentropy}, use it to derive a formula for potential temperature~$\theta$ in Section~\ref{sec:potentialtemperature} and apply the general formula~\eqref{linearstability} to determine the criterion for linear stability in Section~\ref{sec:moistunsatlinearstab}.  We show in Section~\ref{sec:unsaturatedmetastability} that the compressibility $\kappa$ of unsaturated air increases with $w$ because the specific heat capacity of water vapour is greater than that of dry air. However, because $w\ll 1$ in the atmosphere, differences in compressibility are always small, meaning that metastability does not occur with only unsaturated air in practice.

On the other hand, when a parcel of unsaturated air rises and cools sufficiently for vapour to condense (at the so-called lifting condensation level), the newly saturated parcel becomes significantly more compressible than its dry-air surroundings. This is because further decrease in pressure leads to additional cooling and condensation, which releases latent heat and re-warms the parcel somewhat, leading to additional expansion. As a result, the density of the saturated parcel decreases more in response to a change in pressure than the density of the surrounding dry air does. Furthermore, because the specific latent heat of condensation of water is much greater than the typical thermal energy per unit mass of air, differences in compressibility between saturated and unsaturated air can be significant even for $w\ll 1$. In the Earth's atmosphere, cumulonimbus clouds form as the result of nonlinear instability arising from this effect [see, e.g., \citet{RogersYau96}]. We calculate the compressibility of saturated air in Section~\ref{sec:kappasaturated}, after introducing its specific entropy and the liquid-water potential temperature in Sections~\ref{sec:specificentropysaturated} and~\ref{sec:liquidwaterPT}, respectively. We determine the linear-stability criterion for an atmosphere containing saturated air in Section~\ref{sec:moistsatlinearstab}.

\subsection{Case of unsaturated air}

\subsubsection{Specific entropy of unsaturated air\label{sec:specificentropy}}

In what follows, we use the subscripts $\mathrm{d}$, $\mathrm{v}$, $\mathrm{l}$ and $\mathrm{w}$ to refer to dry air, water vapour, liquid water (in Section~\ref{sec:saturated} only) and total water, respectively. For the cases of dry air and water vapour, we have from the first law of thermodynamics applied to an ideal gas that
\begin{equation}
    \dd S_i = \frac{c_i}{M_i}\frac{\dd T}{T} - \frac{R}{M_i}\frac{\dd p_i}{p_i} \implies S_i = S_{0i}+\frac{1}{M_i}\left(c_i\ln \frac{T}{T_0}-R\ln \frac{p_i}{P_0}\right)\label{1stlaw}
\end{equation}where $i \in \{\mathrm{d}, \mathrm{v}\}$ is the species index, $S_i$ specific entropy, $p_i$ partial pressure, $c_i$ molar heat capacity at constant pressure, $M_i$ molar mass, $R$ the universal gas constant, and $S_{0i}$ the specific entropy in the reference state that has temperature and partial pressure equal to $T_0$ and $P_0$, respectively. With $m_i$ the mass of species $i$ in a mixture, the specific entropy $S_{\mathrm{unsat}}$ of a mixture of dry air and water vapour satisfies
\begin{equation}
    (m_\mathrm{d}+m_\mathrm{v})S_{\mathrm{unsat}} = m_\mathrm{d} S_\mathrm{d} + m_\mathrm{v} S_\mathrm{v} \implies (1+w)S_{\mathrm{unsat}} = S_\mathrm{d} + wS_\mathrm{v} \label{s_unsat}
\end{equation}where the water content $w\equiv m_w/m_\mathrm{d}$ is equal to $m_\mathrm{v}/m_\mathrm{d}$ because all water is in the vapour state. Substituting \eqref{1stlaw} into \eqref{s_unsat}, using $ p_i V = n_i R T \implies p_\mathrm{v}/p_\mathrm{d} = w/\varepsilon$ where $\varepsilon \equiv M_\mathrm{d}/M_\mathrm{v}$ and $p_\mathrm{d} + p_\mathrm{v} = P$, we obtain an expression for ${S_{\mathrm{unsat}}=S_{\mathrm{unsat}}(P,T,w)}$:
\begin{multline}
    M_\mathrm{d}(1+w) S_{\mathrm{unsat}} = M_\mathrm{d}(S_{0\mathrm{d}}+w S_{0\mathrm{v}}) + \left(c_\mathrm{d} +\frac{w}{\varepsilon }c_\mathrm{v}\right)\ln \frac{T}{T_0}-R\left(1+\frac{w}{\varepsilon}\right)\ln \frac{P}{P_0}\\+R\left(1+\frac{w}{\varepsilon}\right)\ln \left(1+\frac{w}{\varepsilon}\right) - R \frac{w}{\varepsilon} \ln \frac{w}{\varepsilon}
\end{multline}

\subsubsection{Potential temperature\label{sec:potentialtemperature}}

It is conventional in studies of convection to work not with the entropy directly but rather potential temperature $\theta$, which may be defined as
\begin{align}
    \ln \frac{\theta}{T_0} & \equiv  \frac{1}{c_\mathrm{d}+wc_\mathrm{v}/\varepsilon} \bigg[M_\mathrm{d}(1+w) S_{\mathrm{unsat}} - M_\mathrm{d}(S_{0\mathrm{d}}+w S_{0\mathrm{v}}) \nonumber \\&\hspace{3cm}-R\left(1+\frac{w}{\varepsilon}\right)\ln \left(1+\frac{w}{\varepsilon}\right) + R\frac{w}{\varepsilon} \ln \frac{w}{\varepsilon} \bigg] \label{potentialtemperature}\\ 
    & = \ln \frac{T}{T_0} - \left[1-\frac{1}{\Gamma(w)}\right]\ln \frac{P}{P_0},
\end{align}so that
\begin{equation}
    \theta = T \left(\frac{P}{P_0}\right)^{1/\Gamma(w)-1},\label{thetae}
\end{equation}where the adiabatic index is
\begin{equation}
    \Gamma (w) \equiv \frac{c_\mathrm{d} + c_\mathrm{v} w/\varepsilon}{c_\mathrm{d} - R + (c_\mathrm{v}-R)w/\varepsilon}.
\end{equation}The potential temperature $\theta$ is the temperature of a fluid parcel moved at fixed entropy and water-mixing ratio to the reference pressure $P_0$. We see from its definition~\eqref{potentialtemperature} that $\theta$ is conserved under isentropic displacements that preserve the composition $w$.

\subsubsection{Linear stability of unsaturated air\label{sec:moistunsatlinearstab}}

The linear-stability criterion~\eqref{linearstability} reads
\begin{equation}
    \mathcal{L}\equiv -\frac{\dd \bQ}{\dd z} \bcdot \frac{\p \ln \rho(P,\bQ)}{\p\bQ}>0,\quad \forall z.\label{linearstabilitycopymoist}
\end{equation}We therefore require $\rho = \rho(P,\theta,w)$. It is straightforward to show from the ideal gas law that
\begin{equation}
    P = \rho \frac{RT}{M_\mathrm{d}}\frac{1+w/\varepsilon}{1+w} \implies \rho = \frac{M_\mathrm{d} P_0}{R\theta} \left(\frac{P}{P_0}\right)^{1/\Gamma(w)} \frac{1+w}{1+w/\varepsilon}\label{rhounsaturated}
\end{equation}The partial derivatives in \eqref{linearstabilitycopymoist} are then
\begin{equation}
    \left(\frac{\p \ln \rho }{\p \theta}\right)_{w,P} = -\frac{1}{\theta},\quad \left(\frac{\p \ln \rho }{\p w}\right)_{\theta,P} = -\frac{\Gamma'}{\Gamma^2}\ln \left(\frac{P}{P_0}\right) + \frac{1}{1+w}-\frac{1}{\varepsilon+w},
\end{equation}where $\Gamma' = \dd \Gamma/\dd w<0$. The criterion for linear stability becomes [cf.~\eqref{Lmagnetic}]
\begin{align}
    \mathcal{L}&=\frac{1}{\theta}\frac{\dd \theta}{\dd z} + \left[\frac{\Gamma'}{\Gamma^2}\ln\left(\frac{P}{P_0}\right)+\frac{1-\varepsilon}{(1+w)(\varepsilon+w)}\right]\frac{\dd w}{\dd z}\label{linearmoist1}\\
    & = \frac{\dd \ln T}{\dd z} + \left[\frac{1}{\Gamma(w)}-1\right]\frac{\dd \ln P}{\dd z} + \frac{1-\varepsilon}{(1+w)(\varepsilon+w)}\frac{\dd w}{\dd z}>0,\quad \forall z.\label{linearmoist2}
\end{align}The first two terms become the negative of the gradient of the potential temperature if $w$ is constant in $z$; in that case, the system is stable if $\dd \theta /\dd z>0$. The third term represents the contribution from moisture content: we see that the presence of water vapour is stabilising when $w$ increases with height; this is because the molar mass of water is less than that of dry air ($\varepsilon<1$).

\subsubsection{Metastability of unsaturated air\label{sec:unsaturatedmetastability}}

By~\eqref{rhounsaturated}, the compressibility of unsaturated air is [cf.~\eqref{kappa_mhd}]
\begin{equation}
    \kappa \equiv \frac{\p \ln \rho(P,\theta,w)}{\p \ln P}= \frac{1}{\Gamma(w)}. 
\end{equation}The adiabatic index $\Gamma(w)$ is a decreasing function of $w$ because the specific heat capacity of water vapour is greater than that of dry air. Therefore, air with greater water mixing ratio $w$ is more compressible, and so unsaturated air can be metastable when (i) the atmosphere is sufficiently close to marginal linear stability and (ii) wetter air moves through dryer air. However, because $w$ in the atmosphere is typically smaller than $1\%$, differences in $\kappa$ between different parcels of unsaturated air are small. Metastability in the atmosphere occurs in practice because of the release of latent heat during condensation, as we now describe.

\subsection{Case of saturated air\label{sec:saturated}}

\subsubsection{Specific entropy of saturated air\label{sec:specificentropysaturated}}

Once the partial pressure of water vapour $p_{\mathrm{v}}$ becomes equal to the saturation vapour pressure $p_{\mathrm{sat}}$ (a known function of $T$ stated explicitly in Section~\ref{sec:ClausiusClapeyron}), the vapour condenses to form liquid water. The moist air is then composed of dry air, vapour and suspended liquid water. Its specific entropy $S_{\mathrm{sat}}$ satisfies
\begin{equation}
    (m_\mathrm{d} + m_w)S_{\mathrm{sat}} = m_\mathrm{d} S_\mathrm{d} + m_\mathrm{v} S_\mathrm{v} + m_{\mathrm{l}} S_\mathrm{l} \implies (1+w)S_{\mathrm{sat}}=S_\mathrm{d}+(w-w_\mathrm{v})S_\mathrm{l}+w_\mathrm{v} S_\mathrm{v}\label{s_sat}
\end{equation}where we define $w_\mathrm{v} \equiv m_\mathrm{v}/m_\mathrm{d}$. In the saturated state, $p_\mathrm{v} = p_{\mathrm{sat}}$, so $P=p_\mathrm{d}+p_{\mathrm{sat}}$. Thus, from ${p_iV=n_iRT}$, we have that 
\begin{equation}
    w_\mathrm{v} = \frac{\varepsilon p_{\mathrm{sat}}}{p_\mathrm{d}} = \frac{\varepsilon p_{\mathrm{sat}}}{P-p_{\mathrm{sat}}}.\label{wv}
\end{equation}Using these expressions, we can write $S_{\mathrm{d}}$ and $S_\mathrm{v}$, given by~\eqref{1stlaw}, in terms of $P$, $T$ and $w$. The remaining specific entropy $S_\mathrm{l}$ can be determined by summing the parts that correspond to isobaric heating of vapour from $T_0$ to $T$, isothermal compression from $P_0$ to $p_{\mathrm{sat}}$, and finally condensation with a latent heat per mole of $L(T)$. Thus,
\begin{equation}
    S_\mathrm{l} = S_{0\mathrm{v}} + \frac{1}{M_w}\left[c_\mathrm{v} \ln \left(\frac{T}{T_0}\right)-R\ln \frac{p_{\mathrm{sat}}}{P_0} -\frac{L}{T}\right].
\end{equation}Substituting the three contributions into \eqref{s_sat}, we obtain
\begin{multline}
    (1+w)M_\mathrm{d} S_{\mathrm{sat}}= M_\mathrm{d}(S_{0\mathrm{d}}+w S_{0\mathrm{v}})+  \left(c_\mathrm{d} +\frac{w}{\varepsilon }c_\mathrm{v}\right)\ln \frac{T}{T_0}-R\left(1+\frac{w}{\varepsilon}\right)\ln\frac{P}{P_0}\\+R\left(1+\frac{w}{\varepsilon}\right)\ln\left(1+\frac{w_\mathrm{v}}{\varepsilon}\right)-R\frac{w}{\varepsilon}\ln \frac{w_\mathrm{v}}{\varepsilon} - \frac{L}{\varepsilon T}(w-w_\mathrm{v}).
\end{multline}This constitutes an expression for $S_{\mathrm{sat}}$ as a function of $P$, $T$ and $w$. The functions $p_{\mathrm{sat}}(T)$ and $L(T)$ are given by~\eqref{psat} and~\eqref{latent2}, respectively.

\subsubsection{Liquid-water potential temperature\label{sec:liquidwaterPT}}

Analogously to \eqref{potentialtemperature}, one can define a conserved quantity known as ``liquid-water potential temperature'' by
\begin{align}
    \ln \frac{\theta_{\mathrm{l}}}{T_0} & \equiv  \frac{1}{c_\mathrm{d}+wc_\mathrm{v}/\varepsilon} \bigg[M_\mathrm{d}(1+w) S_{\mathrm{unsat}} - M_\mathrm{d}(S_{0\mathrm{d}}+w S_{0\mathrm{v}}) \nonumber \\&\hspace{3cm}-R\left(1+\frac{w}{\varepsilon}\right)\ln \left(1+\frac{w}{\varepsilon}\right) + R\frac{w}{\varepsilon} \ln \frac{w}{\varepsilon} \bigg] \label{liquidpotentialtemperature}\\ 
    & = \ln \frac{T}{T_0} - \left[1-\frac{1}{\Gamma(w)}\right]\ln \frac{P}{P_0} + \frac{1}{c_\mathrm{d}+wc_\mathrm{v}/\varepsilon} \bigg[R\left(1+\frac{w}{\varepsilon}\right)\ln \left(\frac{1+w_\mathrm{v}/\varepsilon}{1+w/\varepsilon}\right)\nonumber\\&\hspace{3cm} - R\frac{w}{\varepsilon}\ln \frac{w_\mathrm{v}}{w}- \frac{L}{\varepsilon T}(w-w_\mathrm{v})\bigg].\label{liquidpotentialtemperature2}
\end{align}Liquid-water potential temperature is the potential temperature that a parcel of air would have if all the water in it were vaporised ($w_\mathrm{v} = w$). Because the latent heat of condensation of water is large, i.e., $L/\varepsilon RT\sim 40 \gg 1$ (see footnote~\ref{footnotevaluesmoist} for characteristic sizes of these quantities), the term involving $L$ in \eqref{liquidpotentialtemperature2} is typically much larger than the other terms in the second set of square brackets. Neglecting those terms, \eqref{liquidpotentialtemperature2} becomes
\begin{equation}
    \theta_{\mathrm{l}} = T \left(\frac{P}{P_0}\right)^{1/\Gamma(w)-1}\exp \left(-\frac{L}{\varepsilon c_\mathrm{d} T}(w-w_\mathrm{v})\right).\label{thetalapprox1}
\end{equation}

\subsubsection{Change in compressibility on saturation\label{sec:kappasaturated}}

\begin{figure}
    \centering
    \includegraphics[width=0.8\linewidth]{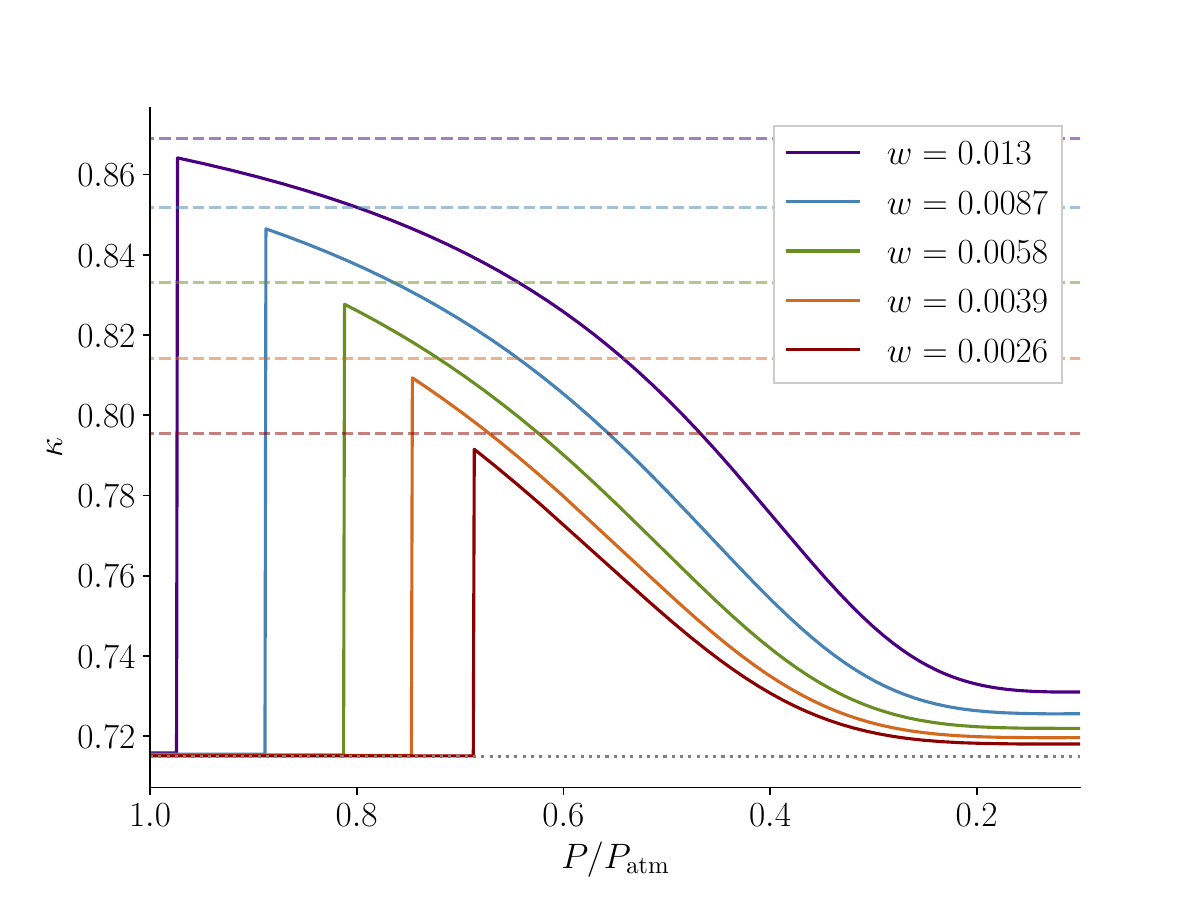}
    \caption{Solid lines show the compressibility $\kappa$ of moist air as a function of total pressure $P$ at fixed $\theta_{\mathrm{l}} = 293\,\mathrm{K}$ and different mixing ratios $w$, calculated numerically from \eqref{wv}, \eqref{liquidpotentialtemperature2} and \eqref{saturateddensity}. Pressure is measured in units of $P_0=P_{\mathrm{atm}} = 101.325\,\mathrm{kPa}$. Dashed horizontal lines show the prediction \eqref{kappasaturated} for the maximum value of $\kappa$ [small discrepancies with the maxima of the solid curves are due to approximations made in deriving \eqref{kappasaturated}, specifically, the neglect of terms involving $w$ that are not multiplied by the large ratio $L/RT$]. The temperatures of saturation $T_{\mathrm{sat}}$ (obtained numerically) are, in order of decreasing $w$, $291\,\mathrm{K}, 283\,\mathrm{K}, 276\,\mathrm{K}, 270\,\mathrm{K}$ and $263\,\mathrm{K}
$ (temperatures smaller than that of the triple point of water,~$\simeq 273\,\mathrm{K}$, correspond to a supercooled liquid state---we neglect any effect of ice formation).}
    \label{fig:moistcompressibility}
\end{figure}

The compressibility $\kappa$ of moist air may be determined by inverting \eqref{liquidpotentialtemperature2} for ${T=T(P, \theta_{\mathrm{l}}, w)}$, substituting the result into
\begin{equation}
    P = \rho \frac{RT}{M_\mathrm{d}}\frac{1+w_\mathrm{v}/\varepsilon}{1+w},\label{saturateddensity}
\end{equation}and then taking the derivative of $\rho$ with respect to $P$ at fixed $\theta_\mathrm{l}$ and $w$. The result of doing so numerically is shown in figure~\ref{fig:moistcompressibility} for a number of different values of $w$, with $L(T)$ and $p_{\mathrm{sat}}(T)$ given by \eqref{latent2} and \eqref{psat} in Section~\ref{sec:ClausiusClapeyron}, respectively.\footnote{\label{footnotevaluesmoist}We take $R = 8.31\, \mathrm{J}\,\mathrm{mol}^{-1}\,\mathrm{K}^{-1}$, $M_\mathrm{d} = 0.0290\, \mathrm{kg}\,\mathrm{mol}^{-1}$, $\varepsilon = 0.622$, $c_\mathrm{d} = 29.2\, \mathrm{J}\,\mathrm{mol}^{-1}\,\mathrm{K}^{-1}$, $c_\mathrm{v} = 33.7\, \mathrm{J}\,\mathrm{mol}^{-1}\,\mathrm{K}^{-1}$, $c_\mathrm{l} = 75.5\, \mathrm{J}\,\mathrm{mol}^{-1}\,\mathrm{K}^{-1}$, $T_c = 283\, \mathrm{K}$, $L_c = 44600\, \mathrm{J}\,\mathrm{mol}^{-1}$, and $p_{\mathrm{sat}}(T_c) = 1230\, \mathrm{Pa}$. These are equivalent to the values quoted by \cite{Stansifer17}.} We observe that there is a significant increase in compressibility when the pressure reaches the critical value at which condensation occurs ($p_\mathrm{v} = p_{\mathrm{sat}}$). The height at which this happens for a rising air parcel is known as the level of lifting condensation (which can be at ground level if $w$ is sufficiently large, as can be the case in tropical regions on Earth). The increase in compressibility that results from saturation means that as the parcel continues to rise, its density decreases relative to that of the surrounding air (provided that the latter is sufficiently close to marginal linear stability). Above the height at which they become equal, known as the level of free convection, the rising air parcel experiences an upwards force and is therefore unstable. The towering cloud formation that forms because of the resulting updraught is known as cumulonimbus~\citep[see, e.g.,][]{RogersYau96}.

The change in compressibility at saturation can be calculated analytically as follows. Due to the temperature dependencies of $L(T)$ [see~\eqref{latent2}] and $w_\mathrm{v}$ through $p_{\mathrm{sat}}(T)$ [see \eqref{wv}], we cannot invert \eqref{thetalapprox1} analytically for ${T=T(P, \theta_{\mathrm{l}}, w)}$, unlike the case for $\theta$ in Section~\ref{sec:moistsatlinearstab}. We therefore restrict attention to the point of saturation, which occurs at pressure $P_{\mathrm{sat}}(\theta_{\mathrm{l}}, w)$ (note that $P_{\mathrm{sat}}$ is not the same as $p_{\mathrm{sat}}$: the former is the total pressure of the air at saturation while the latter is the partial pressure of the water vapour specifically). With ${T_{\mathrm{sat}}\equiv T(P_{\mathrm{sat}}, \theta_{\mathrm{l}}, w)}$, we have, to first order in $\delta P \equiv P - P_{\mathrm{sat}}$,
\begin{multline}
     \theta_{\mathrm{l}} = \left[T_{\mathrm{sat}} + \frac{\p T(P, \theta_{\mathrm{l}},w)}{\p P} \bigg|_{P=P_{\mathrm{sat}}}\delta P\right]\left(\frac{P_s}{P_0}\right)^{1/\Gamma-1}\left[1+\left(\frac{1}{\Gamma}-1\right)\frac{\delta P}{P_{\mathrm{sat}}}\right]\\ \times\left[1+\frac{L^2 w}{\varepsilon c_\mathrm{d} R T_{\mathrm{sat}}^3}\frac{\p T(P, \theta_{\mathrm{l}},w)}{\p P} \bigg|_{P=P_{\mathrm{sat}}}\delta P - \frac{wL}{\varepsilon c_\mathrm{d} T}\frac{\delta P}{P_{\mathrm{sat}}}\right].
\end{multline}It follows that
\begin{equation}
     \frac{\p T(P, \theta_{\mathrm{l}},w)}{\p P} \bigg|_{P=P_{\mathrm{sat}}}= \frac{1/\Gamma - 1 - Lw/\varepsilon c_\mathrm{d} T_{\mathrm{sat}}}{1+L^2 w/\varepsilon c_\mathrm{d} RT_{\mathrm{sat}}^2}.\label{dTdPsat}
\end{equation}Substituting~\eqref{dTdPsat} into~\eqref{saturateddensity}, we have
\begin{equation}
    \kappa(P_{\mathrm{sat}}, \theta_{\mathrm{l}}, w) \equiv \frac{\p \ln \rho (P, \theta_{\mathrm{l}},w)}{\p \ln P}\bigg|_{P=P_{\mathrm{sat}}} = \frac{1}{\Gamma} + \frac{w}{\varepsilon}\frac{L}{c_\mathrm{d} T_{\mathrm{sat}}}\frac{(1-1/\Gamma)L/RT_{\mathrm{sat}} - 1}{1+ L^2 w/\varepsilon c_\mathrm{d} R T_{\mathrm{sat}}^2},\label{kappasaturated}
\end{equation}where we have neglected the contribution to $\kappa$ of the $w_\mathrm{v}$ in~\eqref{saturateddensity}. To leading order in~$w$, the change in compressibility at saturation is
\begin{equation}
    \Delta \kappa \equiv \kappa(P_{\mathrm{sat}}, \theta_{\mathrm{l}}, w) - 1/\Gamma(w) \simeq \frac{w}{\varepsilon} \frac{L^2}{R c_\mathrm{v} T_{\mathrm{sat}}^2} \sim 10^3 w.\label{leadingdeltakappa}
\end{equation}Thus, $\Delta \kappa \sim 0.1$ [which is the largest possible difference in $\kappa$ in MHD, see~\eqref{kappa_mhd}] can be achieved even with a water-mixing ratio of $\sim 10^{-4}$ (we recall from~\eqref{rho_fraction_optimistic} that the fractional difference in density between a fluid parcel moved from pressure $P_1$ to $P_2$ and its surroundings at $P_2$ is $\sim \Delta \kappa /\kappa$ at marginal linear stability). For $w$ much larger than this, the leading-order approximation~\eqref{leadingdeltakappa} is inaccurate and one must use \eqref{kappasaturated} instead. 

\subsubsection{Linear stability of saturated air\label{sec:moistsatlinearstab}}

We note that the linear-stability criterion \eqref{linearmoist2} is modified if the air is saturated. It is straightforward to show from \eqref{linearstability_alternative} that the criterion becomes
\begin{equation}
    \mathcal{L} = \frac{\dd \ln T}{\dd z} +\left[\kappa(P(z),\theta_{\mathrm{l}}(z),w(z))-1\right]\frac{\dd \ln P}{\dd z}>0,
\end{equation}provided that $w_\mathrm{v}$ and $w$ can be neglected when compared with $1$ in \eqref{saturateddensity}. The compressibility $\kappa$ must be evaluated numerically, as explained in Section~\ref{sec:kappasaturated}.

\subsubsection{Saturation pressure and latent heat of condensation of water vapour\label{sec:ClausiusClapeyron}}

For completeness, we note that the saturation pressure of water vapour $p_{\mathrm{sat}}(T)$ can be obtained from the Clausius--Clapeyron equation
\begin{equation}
    \frac{\dd \ln p_{\mathrm{sat}}}{\dd \ln T} = \frac{L}{RT}\label{CCE}.
\end{equation}The latent heat of condensation $L(T)$ can be determined by considering a reversible cycle in which vapour on the coexistence curve is heated isobarically from reference temperature $T_c$ to $T$, compressed isothermally from $p_{\mathrm{sat}}(T_c)$ to $p_{\mathrm{sat}}(T)$, condensed to the liquid phase, cooled isobarically from $T$ to $T_c$, allowed to expand isothermally from $p_{\mathrm{sat}}(T) $ to $p_{\mathrm{sat}}(T_c)$ and then finally evaporated, thus returning to the initial state. For the total entropy change to be zero, we must have
\begin{equation}
    \frac{L}{T} = \frac{L_c}{T_c} - R\ln \frac{p_{\mathrm{sat}}(T)}{p_{\mathrm{sat}}(T_c)}+(c_\mathrm{v}-c_{\mathrm{l}})\ln \frac{T}{T_c}.\label{latent1}
\end{equation}Substituting this into~\eqref{CCE} and integrating, we find that
\begin{equation}
    R\ln \frac{p_{\mathrm{sat}}(T)}{p_{\mathrm{sat}}(T_c)}=\frac{L_c}{T_c}\left(1-\frac{T_c}{T}\right)+(c_\mathrm{v}-c_{\mathrm{l}})\left(\ln \frac{T}{T_c}-1+\frac{T_c}{T}\right),\label{psat}
\end{equation}while by eliminating $p_{\mathrm{sat}}$ between \eqref{latent1} and \eqref{psat}, we have
\begin{equation}
    L = L_c + (c_\mathrm{v}-c_{\mathrm{l}})(T-T_c).\label{latent2}
\end{equation}

\section{Stability analysis to quadratic order \label{quadratic}}

In this appendix, we present an expansion of the buoyancy force on a displaced fluid element,~\eqref{F2pressure}, for an equilibrium close to marginal linear stability, to quadratic order in the displacement. We show that the results are consistent with the conclusions of Section~\ref{sec:metastabilityMHD}. 

We take the typical scales of variation of $s$, $\chi$, $P$ and $\rho$ in $z$ to be the same, denoted $H$, but we take the two terms that appear in the scalar product in~\eqref{F2pressure} to have opposite signs and mostly cancel, so that the equilibrium is close to marginal linear stability, i.e.,
\begin{equation}
    \mathcal{L}\sim \epsilon \frac{\rho}{ H}>0,\label{marginal}
\end{equation}where $0<\epsilon \ll 1$. Expanding~\eqref{F2pressure} in ${\delta z=z_2-z_1}$ yields
\begin{equation}
    \frac{F}{gV_2} = -\mathcal{L}\,\delta z + \mathcal{N}\,\delta z^2 + \mathcal{O}(\delta z^3),\label{expansion}
\end{equation}where
\begin{align}
    \mathcal{N} & =
     -\frac{\dd \mathcal{L}}{\dd z} + \frac{\dd P}{\dd z}\left(\frac{\dd s}{\dd z}\frac{\p^2 \rho}{\p P \p s}+\frac{\dd \chi}{\dd z}\frac{\p^2 \rho}{\p P \p \chi}\right) \nonumber \\
     & = \frac{\dd P}{\dd z}\left(\frac{\dd s}{\dd z}\frac{\p^2 \rho}{\p P \p s}+\frac{\dd \chi}{\dd z}\frac{\p^2 \rho}{\p P \p \chi}\right)+ \mathcal{O}(\epsilon),\label{N}
\end{align}where we have used that ${\dd \mathcal{L}/\dd z \sim \epsilon \rho /H^2}$ by~\eqref{marginal}.

Unless both $s$ and $\chi$ increase with height [in which case~\eqref{marginal} demands that their gradients each be small], ${\mathcal{N}\sim \rho /H^2}$ and does not generally vanish at marginal stability. The equilibrium is therefore stable to linear perturbations with ${\delta z \ll \mathcal{L}/\mathcal{N}\sim \epsilon H}$ but unstable to nonlinear ones with ${\epsilon H \ll \delta z \ll H}$ [the latter condition ensuring that the $\mathcal{O}(\delta z^3)$ terms in~\eqref{expansion} are negligible]. Replacing partial derivatives by their expressions in 2D MHD,~\eqref{N} becomes
\begin{equation}
    \mathcal{N} = - \frac{\rho g c_s^2}{c^4}(2-\gamma)\frac{\dd \ln s}{\dd z}+\mathcal{O}(\epsilon),\label{Nmagnetic}
\end{equation}which implies that $\mathcal{N}<0$ for a stabilising entropy gradient $\dd \ln s /\dd z>0$ (and $\dd \ln \chi /\dd z<0$), and therefore the atmosphere is metastable to \textit{downwards} displacements (provided $\gamma<2$). Conversely, the atmosphere is metastable to \textit{upwards} displacements if it has a destabilising entropy gradient but stabilising gradient of magnetic flux. These are the same conclusions as would be obtained by determining the direction of metastable displacements as the ones in the direction that the ratio $s/\chi$ decreases, as should be the case according to the argument in Section~\ref{sec:metastabilityMHD}. 

\section{Available energy of a two-phase atmosphere\label{sec:anaytic_ground_state}}

\begin{figure*}
    \centering
    \includegraphics[width=0.9\textwidth]{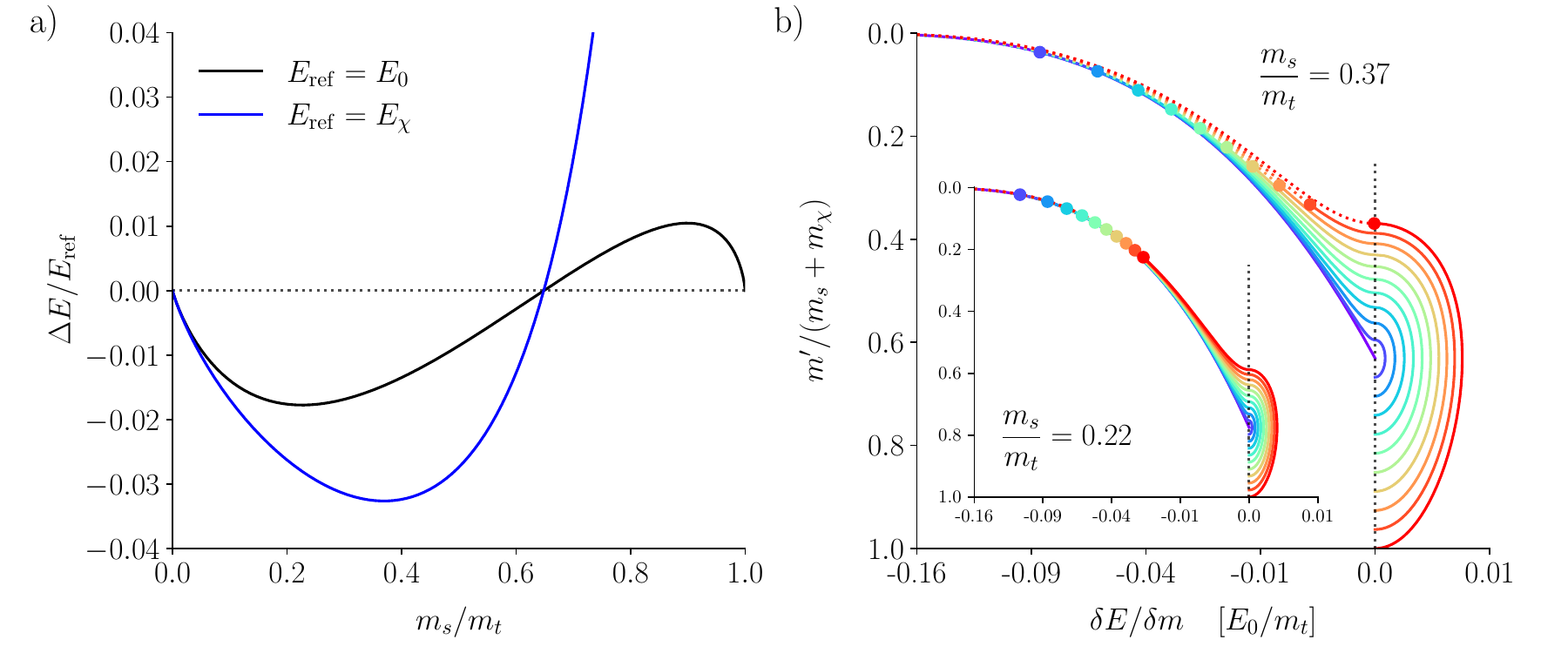}
    \caption{Panel (a): the available energy of the ``two-phase'' atmosphere considered in Appendix~\ref{sec:anaytic_ground_state}, as a function of the mass $m_s$ of fluid with $\beta = \infty$. The black line normalises the available energy by the total initial potential energy, the blue line by the initial potential energy of the fluid with $\beta = 0$ only. Panel (b): The energy liberated per unit mass of moving fluid when a slice moves from its initial position to a new location where the supported mass is $m'$. The main plot shows the case of $m_s/m_t=0.37$, which corresponds to the largest possible available energy [minimum of the blue line in panel~(a)]; the inset shows the case of $m_s/m_t=0.37$, which corresponds to the largest possible fractional available energy [minimum of the black line in panel~(a)].}
    \label{fig:energy_analytically}
\end{figure*}

In this appendix, we consider the simple case of a ``two-phase'' atmosphere in which a mass $m_s$ of fluid with ${\chi = 0}$ ($\beta = \infty$) is situated initially below a mass $m_{\mathrm{\chi}}$ of fluid with $s=0$ ($\beta = 0$), for which the available energy can be determined analytically. This case is of pedagogical value as it illustrates why the available energy is always a small fraction of the total.

According to~\eqref{metastability_compressibility}, the large-$\beta$ fluid experiences a destabilising force when displaced upwards into the small-$\beta$ fluid (indeed, the destabilising force is the greatest possible, as the difference in compressibility is maximal). Intuitively, therefore, we expect the minimum-energy state to be obtainable by re-stacking the atmosphere such that the large-$\beta$ fluid sits above the small-$\beta$ fluid. We consider the most optimistic case from the perspective of available energy---that of an atmosphere initially at marginal linear stability, i.e., $\mathcal{L}=0$---which means that, by~\eqref{linearstability}, $s$ and $\chi$ are piecewise-constant functions of~$m$:
\begin{equation}
    s = \begin{cases}
  0,  & m<m_\chi, \\
  s_0, & m_\chi< m < m_t,
\end{cases} \quad 
\chi = \begin{cases}
  \chi_0,  & m<m_\chi, \\
  0, & m_\chi< m < m_t,
\end{cases}
\end{equation}where $m_t=m_s+m_{\chi}$ is the total mass of the atmosphere. The interface of the two phases at $m=m_\chi$ is stable provided the density of the fluid above it is smaller than the density of the fluid below---the case of marginal linear stability is the one where the fluid density is continuous at the interface. By~\eqref{equilibrium_rho}, this implies that
\begin{equation}
    \frac{s_0}{\chi_0}=\frac{1}{\sqrt{2}}(m_\chi g)^{1/\gamma - 1/2}.
\end{equation}
The change in energy $\Delta E$ when the entire mass $m_s$ of large-$\beta$ fluid is moved from the bottom to the top may then be shown straightforwardly to satisfy
\begin{equation}
    \frac{\Delta E}{E_\chi} = C(r^{2-1/\gamma}-(1+r)^{2-1/\gamma}+1)+(1+r)^{3/2}-1-r^{3/2},\label{deltaE_analytic}
\end{equation}where $r=m_s/m_\chi$, $C = 3\gamma^2/4(\gamma-1)(2\gamma-1)$ and
\begin{equation}
    E_\chi = \frac{2}{3}(2 m_{\chi}^3 g)^{1/2}\chi_c \label{Echi}
\end{equation}is the total energy of the small-$\beta$ fluid at the top of the atmosphere, which is related to the total energy of the atmosphere $E_{0}$ via
\begin{equation}
    \frac{E_0}{E_\chi} = 1+C[(1+r)^{2-1/\gamma}-1]. \label{E0_on_Echi}
\end{equation}

In the limit of $r\to 0$,~\eqref{deltaE_analytic} and~\eqref{E0_on_Echi} give
\begin{equation}
    \frac{\Delta E}{E_{\chi}} = \frac{\Delta E}{E_0}= -\frac{3r}{8} = -\frac{3m_s}{8m_{t}}.
\end{equation}Thus, the fractional energy change of a small-$\beta$ atmosphere at marginal linear stability when a small parcel of large-$\beta$ fluid rises through it is proportional to the mass fraction of the fluid moved, with a proportionality factor of order unity. However, returns in the fractional energy change diminish rapidly as the amount of mass moved increases. In figure~\ref{fig:energy_analytically}(a), we visualise $\Delta E$ normalised by~$E_0$ (black line) and~$E_{\chi}$ (blue line) as a function of the mass fraction $m_s/m_t$ of large-$\beta$ fluid in the atmosphere. We observe that, despite its order-unity initial gradient of~$-3/8$, $\Delta E/E_0$ shallows rapidly, reaching a minimum of $-1.8\%$ where the large-$\beta$ fluid has a mass fraction of $22\%$. If the mass fraction of the large-$\beta$ is greater than this, the fractional energy liberated in moving the large-$\beta$ fluid is smaller, and, for a mass fraction of greater than $65\%$, moving the large-$\beta$ fluid incurs an energetic cost.

To elucidate the reason for the diminishing return, we consider the rearrangement as a sequence of displacements of parcels of the large-$\beta$ fluid from just below the interface to new positions at the top of the atmosphere. Figure~\ref{fig:energy_analytically}(b) shows, for a sequence of parcel displacements represented by lines coloured from purple (the first parcel to move) to red (the last parcel to move), the energy $\delta E$ that is liberated when a mass $\delta m$ of the large-$\beta$ fluid is moved from the (evolving) position of the interface to a new, smaller, supported mass $m'$ (end points of the particle motions are shown as filled circles). The decrease in the magnitude of $\delta E/\delta m$ for subsequent parcels is seen to occur because fluid parcels obey an exclusion principle---two slices cannot have the same supported mass. This means that a parcel of fluid that has already risen from the bottom to the top of the atmosphere will prevent the next parcel from rising to the same height, thus limiting the difference in total pressure that this parcel experiences over its motion, and so reducing the work done on the parcel by the buoyancy force (recall that for an atmosphere at marginal linear stability, the latter is proportional to the ratio of pressures at the initial and final positions).

A second effect visible in figure~\ref{fig:energy_analytically}(b) is that the motion of the first fluid parcel stabilises the atmosphere somewhat, because it moves the interface between the large- and small-$\beta$ fluid downwards, where the total pressure is greater (the constants $s_0$ and $\chi_0$ were chosen such that the large- and small-$\beta$ fluids had the same density at $P = m_{\chi} g$; at larger pressure, the more compressible, large-$\beta$ fluid underneath the interface is denser). The buoyancy force on a rising parcel is downwards (restoring) until it passes the point of neutral buoyancy at the original position of the interface. If the work required to overcome the restoring force is greater than the energy liberated when the parcel moves to the greatest height permitted by the exclusion principle, then the reassignment incurs a net energetic cost. This situation occurs for $m_s\gtrsim 0.37 m_t$: for $m_s$ larger than this, $\Delta E/E_{\mathrm{\chi}}$ rises with $m_s$ [figure~\ref{fig:energy_analytically}(a)], becoming positive for $m_s\gtrsim 0.65 m_t$. $\Delta E/E_{0}$ reaches a minimum at smaller $m_s/m_{t}$ because the small increase in available energy is outweighed by the energetic cost incurred by increasing the total mass of the atmosphere.

\section{A necessary condition for ``one-to-many'' optimal assignment\label{app:onetomany}}

We can deduce a necessary condition for the one-to-many optimal assignment described in Section~\ref{sec:invm} in a similar manner to the one in which we deduced the equal-density condition~\eqref{samerho} for ``many-to-one'' assignments in Section~\ref{sec:selfsimilar}. Let us suppose that we are given the optimal assignment of all slices apart from those with initial supported mass $m_{i}=m_a+\delta m_{i}$. Their contribution to the total energy of the atmosphere is
\begin{align}
    \delta E & = \Delta m \sum_{i} \mcE(m_{\sigma(i)}, m_a+\delta m_i) \nonumber \\ & = \Delta m \sum_{i} \bigg[ \mcE(m_{\sigma(i)}, m_a) + \delta m_i \frac{\p \mcE}{\p \mu}(m_{\sigma(i)}, m_a) +\mathcal{O}(\delta m_{i}^2)\bigg],\label{deltaE_invm}
\end{align}where
\begin{equation}
    \frac{\p \mcE}{\p \mu} = \frac{1}{\gamma - 1}\rho^{\gamma -1}\frac{\dd s^{\gamma}}{\dd \mu}+\frac{1}{2}\rho\frac{\dd \chi^2}{\dd \mu}.
\end{equation}Thus, a necessary (but not sufficient) condition for the optimal assignment to be one-to-two is that $\p \mcE/\p \mu$ must be equal (up to a difference proportional to $\Delta m$) when evaluated [with ${s=s(\mu)}$ and ${\chi=\chi(\mu)}$] at each of the two different supported masses $m_{\sigma(i)}$ to which slices in the vicinity of $m_a$ are to be assigned. If this is not the case for any $\delta m_i$ in the given range, then~\eqref{deltaE_invm} is minimised to leading order in $\delta m$ by assigning the slice with largest initial supported mass to the new supported mass for which $\p \mcE/\p \mu$ is smallest, and so on---this will always be a one-to-one assignment, as $\p \mcE/\p \mu$ is continuous in $m_{\sigma(i)}$ for fixed $m_a$.

Denoting the density of slice $i$ at the two values of $m_{\sigma(i)}$ by $\rho$ and $\rho'$, the condition for a one-to-two mapping is [cf.~\eqref{samerho}]
\begin{equation}
    \frac{1}{\gamma - 1}(\rho'^{\gamma -1}-\rho^{\gamma -1})\frac{\dd s^{\gamma}}{\dd m}+\frac{1}{2}(\rho'-\rho)\frac{\dd \chi^2}{\dd m} = \mathcal{O}(\Delta m).\label{one-to-two_condition}
\end{equation}From this expression, we see that, for any given $\rho$, there can be at most one solution for $\rho'$ in addition to the trivial $\rho'=\rho$. The existence of a second solution requires that gradients of $s$ and $\chi$ with $m$ have opposite signs in the initial profile. Thus, while an optimal assignment may in general be one-to-two, as in figure~\ref{fig:downwards_epilons}, it cannot be one-to-$X$ with ${X>2}$. 

\section{Viscous relaxation in the case of downwards metastability,~\eqref{invm}\label{sec:2D_sims_down}}

\begin{figure*}
    \centering
    \includegraphics[width=\textwidth]{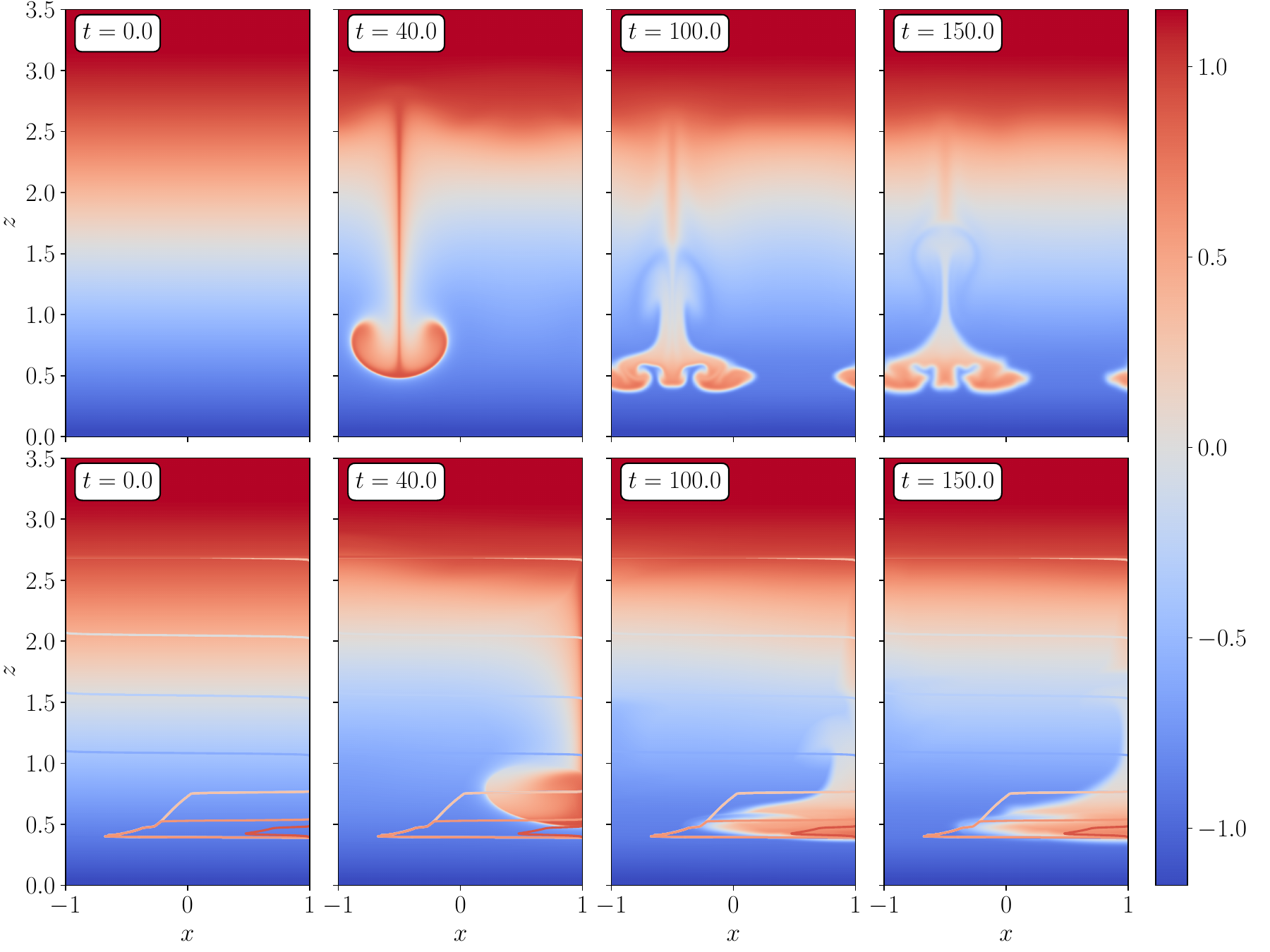}
    \caption{As for figure~\ref{fig:upwards_2D_sim}, but for the equilibrium defined by~\eqref{invm}.\label{fig:downwards_2D_sim}}
\end{figure*}

\begin{figure*}
    \centering
    \includegraphics[width=.85\textwidth]{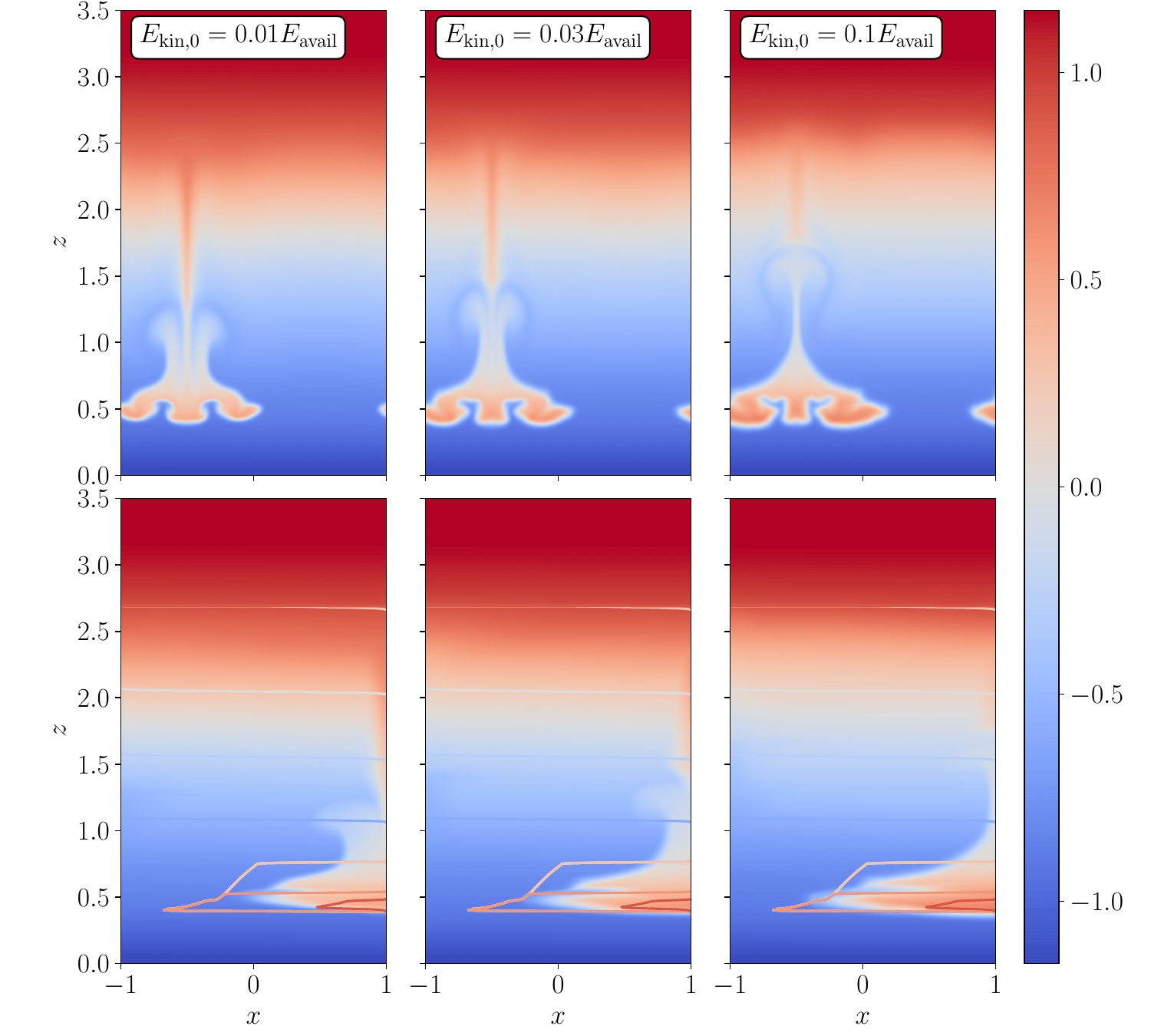}
    \caption{As for figure~\ref{fig:upwards_2D_v0_comparison}, but for the equilibrium defined by~\eqref{invm}.\label{fig:downwards_2D_v0_comparison}}
\end{figure*}

Figures~\ref{fig:downwards_2D_sim} and~\ref{fig:downwards_2D_v0_comparison} are analogues of figures~\ref{fig:upwards_2D_sim} and~\ref{fig:upwards_2D_v0_comparison} for the equilibrium defined by~\eqref{invm}. The dimensionless numbers $\mathrm{Re}$, $\mathrm{Pr}_m$ and $\mathrm{Pr}_t$ are the same as in Section~\ref{sec:LowReNumerics}, but $z_0=2.25$ in~\eqref{u0}. In figure~\ref{fig:downwards_2D_v0_comparison}, We again compare $u_0=0.1$, $u_0=0.05$, and $u_0=0.025$; in this case, this corresponds to $E_{\mathrm{kin,0}}\simeq 0.1 E_{\mathrm{avail}}$, $E_{\mathrm{kin,0}}\simeq 0.03 E_{\mathrm{avail}}$ and $E_{\mathrm{kin,0}}\simeq 0.01 E_{\mathrm{avail}}$, respectively. Again, we observe the formation of a long-lived 2D state (subject to slow diffusion) that is consistent with the minimum-energy state obtained in Section~\ref{sec:invm} (see lower panels of figures~\ref{fig:downwards_2D_sim} and~\ref{fig:downwards_2D_v0_comparison}), with the quality of agreement between theory and simulation being better at $E_{\mathrm{kin,0}}\simeq 0.1 E_{\mathrm{avail}}$ than $E_{\mathrm{kin,0}}\simeq 0.01 E_{\mathrm{avail}}$.

\section{Alternative formulation of the Lynden-Bell statistical mechanics\label{sec:alternative}}

For completeness, we note in this appendix how Lynden-Bell statistical mechanics may be formulated for the joint probability-distribution function for $s$ and $\chi$, rather than for $\mu$ as in Section~\ref{sec:LyndenBellEquations}. In that case, we define $\mathcal{P}(m, s, \chi)\dd s \dd \chi$ to be the probability that a fluid element at supported mass~$m$ has specific entropy and flux in the ranges~$s$ to~$s+\dd s$ and $\chi$ to~$\chi+\dd \chi$, respectively. We obtain $\mathcal{P}(m, s, \chi)$ by maximising the number of microstates with which it is consistent, after coarse graining. This corresponds to maximising the mixing entropy~\citep{RobertSommeria91}
\begin{equation}
    S = -\int \dd s \int \dd \chi \int \dd m \mathcal{P}(m, s, \chi) \ln \mathcal{P}(m, s, \chi).\label{entropy_appendix}
\end{equation}Maximisation of $S$ is subject to the constraints of fixed total probability (i.e., the normalisation of $P$),
\begin{equation}
    \int \dd s \int \dd \chi \, \mathcal{P}(m, s, \chi) = 1,\label{lambda_appendix}
\end{equation}fixed potential energy $E_{\mathrm{pot}}$,
\begin{equation}
    \int \dd m\int \dd s \int \dd \chi \, \mcE(mg,s,\chi) \mathcal{P}(m, s, \chi) = E_{\mathrm{pot}},\label{betaT_appendix}
\end{equation}and, for all $s$ and $\chi$, a fixed mass of fluid $M(s, \chi) \dd s \dd \chi$ having both a value of $s$ in the range $s$ to $s+\dd s$ and a value of $\chi$ in the range $\chi$ to $\chi+\dd \chi$,
\begin{equation}
    \int \dd m \mathcal{P}(m, s, \chi) = M(s,\chi).\label{mu_appendix}
\end{equation}The equivalence of the above constraints to the corresponding ones in Section~\ref{sec:LyndenBellEquations} [viz.,~\eqref{lambda},~\eqref{betaT} and~\eqref{mu}] is readily demonstrated by substituting 
\begin{equation}
    \mathcal{P}(m,s,\chi)=\int \dd \mu \mathcal{P}(m,\mu)\delta (s-s(\mu))\delta (\chi-\chi(\mu))\label{P(s,chi)_P(m_0)}
\end{equation}and evaluating integrals over $s$ and $\chi$. Substitution of~\eqref{P(s,chi)_P(m_0)} into the expression~\eqref{entropy_appendix} for the thermodynamic entropy $S$ yields
\begin{equation}
     S = -\int \dd \mu \int \dd m \mathcal{P}(m, \mu) \ln \int \dd \mu' \mathcal{P}(m, \mu')\delta (s(\mu)-s(\mu'))\delta (\chi(\mu)-\chi(\mu')).
\end{equation}This reduces to~\eqref{entropy} provided that each value of $\mu$ has a different pair of values of $s$ and $\chi$: in that case, both delta functions on the second line are simultaneously non-zero only when ${\mu'=\mu}$. $\mathcal{P}(m,\mu)$ can then be brought outside the $\mu'$ integral, leaving [up to an additive constant that, after application of~\eqref{mu}, does not depend on $\mathcal{P}(m,\mu)$]
\begin{equation}
     S = -\int \dd \mu \int \dd m \mathcal{P}(m, \mu) \ln \mathcal{P}(m, \mu),
\end{equation}which is~\eqref{entropy}. The solution of the constrained maximisation is [cf.~\eqref{P}]
\begin{equation}
    \mathcal{P}(m, s, \chi) = \frac{\displaystyle e^{\displaystyle-\beta_T [\mathcal{E}(mg, s, \chi)-\psi(s,\chi)]}}{\displaystyle\int \dd s' \int \dd \chi' e^{\displaystyle-\beta_T [\mathcal{E}(m, s', \chi')-\psi(s',\chi')]} },\label{P_appendix}
\end{equation}where the Lagrange multipliers $\beta_T$ and $\psi (s,\chi)$ are determined by~\eqref{betaT} and~\eqref{mu}, respectively, for given $E$ and $M(s,\chi)$.

\section{The minimum-energy state of Section~\ref{sec:ground_state_theory} as the ${\beta_T\to\infty}$ limit of Lynden-Bell statistical mechanics\label{sec:LyndenBelltoLSA}}

On physical grounds, we expect that the $\beta_T\to \infty$ limit of~\eqref{P}, i.e., the limit of zero statistical mechanical temperature, corresponds to the smallest energy permitted by the system, i.e., the result of solving the linear sum assignment (LSA) problem (Section~\ref{sec:ground_state_theory}). In this appendix, we verify that the solution to the LSA problem is indeed recoverable from~\eqref{P} in the $\beta_T\to\infty$ limit.

\subsection{Discrete problem: non-degenerate case\label{app:LB_to_LSA}}

Because the LSA problem is discrete, it is strictly the $\beta_T\to \infty$ limit of~\eqref{lambda},~\eqref{betaT},~\eqref{mu} and~\eqref{P} only after they are discretised over a small but finite scale $\Delta m$ (we consider reversing the order of the $\Delta m\to0$ and $\beta_T\to\infty$ limits in Section~\ref{sec:continuouslimit}). We adopt the economical notation ${\mathcal{P}_{ij}\equiv \mathcal{P}(m_j, m_{i})}$, $\mathcal{E}_{ij}\equiv\mathcal{E}(m_j, s_i, \chi_i)=\mathcal{E}(m_j, m_{i})$ and $\psi_i \equiv \psi(m_{i})$. Converting integrals to sums,~\eqref{lambda},~\eqref{mu}~\eqref{betaT} and~\eqref{Pfull} become
\begin{equation}
    \Delta m \sum_i \mathcal{P}_{ij} = 1,\label{lambda_discrete}
\end{equation}
\begin{equation}
    \Delta m \sum_j \mathcal{P}_{ij} = 1,\label{mu_discrete}
\end{equation}
\begin{equation}
    \Delta m^2 \sum_{ij} \mcE_{ij} \mathcal{P}_{ij} = E_{\mathrm{pot}}\label{betaT_discrete}
\end{equation}and
\begin{equation}
    \mathcal{P}_{ij} = \frac{1}{\Delta m}\frac{\displaystyle e^{\displaystyle-\beta_T (\mathcal{E}_{ij}-\psi_i)}}{\displaystyle\sum_k e^{\displaystyle-\beta_T (\mathcal{E}_{kj}-\psi_k)} },\label{P_discrete_notilde}
\end{equation}respectively. We define ${\tilde{\mcE}_{ij} = \mcE_{ij} - a_{j} - b_{i}}$ and ${\tilde{\psi}_i = {\psi_i - b_{i}}}$, whence
\begin{equation}
    \mathcal{P}_{ij} =\frac{1}{\Delta m}\frac{\displaystyle e^{\displaystyle-\beta_T (\tilde{\mathcal{E}}_{ij}-\tilde{\psi}_i)}}{\displaystyle\sum_k e^{\displaystyle-\beta_T (\tilde{\mathcal{E}}_{kj}-\tilde{\psi}_k)} }.\label{P_discrete}
\end{equation}The advantage to introducing the ``tilded'' variables in~\eqref{P_discrete} is that we may always choose the functions $b_{i}$ and $a_{j}$ to be such that $\tilde{\mcE}_{ij}$ is the normal form of the cost matrix ${\mcE}_{ij}$ [see~\eqref{normalform}]. As explained in Section~\ref{sec:HungarianTheory}, this means that $\tilde{\mcE}_{ij}\geq 0$ and has at least one zero in each row and column, or, equivalently, there exists at least one bijection $\sigma$ such that $\tilde{\mcE}_{i\sigma(i)} = 0$ for all~$i$ [$j = \sigma(i)$ constitutes an optimal assignment for the LSA problem]. 

In the simplest case, $\tilde{\mcE}_{ij}$ has exactly one zero in each row and column. We expect this to be the generic case, and, indeed, this is true for all the profiles examined in Section~\ref{sec:HungarianResults}. These zeros define the unique optimal assignment $j=\sigma (i)$---taking $\tilde{\psi}_i=0$ [the natural choice, as with only one zero in each row and column, the ``exclusion principle'' is obeyed automatically and the constraint~\eqref{mu_discrete} becomes superfluous] then implies that
\begin{equation}
    \mathcal{P}_{ij} \to \begin{cases}
  1/\Delta m  & \mathrm{if}\,\,\tilde{\mcE}_{ij}=0, \\
  0  & \mathrm{otherwise}.\label{Pmu=0}
\end{cases}
\end{equation}Thus, in the $\beta_T\to\infty$ limit, the system is certain to found in the state given by the optimal assignment $\sigma$ for the LSA problem. Equation~\eqref{Pmu=0} straightforwardly satisfies~\eqref{mu_discrete}---the sum picking out the single value of $j$ for which $\mathcal{P}_{ij}\neq 0$ for any given $i$---which justifies formally our setting ${\tilde{\psi}_i=0}$. Substituting~\eqref{Pmu=0} into~\eqref{betaT_discrete} yields 
\begin{equation}
    E=\Delta m \sum_i \mcE_{i\sigma(i)}=E_{\mathrm{min}}.
\end{equation}

\subsection{Discrete problem: degenerate case\label{app:LB_to_LSA_degenerate}}

In principle, $\tilde{\mcE}_{ij}$ may have more than one zero in each row and column: these may either be part of other optimal assignments or not be part of any optimal assignment. In such cases,~\eqref{Pmu=0} becomes
\begin{equation}
    \mathcal{P}_{ij} \to \frac{1}{\Delta m}\frac{\displaystyle a_{ij}F_i}{\displaystyle\sum_k a_{kj} F_k}\,\, \mathrm{as}\,\,\beta_T\to\infty\label{P_discrete_zeros}
\end{equation}where $a_{ij}=1$ if $\tilde{\mcE}_{ij}=0$ and $a_{ij}=0$ otherwise, and ${F_i = e^{\beta_T \tilde{\psi}_i}}$ is the quantity sometimes known as ``fugacity'' [in writing~\eqref{P_discrete_zeros}, we assumed that $\tilde{\psi}_i<0$; we were free to do so because taking $\tilde{\psi}_i\to \tilde{\psi}_i+C$ for all $i$ with $C$ a constant leaves~\eqref{P_discrete} unchanged]. The values of $F_i$ are determined by~\eqref{mu_discrete}, which requires that
\begin{equation}
    \sum_{j} \frac{\displaystyle a_{ij}F_i}{\displaystyle\sum_k a_{kj} F_k} = 1,\quad \forall i,\label{mu_soln}
\end{equation}an equation that may be solved iteratively for $F_i$, and hence $\mathcal{P}_{ij}$. The possible outcomes of such a procedure are constrained by the \textit{Birkhoff--von Neumann theorem}, which states that any doubly stochastic matrix---one whose rows and columns all sum to unity---can be expressed as a sum of permutation matrices with positive weights that also sum to unity. From~\eqref{lambda_discrete} and~\eqref{mu_discrete}, $\mathcal{P}_{ij}$ is a doubly stochastic matrix, so that
\begin{equation}
    \mathcal{P}_{ij} = \frac{1}{\Delta m}\sum_k \theta_k S^{(k)}_{ij}, \quad \theta_k>0,\quad \sum_k\theta_k=1,\label{bvn}
\end{equation}where $S^{(k)}_{ij}$ are permutation matrices.  In the $\beta_T\to\infty$ limit, $\mathcal{P}_{ij}\propto a_{ij}$ [by~\eqref{P_discrete_zeros}], so $\{S^{(k)}_{ij}\}$ is the maximal set of permutations that vanish wherever $\tilde{\mcE}_{ij}> 0$, which we recognise as the optimal assignments of the LSA problem: with $\sigma^{(k)}$ the $k$th optimal assignment, $S^{(k)}_{ij}$ is the matrix representation of  $\sigma^{(k)}$, i.e., $S^{(k)}_{ij}= \delta_{j\sigma^{(k)}(i)}$. It follows that $\mathcal{P}_{ij}=0$ for any $ij$ that is not part of an optimal assignment, even if $\tilde{\mcE}_{ij}=0$, and that
\begin{equation}
    E_{\mathrm{pot}} = \Delta m \sum_{ijk} \mcE_{ij} \theta_k S^{(k)}_{ij}= \Delta m \sum_k \theta_k\sum_{i}\mcE_{i\sigma^{(k)}(i)} = E_{\mathrm{min}}.
\end{equation}

\subsection{Continuous problem\label{sec:continuouslimit}}

We have so far determined that the discretised probability-density function $\mathcal{P}_{ij}$ reproduces the solution of the LSA problem (or some weighted combination of its solutions if they are not unique) as~$\beta_T\to \infty$ at fixed~$\Delta m$. This is not, however, the large-$\beta_T$ limit of the continuous system~\eqref{betaT},~\eqref{mu} and~\eqref{P}: discretising the integrals is valid only if $\Delta m$ is small compared to the scale of variation of the continuous probability density~$\mathcal{P}(m,\mu)$, which shrinks to zero as ${\beta_T\to \infty}$. Nonetheless, the $\beta_T\to \infty$ limit of the continuous Lynden-Bell equations is equal to the $\Delta m \to 0$ limit of the \textit{coarse-grained} solution of the LSA problem, provided it is unique, as follows. 

Let $\tilde{\mcE}(m,\mu)$ be the continuous limit of $\tilde{\mcE}_{ij}$ obtained by taking $\Delta m\to 0$. As $\beta_{T}\to \infty$, it is clear that
\begin{equation}
    \mathcal{P}(m,\mu)\propto e^{\displaystyle-\beta_T \tilde{\mcE}(m,\mu)}
\end{equation}becomes increasingly localised to the curve in $(m,\mu)$ space on which $\tilde{\mcE}(m,\mu)=0$. The same is true for the coarse-grained discrete solution $\mathcal{P}_{ij}$ of the LSA problem, which we define by
\begin{equation}
    \langle \mathcal{P}\rangle \equiv \frac{1}{2n}\int^{m+n\Delta m}_{m-n\Delta m} \dd m' \int^{\mu+n\Delta m}_{\mu-n\Delta m} \dd \mu'\sum_{ij} \mathcal{P}_{ij} \delta (\mu-m_{i})\delta (m-m_{j}),\label{coarsegrainP}
\end{equation}with $n$ a constant satisfying $1\ll n \ll N$. As $\beta_T\to\infty$, both $\mathcal{P}(m,\mu)$ and $\mathcal{P}$ tend to a delta distribution centred on the optimal-solution curve. The most general form such a distribution can take is
\begin{equation}
    A(m,\mu)\delta (F(m,\mu))\label{generalP}
\end{equation}where $F$ vanishes along the optimal-solution curve and, without loss of generality, can be taken to have~${|\bnabla F|=1}$ there [$F$ contributes no further degrees of freedom in the expression~\eqref{generalP}]. However, as we show in Appendix~\ref{app:geometry}, the function $A(m,\mu)$ is uniquely determined by the conditions~\eqref{lambda_discrete} and~\eqref{mu_discrete}, which must be satisfied by both $\langle P\rangle$ and $P$. It follows that their large-$\beta_T$ limits are the same. A practical consequence of this is that one may determine the horizontal composition of 2D minimum-energy ground states from the $\beta_T\to \infty$ limit of $\mathcal{P}(m,\mu)$ (as in figure~\ref{fig:2Drestacking}).

\section{Uniqueness of doubly stochastic delta distributions\label{app:geometry}}

In this appendix, we show that the delta distribution
\begin{equation}
    \mcP(x,y) = A(x,y)\delta (F(x,y)),\label{app_generalP}
\end{equation}where $A>0$, and $F(x,y)=0$ on a piecewise-once-differentiable curve $\mathcal{C}$, is unique under the conditions that
\begin{enumerate}[(i)]
    \item $\mathcal{C}$ intersects every horizontal and vertical line in the unit square at least once;
    \item $\mathcal{C}$ does not admit any cycles formed by traversing horizontal and vertical lines (in the context of~\eqref{coarsegrainP}, this corresponds to uniqueness of the optimal assignment);
    \item $\mcP(x,y)$ satisfies the integral constraints
\begin{equation}
    \int_0^1 \dd x \mcP(x,y)= 1,\label{app_intPx}
\end{equation}and
\begin{equation}
    \int_0^1 \dd y \mcP(x,y) = 1.\label{app_intPy}
\end{equation}
\end{enumerate}

Without loss of generality, we can take $|\bnabla F|=1$; any other choice amounts to a redefinition of $A$. Then,~\eqref{app_intPx} and~\eqref{app_intPy} become
\begin{equation}
    \sum_i \frac{A(x_i(y), x)}{ |t_x(x_i(y), y)| }= 1,\quad\sum_i \frac{A(x, y_i(x))}{ |t_y(x, y_i(x))| }= 1,\label{app_sumsA}
\end{equation}respectively, where $y_i(x)$ is the $y$-coordinate of the $i$th intersection of $\mathcal{C}$ with the vertical line through ($x$,0), $x_i(y)$ is the $x$-coordinate of the $i$th intersection of $\mathcal{C}$ with the horizontal line through $(0,y)$, and 
\begin{equation}
    (t_x, t_y) =\left(\frac{\p F}{\p y},-\frac{\p F}{\p x}\right)
\end{equation}is the unit tangent vector to $\mathcal{C}$.

For each point on $\mathcal{C}$,~\eqref{app_sumsA} each provide an equation relating $A$ at that point to the value of $A$ at each other point on $\mathcal{C}$ with the same $x$ or $y$ coordinate. In turn,~\eqref{app_sumsA} provide one additional equation for each such point of intersection, relating $A$ evaluated there to $A$ evaluated at other points on $\mathcal{C}$ with the same~$x$ or $y$-coordinate, and so on for the points with which those points share an~$x$ or~$y$~coordinate. There are three possible outcomes for this proliferation of coupled equations: (a) a stage of the process is reached when all new equations only involve a single point on $\mathcal{C}$, and so the proliferation is arrested when $n+1$ independent (prior to the specification of the tangent vectors) linear equations couple the values of $A$ at $n$ different points; (b) proliferation of equations ceases, but the $n$ points on $\mathcal{C}$ form one or more cycles of the sort envisioned in (ii) and are therefore coupled by fewer than $n+1$ equations; (c) proliferation of equations never ceases---in this case, infinitely many points are coupled, some of them arbitrarily close to each other (infinite cycle). By condition~(ii), we restrict attention to case (a). Then, the first $n$ of the $n+1$ independent linear equations can be solved for the values of $A$ at each of the $n$ points, so the function $A(x,y)$ is indeed unique given conditions (i)-(iii), as claimed. The final equation gives a constraint on the tangent vectors, which must be satisfied in order for a delta distribution~\eqref{app_generalP} satisfying conditions (i)-(iii) to exist.

\begin{figure}
    \centering
    \includegraphics[width=.6\columnwidth]{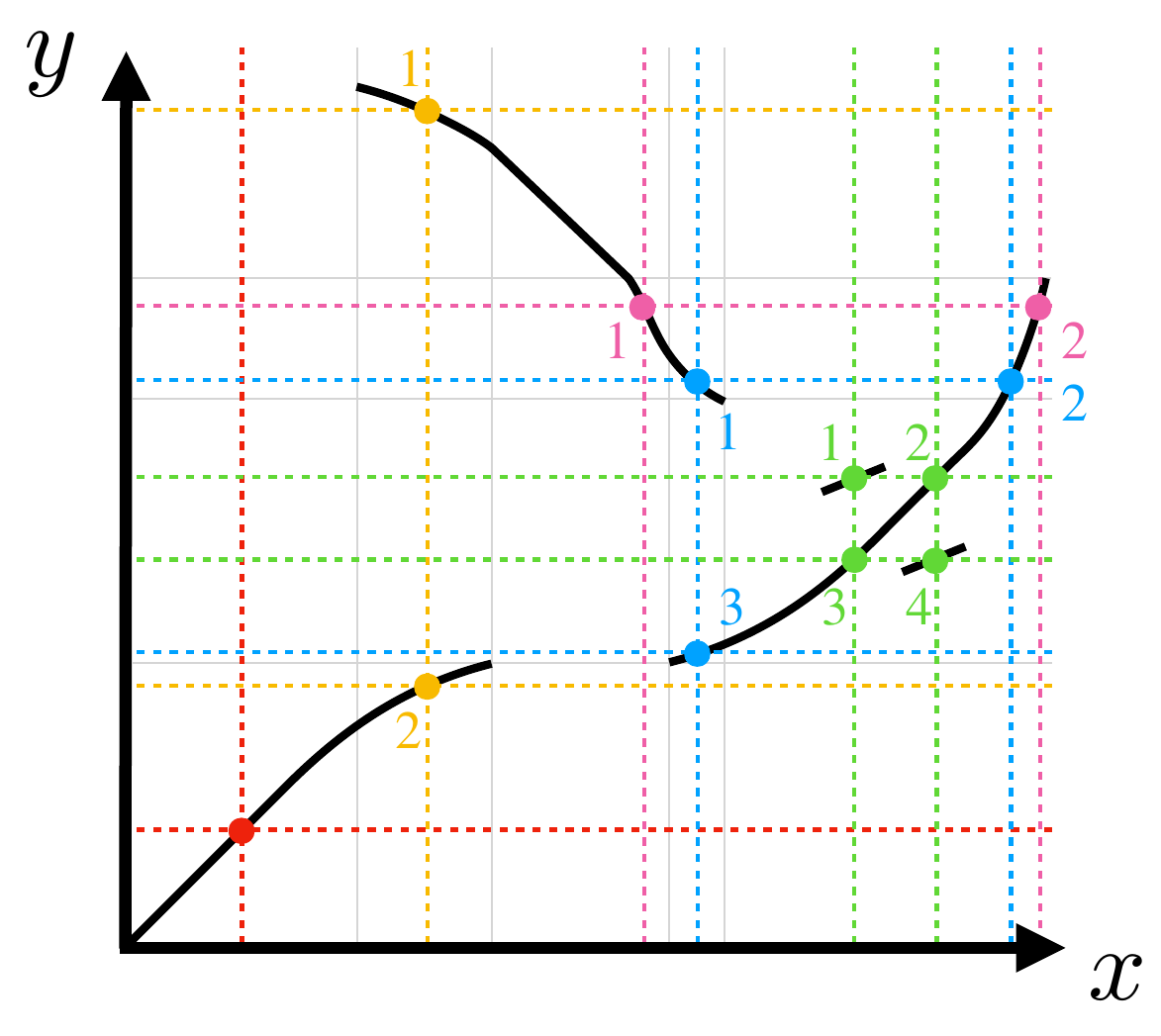}
    \caption{Visualisation of the construction of the delta distribution~\eqref{app_generalP} for the curve $\mathcal{C}$ by solving~\eqref{app_sumsA}. Coloured dashed lines indicate the relevant vertical and horizontal lines used in the evaluation of~\eqref{app_sumsA}. For ease of comparison between different parts of $\mathcal{C}$, the grey lines mark certain points at which $\mathcal{C}$ has discontinuities.}
    \label{fig:stochasticdelta}
\end{figure}

For clarity, we illustrate the procedure described above for the example curve $\mathcal{C}$ shown in figure~\ref{fig:stochasticdelta}.

\subsection{Case where vertical and horizontal lines intersect $\mathcal{C}$ once only---red point in figure~\ref{fig:stochasticdelta}}

In this case,~\eqref{app_sumsA} become
\begin{equation}
    \frac{A}{|t_y|} = 1,\quad \frac{A}{|t_x|}=1\implies A = |t_x| = |t_y| = \frac{1}{\sqrt{2}},
\end{equation}from which it follows that
\begin{equation}
    \frac{\dd y}{\dd x}=\pm 1
\end{equation}must be satisfied at the point in question. As claimed, we find the unique value of $A$ and a condition on the tangent vector.

\subsection{Case where one vertical line intersects $\mathcal{C}$ twice---yellow points in figure~\ref{fig:stochasticdelta}}

With subscripts indicating evaluation at points according to the labelling scheme in figure~\ref{fig:stochasticdelta},~\eqref{app_sumsA} yield
\begin{equation}
    \frac{A_1}{|t_{y1}|} = 1,\quad \frac{A_2}{|t_{y2}|} = 1,\quad \frac{A_1}{|t_{x1}|}+\frac{A_2}{|t_{x2}|} = 1,
\end{equation}from which it follows that
\begin{equation}
    A_1 = |t_{y2}|,\quad A_2=|t_{y2}|,\quad \bigg|\frac{\dd y}{\dd x}\bigg|_1+\bigg|\frac{\dd y}{\dd x}\bigg|_2=1.
\end{equation}As claimed, we find the unique values of $A_1$ and $A_2$ and a condition on the tangent vector at the relevant points.

\subsection{Case where one horizontal line intersects $\mathcal{C}$ twice---pink points in figure~\ref{fig:stochasticdelta}}

In this case,~\eqref{app_sumsA} yield
\begin{equation}
    \frac{A_1}{|t_{x1}|} = 1,\quad \frac{A_2}{|t_{x2}|} = 1,\quad \frac{A_1}{|t_{y1}|}+\frac{A_2}{|t_{y2}|} = 1,
\end{equation}from which it follows that
\begin{equation}
    A_1 = |t_{x2}|,\quad A_2=|t_{x2}|,\quad \bigg|\frac{\dd y}{\dd x}\bigg|^{-1}_1+\bigg|\frac{\dd y}{\dd x}\bigg|^{-1}_2=1.
\end{equation}Again, we find the unique values of $A_1$ and $A_2$ and a condition on the tangent vector at the relevant points.

\subsection{Case where both one vertical and one horizontal line intersects $\mathcal{C}$---blue points in figure~\ref{fig:stochasticdelta}}

In this case, three points are coupled by~\eqref{app_sumsA}, which yield
\begin{equation}
    \frac{A_2}{|t_{x2}|} = 1,\quad \frac{A_3}{|t_{y3}|} = 1,\quad \frac{A_1}{|t_{x1}|}+\frac{A_3}{|t_{x3}|} = 1,\quad \frac{A_1}{|t_{y1}|}+\frac{A_2}{|t_{y2}|} = 1,
\end{equation}from which it follows that the values of $A_i$ are
\begin{equation}
    A_1 = \left(1-\frac{|t_{x2}|}{|t_{y2}|}\right)|t_{y1}|,\quad A_2=|t_{x2}|,\quad A_3=|t_{y3}|
\end{equation}and the condition on $\mathcal{C}$ is
\begin{equation}
    \bigg|\frac{\dd y}{\dd x}\bigg|_3 = \bigg|\frac{\dd y}{\dd x}\bigg|_1\left[ \bigg|\frac{\dd y}{\dd x}\bigg|^{-1}_1+\bigg|\frac{\dd y}{\dd x}\bigg|_2^{-1}-1 \right].
\end{equation}\newline

\subsection{Case in which $\mathcal{C}$ has a simple cycle---green points in figure~\ref{fig:stochasticdelta}}

To illustrate how uniqueness fails if condition (ii) is not satisfied, we consider as a final example the case where four points form a cycle along horizontal and vertical lines. Equations~\eqref{app_sumsA} yield
\begin{equation}
    \frac{A_1}{|t_{y1}|}+\frac{A_2}{|t_{y2}|} = 1,\quad \frac{A_1}{|t_{x1}|}+\frac{A_3}{|t_{x3}|} = 1,\quad \frac{A_3}{|t_{y3}|}+\frac{A_4}{|t_{y4}|} = 1,\quad \frac{A_2}{|t_{x2}|}+\frac{A_4}{|t_{x4}|} = 1.
\end{equation}These equations can be reduced to the matrix equation
\begin{equation}
    \begin{pmatrix}
\displaystyle\frac{1}{|t_{y1}|} & -\displaystyle\frac{|t_{x2}|}{|t_{y2}||t_{x4}|}\vspace{2mm}\\
-\displaystyle\frac{|t_{x3}|}{|t_{y3}||t_{x1}|} & \displaystyle\frac{1}{|t_{y4}|}
\end{pmatrix}
\begin{pmatrix}
A_1\\
A_4
\end{pmatrix}
=
\begin{pmatrix}
1-\displaystyle\frac{|t_{x2}|}{|t_{y2}|}\vspace{2mm}\\
1-\displaystyle\frac{|t_{x3}|}{|t_{y3}|}
\end{pmatrix},
\end{equation}which has unique solutions for $A_1$ and $A_4$ unless the determinant of the matrix is zero, which requires
\begin{equation}
    \bigg|\frac{\dd y}{\dd x}\bigg|_1\bigg|\frac{\dd y}{\dd x}\bigg|_4=\bigg|\frac{\dd y}{\dd x}\bigg|_2\bigg|\frac{\dd y}{\dd x}\bigg|_3.\label{derivatives}
\end{equation}In this case, the $A_i$ are not determined uniquely. It is straightforwardly verified that~\eqref{derivatives} is precisely the condition for the four points separated by an infinitesimal distance along $\mathcal{C}$ from the green ones to \textit{also} be on a cycle of horizontal and vertical lines. Thus, it appears that condition (ii) is actually too strong a condition to guarantee uniqueness of~\eqref{app_generalP}: cycles are permissible if they only exist in a set of measure zero in $x$ and $y$. It follows that cases where the optimal solution to the assignment problem is non-unique on a set of measure zero only do not violate the conclusion of Section~\ref{sec:continuouslimit}. 

\bibliographystyle{jpp}
\bibliography{bibliography}

\end{document}